\pdfoutput=1
\documentclass[useAMS,usenatbib]{mn2e}
\usepackage{graphicx}  
\usepackage{dcolumn}  
\usepackage{bm}       
\usepackage{amssymb,amsmath}  
\usepackage{color}

\newcommand{\beq}{\begin{equation}}
\newcommand{\eeq}{\end{equation}}
\newcommand{\lexp}{\mathop{\langle}}
\newcommand{\rexp}{\mathop{\rangle}}
\newcommand{\mg}{\big<}
\newcommand{\md}{\big>}
\newcommand{\Dirac}{\delta_{\rm D}}

\newcommand{\beqa}{\begin{eqnarray}}
\newcommand{\eeqa}{\end{eqnarray}}

\def\vk{{\hbox{\bf k}}}
\def\vq{{\hbox{\bf q}}}

\def\vp{{\hbox{\bf p}}}

\def\Mpc{\, h^{-1} \, {\rm Mpc}}

\def\Gpccube{\, h^{-3} \, {\rm Gpc}^3}
\def\kvecMpc{\, h \, {\rm Mpc}^{-1}}
\def\Msun{\,h^{-1}\,{\rm M_{\odot}}}
\def\ltsima{$\; \buildrel < \over \sim \;$}   
\def\gtsima{$\; \buildrel > \over \sim \;$}   
\def\simlt{\lower.5ex\hbox{\ltsima}}   
\def\simgt{\lower.5ex\hbox{\gtsima}}

\newcommand{\gapprox}{{_>\atop^\sim}} 
\newcommand{\textfb}[1]{\textcolor{black}{#1}}
\newcommand{\sigmav}{\sigma_{\rm d}}
\newcommand{\MPTbreeze}{{\texttt{MPTbreeze}}}
\newcommand{\RegPT}{{\texttt{RegPT}}}

\title[MPTbreeze]
{\MPTbreeze: A fast renormalized perturbative scheme}
\author[Crocce, Scoccimarro \& Bernardeau]{Mart\'in
  Crocce$^1$\thanks{E-mail:martincrocce@gmail.com},
  Rom\'an Scoccimarro$^2$ and Francis Bernardeau$^3$ \\ \\
  $^1$Institut de Ci\`encies de l'Espai (IEEC-CSIC), 
  E-08193 Bellaterra (Barcelona), Spain. \\
  $^2$ Center for Cosmology and Particle Physics, Department of
  Physics, New York University, New York, NY 10003, USA. \\
  $^3$ Institut de Physique Th\'eorique, CEA, IPhT, F-91191
  Gif-sur-Yvette, France.\\
  \ \ CNRS, URA 2306, F-91191, Gif-sur-Yvette, France.
 \\}
\begin{document}

\date{\today}
\pagerange{1--10} \pubyear{2012}
\maketitle

\begin{abstract} \\
We put forward and test a simple description of multi-point
propagators (MP), which serve as building-blocks to calculate the
nonlinear matter power spectrum. On large scales these propagators
reduce to the well-known kernels in standard perturbation theory,
while at smaller scales they are suppresed due to nonlinear couplings.
Through extensive testing with numerical simulations we find that this
decay is characterized by the same damping scale for both two and
three-point propagators. In turn this transition can be well modeled 
with resummation results that exponentiate one-loop computations. 
For the first time, we measure the four components of the non-linear (two-point) propagator 
using dedicated simulations started from two independent random Gaussian fields for positions 
and velocities, verifying in detail the fundamentals of propagator resummation. 

We use these results to develop an implementation of the MP-expansion
for the nonlinear power spectrum that only requires seconds to
evaluate at BAO scales. To test it we construct six suites of large
numerical simulations with different cosmologies. From these and {\tt
  LasDamas} runs we show that the nonlinear power spectrum can be
described at the $\lesssim 2\%$ level at BAO scales  for redshifts in
the range $\left[0-2.5\right]$. We make a public release of the {\MPTbreeze} 
code with the hope that it can be useful to the community.

\end{abstract}


\section{Introduction}

Ongoing and future galaxy redshift surveys will render the large scale
structure of the Universe with unprecedented detail thanks to a
combination of redshift depth and large survey area. Among them are the
Sloan Digital Sky Survey III\footnote{\tt www.sdss3.org}, the WiggleZ
survey\footnote{\tt wigglez.swin.edu.au}, the    
Dark Energy Survey\footnote{\tt www.darkenergysurvey.org},  the Physics of the
Accelerating Universe collaboration\footnote{\tt www.pausurvey.org} and 
ESA/Euclid survey \footnote{\tt www.euclid-ec.org}.
The main driver underneath this effort is to seed light into the present cosmic
acceleration. Various probes exist that connect different statistical aspects 
of galaxies properties to cosmological parameters, in particular to those
related to acceleration, such as the Baryon Acoustic Oscillations (BAO), Redshift
Space Distortions (RSD) or Weak Lensing (WL). However in order to maximize the
scientific outcome from this data we need to put forward precise
theoretical and/or numerical predictions, for example for 
two-point statistics. The difficulty arise because 
the most rewarding range of scales lie in the nonlinear regime
of structure formation.

\begin{table*}
\begin{center}
\begin{tabular}{lllllllllll}
\hline \\
Run   &  $\Omega_m$  &  $\Omega_b$ &   h  &  $\sigma_8$ & $n_s$ &
$L_{box}(\Mpc)$  & $N_{runs}$  \\
      &&&&&&&&       \\
FID               &  0.27   &  0.04    & 0.7  &    0.9  & 1      & 1280   & 50 \\
tilt              &         &          &      &         & 0.9  &    
1250   & 4        \\  
WMAP3             &  0.2383 &  0.0418  & 0.73 &    0.74 & 0.95 &    
1250 & 4    \\
Low-$\Omega_m$    &  0.20   &          &      &         &      &    
1250 & 4     \\
Mid-$\sigma_8$    &         &          &      &    0.8  &      &    
1250 & 4   \\
Low-$\sigma_8$    &         &          &      &    0.7  &      &    
1250& 4      \\
{\tt LasDamas}             &  0.25   &  0.04   & 0.7  &    0.8  & 1     & 2400   & 4 & \\
\\ \hline
\end{tabular}
\end{center}
\caption{Our suite of N-body simulations spanning different cosmological  models. The top entry corresponds to our largest ensemble used as a
  benchmark for testing different components of the model. The last
  entry refers in particular to four {\it Oriana} runs,   the largest
  box-size within the {\tt LasDamas} simulations. Null entries
  indicate the same value as the FID run.}
\label{cosmologies}
\end{table*}

For BAO (and RSD) these nonlinearities, 
due in large part to gravitational instabilities, are not
strong \citep{2007ApJ...664..660E,2008PhRvD..77b3533C,2008MNRAS.383..755A}. Hence the problem can be addressed using perturbative schemes
of the equations of motion in addition to numerical N-body simulations.
This is then the main goal of this paper, to 
 \textfb{propose} 
an implementation of a (resummed) perturbative expansion for
the matter power spectrum that is both accurate (percent level) and
practical (few seconds of evaluation) on large BAO scales.
And to have it tested as much as possible against numerical simulations
of different cosmological models.

The process of nonlinear structure formation can be very well traced by
gravitational N-body codes once certain requirements are met
(e.g. \cite{2010ApJ...715..104H} and references therein). These
involve, among others, good mass resolution (i.e. particle load),
fine time stepping, high starting redshift, large box-size (to include long
wavelength modes), etc. A percent level estimate of the power spectrum puts
strong constrains in these parameters which generally slow down
the numerical solver. In addition it is numerically very expensive to
explore cosmological parameter space with large high resolution simulations, although efforts in this direction
are currently ongoing \citep{2009ApJ...705..156H}.

In turn, cosmological perturbation theory (PT) is a well
defined formalism that can be applied without extra cost to any LCDM model (and even
beyond) but leads to poorly convergent results (see \cite{2002PhR...367....1B} for a review). One step beyond 
this issue was put forward by \cite{2006PhRvD..73f3520C,2006PhRvD..73f3519C}
through a systematic re-organization of the perturbative series so
called Renormalized Perturbation Theory. This
led to a better behaved expansion and more robust results \citep{2008PhRvD..77b3533C}. A 
number of similar studies with alternative methods to resum the PT expansion
quickly followed,
e.g. \cite{2007JCAP...06..026M,2008ApJ...674..617T,2008PhRvD..77f3530M,2008PhRvD..78j3521B,2008PhRvD..78h3503B,2011JCAP...06..015A}
(and more recently \cite{2011PhRvD..84d3501S,2011MNRAS.416.1703E,2012PhRvD..86d3508W} for the case of biased tracers). Overall the resulting conclusion of these works is that the matter $P(k)$
can be modeled at the percent level accuracy on weakly nonlinear scales ($k \lesssim
0.2 -0.4 \kvecMpc$ depending on redshift), improving over standard PT. Nonetheless the structure
of the solutions are complex, generally involving a set of couple
integro-differential equations or multi-dimensional integrals that in
any case require a time-scale of hours to evaluate.

In this paper we try to overcome this problem using an effective
description of the multi-point propagators introduced in
\cite{2008PhRvD..78j3521B}. The multi-point propagators are formally
defined as the infinitesimal variation of cosmic fields with respect to
the initial conditions. For Gaussian initial conditions they are
equivalent to a measure of the cross correlation of final fields with
initial configurations. On large scales, where PT is valid, the
propagators coincide with the standard kernels in the PT expansion. Towards small
scales nonlinear effects drive the propagators to zero \citep{2008PhRvD..78j3521B}. The full
dependence with time and scale is then highly non-trivial (besides the
fact that formally they have a matrix structure).
However the importance of the MP reside in the fact that they can
be used as a well behaved expansion basis for equal time correlators such as the
power spectrum, bispectrum, etc. Our effective description for the MP 
accounts only for most growing contributions. This, in turn, allows for
a rapid evaluation of the first few terms in the MP expansion of the
power spectrum. We have \textfb{thoroughly} 
tested against a large set of dedicated N-body simulations both the
prescription for the different MP (and the matrix structure) as well as the resulting prediction
for the nonlinear power spectrum. For all cosmologies investigated we
find that our approach is able
to reproduce N-body measurements at BAO scales at the $\sim 2\%$ level from low to
high redshift, with evaluation times of about ten seconds. 

This paper is organized as follows. In Sec.~\ref{sec:nbody} we
present the sets of large N-body simulations of different cosmological
models used throughout the paper. In Sec.~\ref{sec:MPexpansion} we
briefly recall the concept of multi-point propagators and their use as
expansion basis for the nonlinear matter spectrum. Section ~\ref{sec:MP} goes into
more detail with the MP, comparing our effective modeling against
measurements of two and three point propagators in our N-body
simulations. In Sec.~\ref{sec:powerspectrum} we use these results to compute the nonlinear
$P(k)$ and compare it to measurements in our fiducial ensemble, with a
detailed discussion of the code performance (evaluation time,
integration accuracy, etc). Sec.~\ref{sec:cosmoperformance} extends this comparison to power
spectrum measurements at various redshifts in our six (6) different cosmological models. 
Lastly, Sec.~\ref{sec:conclusions} contains our conclusions. 

We leave for Appendix \ref{sec:EdS} an important discussion regarding
the validity of using PT
techniques  derived within an Einstein - de Sitter cosmology
to describe arbitrary LCDM models. We carry this out with a novel approach
using numerical simulations.


\section{N-body Simulations}
\label{sec:nbody}

We now describe the set of large N-body simulations that we developed to test our theoretical predictions.
We have developed a large ensemble of high statistical significance for a fixed cosmological model and we have also carried out a set of smaller ensembles for different cosmological models. To this later set we add measurements of $P(k)$ from some of the 
${\tt LasDamas}$ simulations\footnote{http://lss.phy.vanderbilt.edu/lasdamas}.

All simulations used {\tt Gadget2} \citep{2001NewA....6...79S} to compute the gravitational evolution, and
2nd order Lagrangian Perturbation Theory (2LPT) to set up the initial conditions at $z_i=49$ \citep{1998MNRAS.299.1097S,2006MNRAS.373..369C}.

\subsection{Fiducial Cosmology}

The core testing will be done against measurements in a set of $50$ N-body simulations, each of comoving
box-size $L_{box}=1280 \Mpc$ and $640^3$ particles.  This set then
constitutes more than $100\Gpccube$ of simulated volume and will be particularly
important to test our model assumptions beyond two-point
statistics, i.e. the three-point propagator that we discuss below. 

We will refer to this set as the {\it fiducial cosmology} (FID).
The cosmological parameters and relevant information are given in
Table~\ref{cosmologies} (see \cite{2008PhRvD..77b3533C}
for more details on the simulations). The
corresponding particle mass was $m_p = 6 \times 10^{11}\Msun$. We note that both
the mass resolution and the settings employed to run {\tt Gadget2} and
{\tt 2LPT} ensure that we achieve unbiased measurements of the power
spectrum at the scales of the baryon acoustic oscillations
(\cite{2010ApJ...715..104H}).

\subsection{Cosmological Suite}

The main goal of this paper is to provide an efficient prediction for
$P(k)$ that serves across cosmological parameter space. In order to
explore this we implemented numerical simulations of 5 different cosmological
models in addition to the FID case, changing the parameters that are of most importance to large
scale clustering such as matter density $\Omega_m$, spectral tilt
$n_s$, linear amplitude of fluctuations $\sigma_8$ and more. Full
details are listed  in Table~\ref{cosmologies}. 

Each cosmological model was
simulated with four (4) runs of comoving box-size $L_{box}=1250\Mpc$ and $640^3$
particles ({\tt Gadget2} and {\tt 2LPT} settings as in the FID case above).

In addition we use some of the {\tt LasDamas} simulations (McBride et al. 2012, in preparation).  {\tt LasDamas} is a collaborative effort that run 50 boxes of 4 different resolutions each, tailored to describing the clustering of different galaxy samples in the SDSS-II survey. Here since we are interested in large-scale clustering, we only use 4 of their largest boxes, named {\it Oriana}, each with $1280^3$ particles within $L_{box}=2400\Mpc$ (particle mass $m_p=4.57 \times 10^{11}$). The outputs for {\it Oriana} sample a wide redshift range ($z=0,0.34,0.52,0.97,1.50$ and $2.52$), allowing us to test how the performance of our power spectrum prescription evolves with redshift.

\begin{figure*}
\begin{center}
\includegraphics[trim = 0cm 0cm 0cm 0cm, width=0.32\textwidth]{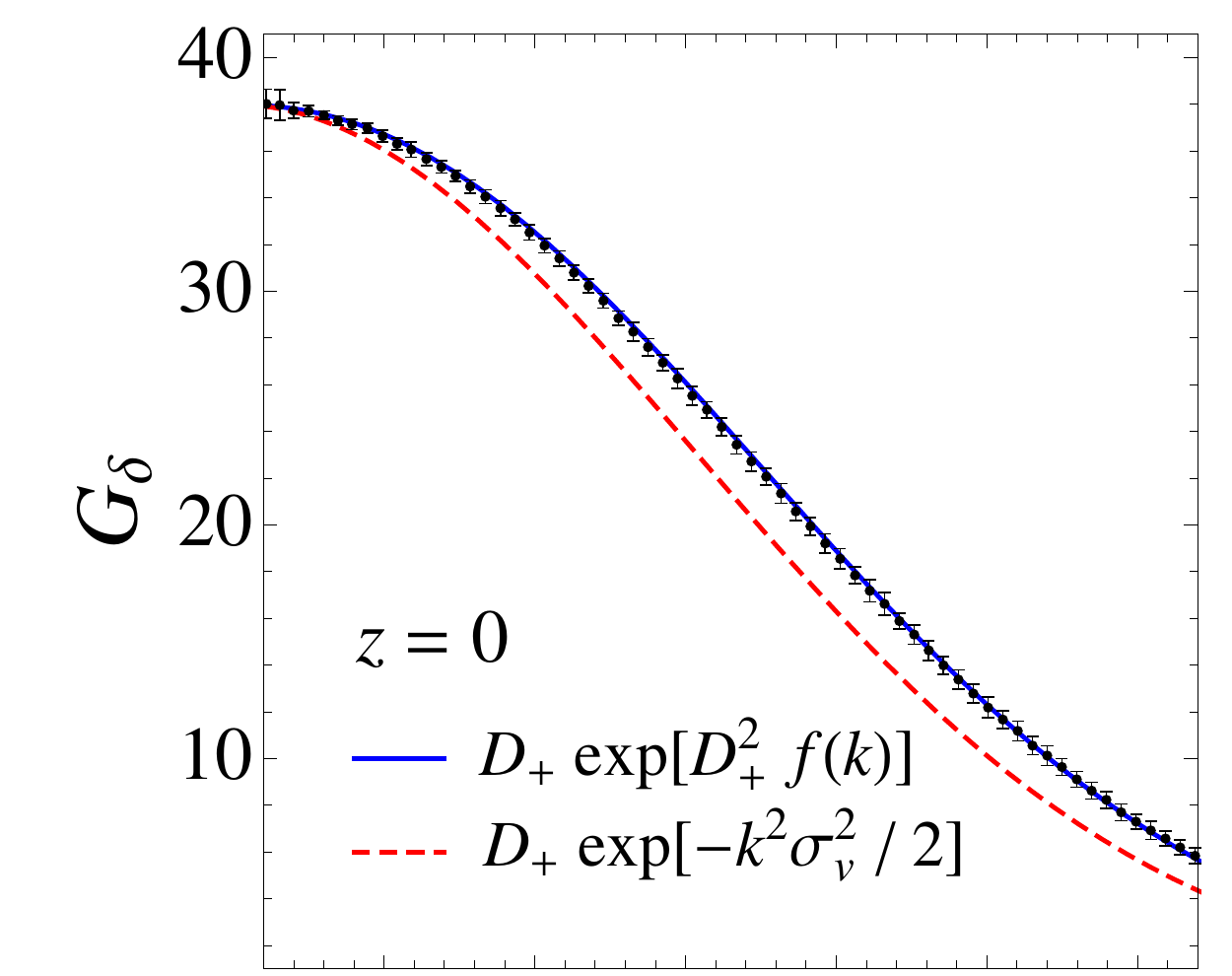}
\includegraphics[trim = 0cm 0cm 0cm 0cm, width=0.32\textwidth]{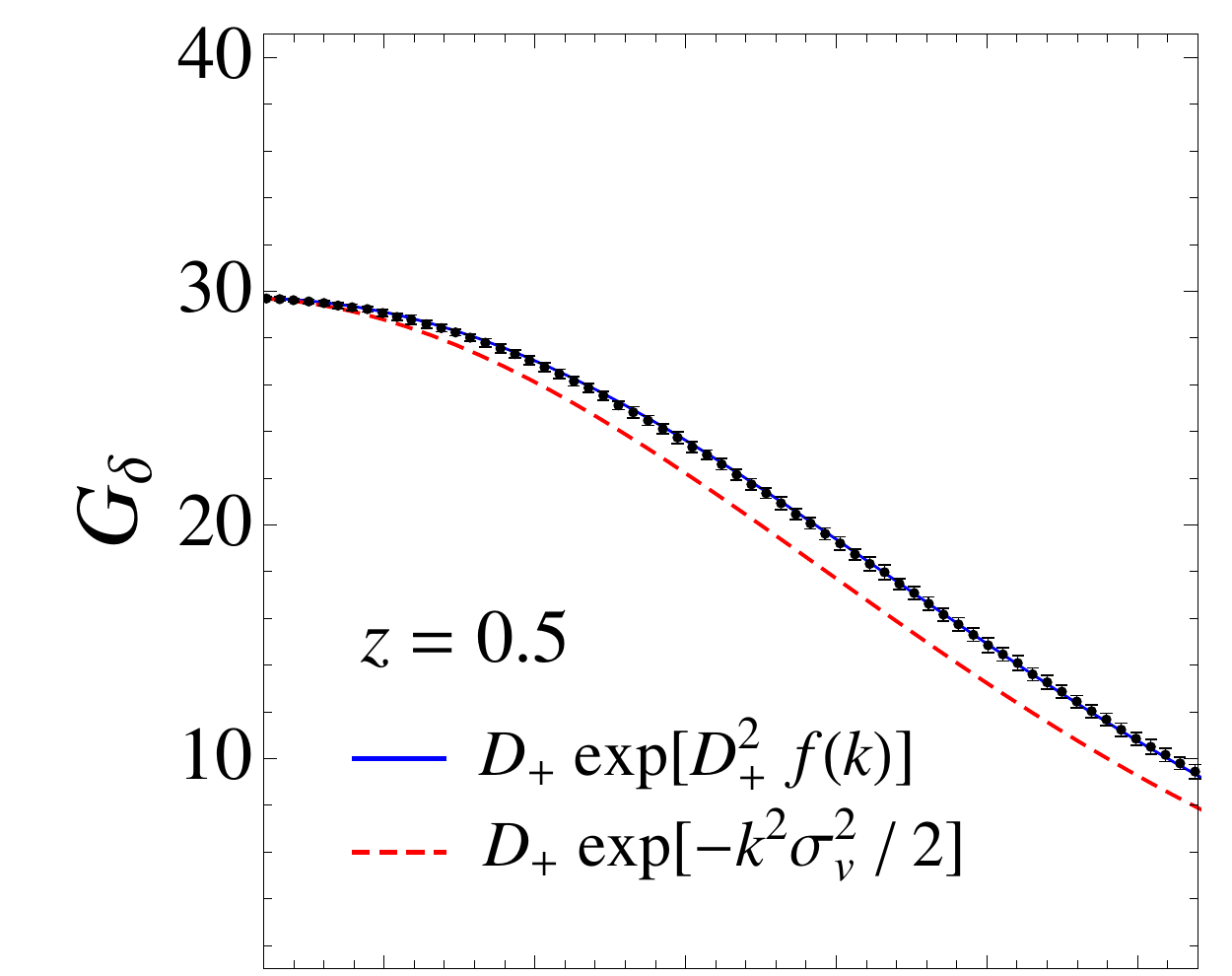}
\includegraphics[trim = 0cm 0cm 0cm 0cm, width=0.32\textwidth]{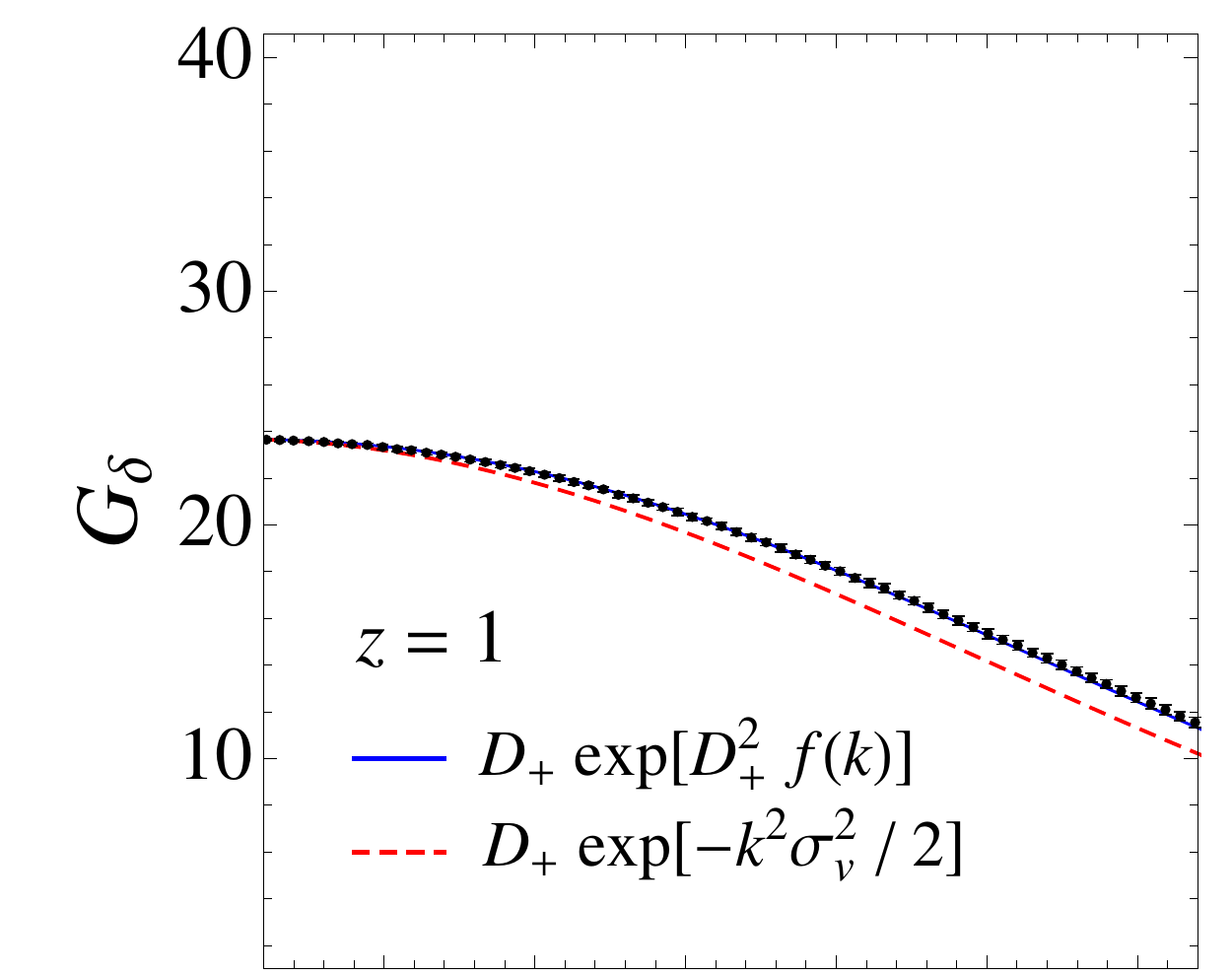} \\
\includegraphics[width=0.32\textwidth]{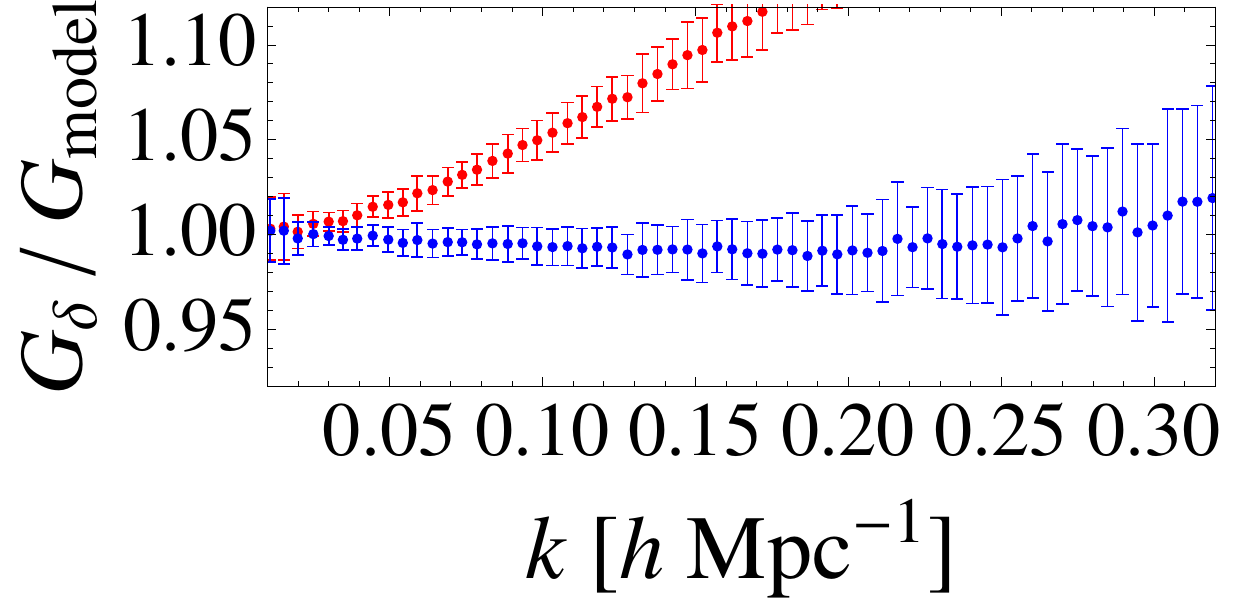}
\includegraphics[width=0.32\textwidth]{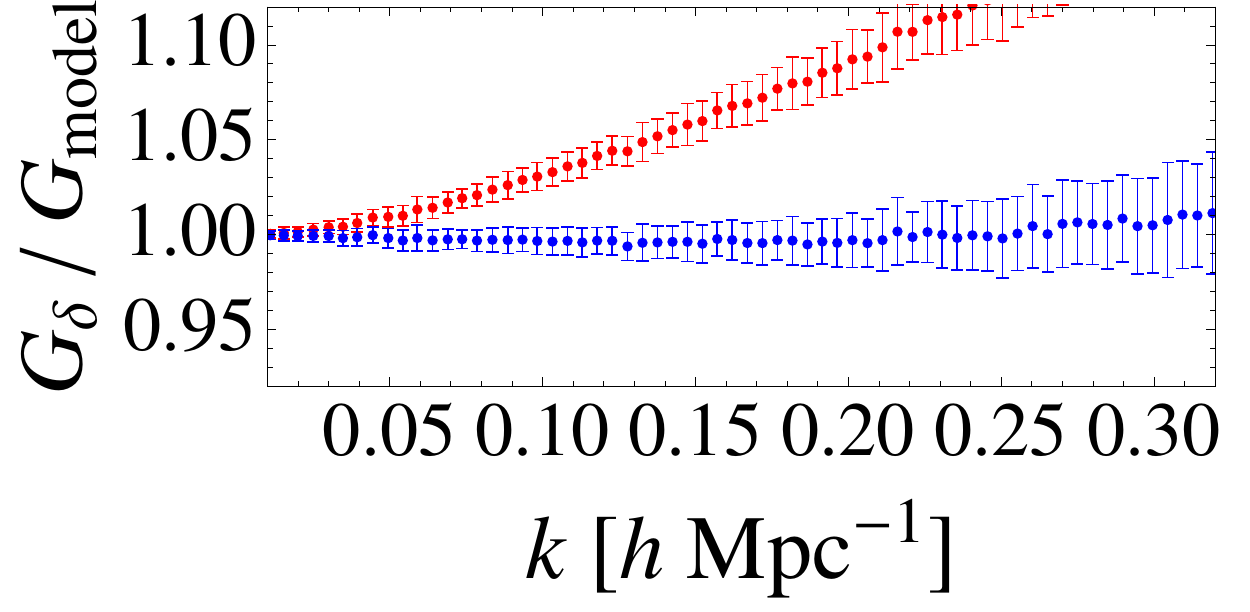}
\includegraphics[width=0.32\textwidth]{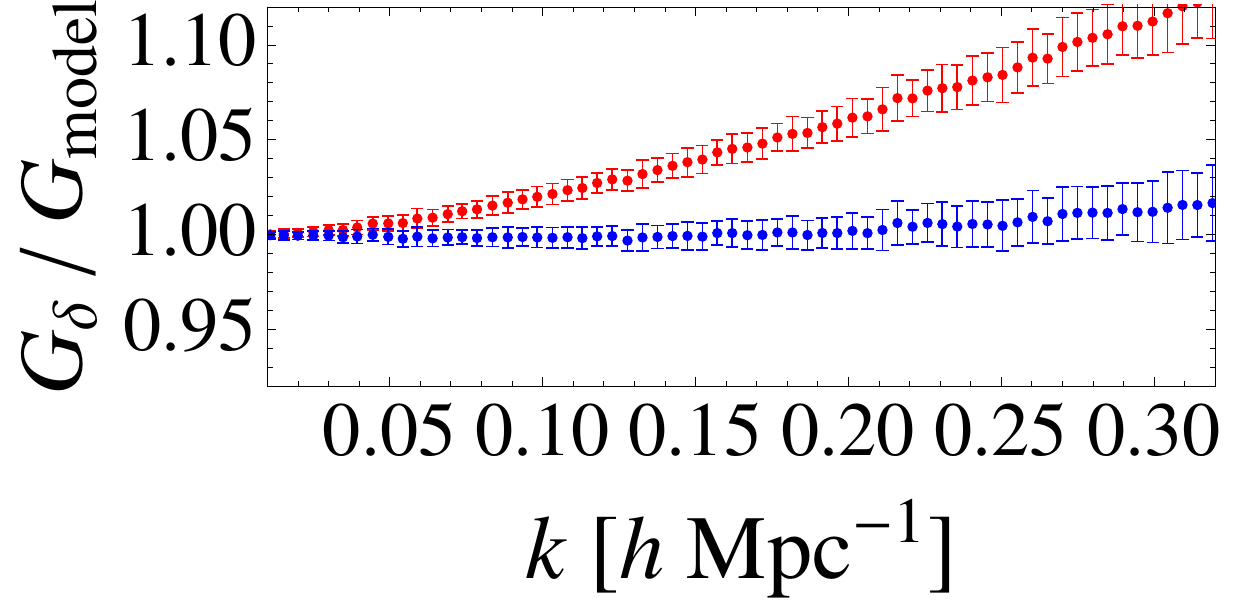}
\caption{{\it Two-point (nonlinear) propagator for the density field: } model vs. measurements
    in N-body simulations at $z=0,0.5,1$. The model in Eq.~(\ref{eq:prop}) performs
    remarkably well at all redshifts shown. The dashed line shows the corresponding
    high-$k$ limit (which is only reached at very high-$k$, not shown here). Lower panels show the ratio of the measurements to
  the two different analytic descriptions and stress that the accuracy
  of Eq.~(\ref{eq:prop}) is at the percent level. } 
\label{fig:prop}
\end{center}
\end{figure*}

\subsection{Simulations with Independent Initial Positions and Velocities}
\label{sec:mixmodeIC}

Initial conditions in an N-body simulation of a given cosmological model are fully specified
by initial (random) values of particle positions and velocities, i.e. density ${ \delta}$ and velocity ${ {\bf v}}$ perturbations\footnote{By velocity perturbations we mean the  velocity field after the Hubble flow at the given position has been  subtracted off (the so-called peculiar velocities).}.
Since in linear evolution the vorticity component of ${\bf v}$ decays in time,  the only relevant
component of the velocity field is its divergence $\theta = -\nabla \cdot {\bf v} / {\mathcal H}
f$. Moreover the linear evolution of $\delta$ and $\theta$
admits two solutions: one that grows
in time ($\propto$ to the linear growth factor)
another that decays away as $H(t)$ in $\Lambda$CDM cosmology.
Only the first solution survives in the long-time limit. Hence cosmological
N-body simulations are always initialized already in the ``growing mode'', that
is, setting to zero the decaying solution by construction. This is
achieved in practice by requiring that initially (minus) the divergence of the
peculiar velocity field is in phase with density
perturbations, i.e. ${ \delta}_{\rm init} = { \theta}_{\rm init}={ \phi} $. Hence only
one scalar field ${ \phi}$ (randomly sampled from the linear power
spectrum) determines the full realization of initial perturbations.

Here we depart from this standard practice, for the following reason. 
Our theoretical framework is built upon the multi-point propagators,
which can be measured in simulations as the cross correlations of
final and initial fields (\cite{2006PhRvD..73f3520C}, \cite{2008PhRvD..78j3521B},
\cite{2012PhRvD..85l3519B})\footnote{We are assuming Gaussian initial conditions in this paper, otherwise the relationship between propagators and cross-correlations is more complicated, see \cite{2010PhRvD..82h3507B}.}. However if initial fields are in the
growing mode only certain combinations in the cross
correlations can be tested. For instance, the 2-point
cross-correlation $r_{ab}=\langle a b \rangle$ (related to the 2-point or
nonlinear propagator $G_{ab}$) should a priori have 4 independent
combinations out of $a=(\delta_f,\theta_f)$ and $b=(\delta_i,\theta_i)$; but
in practice only the projection along $\delta_i = \theta_i$ is
measurable if the initial conditions are in the growing mode: $G_a=G_{ab}u_b=G_{a\delta_i}+G_{a\theta_i}$, where $u_b=(1,1)$ indicates that the initial conditions are given by $(\delta_i,\theta_i)=(1,1)\, \phi$. Thus, to date only
the ``density'' $G_{\delta}$ or ``velocity'' $G_{\theta}$ two-point propagators
  have been tested against simulations
  \citep{2006PhRvD..73f3520C,2008PhRvD..78j3521B,2012PhRvD..85l3519B}.

In order to measure all the components of the propagator, and
hence test our analytic predictions in much more detail, we
have performed for the first time simulations with independent particle positions and velocities, leading to independent density and velocity perturbations.  This was done as follows.
We first generate two Gaussian random fields, ${\phi}_1(\vk)$ and
${\phi}_2(\vk)$, out of the same linear power spectrum $P_0$. 
We then run two simulations mixing the initial conditions in densities
and velocities:
\beq {\rm run \, 1 \ }\left\{
  \begin{array}{ c }
     {\delta}_{\rm init}(\vk) = {\phi}_1(\vk) \\
     {\theta}_{\rm init} (\vk) = {\phi}_2(\vk)
  \end{array} \right.,
\label{eq:runsmixmode}
\eeq
and the opposite with run 2. In this way each run has initial
conditions which are general in their initial values of density and
velocity perturbations (hence a linear combination of growing and
decaying mode solutions), such that $P_{\delta_i
  \delta_i}(k)=P_{\theta_i \theta_i}(k)=P_0 (k)$ while $P_{\delta_i \theta_i}(k)=0$.

At the practical level the procedure described above is equivalent to
use any standard initial condition generator to produce two random
sets of initial positions and velocities (out of the same linear
$P_0$), and then combine the initial positions of the first set with
the initial velocities of the second to perform the first simulation
run (and viceversa for the second run). See for instance \cite{1998MNRAS.299.1097S} for a
more detailed explanation on how initial positions and velocities are set in
simulations.

To some extent these two runs are not fully independent from each other, since initial positions and velocities are interchanged. We repeat the process twice so we end up with four runs of simulations with such initial conditions, our measurements below average over such four realizations.

\section{Multi-point Propagator Expansion}
\label{sec:MPexpansion}

In this paper we will work with the so called multi-point propagator
expansion of (equal-time) correlators of cosmic fields
\citep{2008PhRvD..78j3521B}.
In particular we are interested in the  nonlinear 
density power spectrum, which in this framework is given by
\begin{eqnarray}
P_{\delta \delta }(k,z)&=&\sum_{r\ge 1} r! \int \delta_{\rm D}(\vk-\vq_{1\ldots r})
\left[\Gamma^{(r)}_\delta(\vq_1,\ldots,\vq_r;z)\right]^2  \nonumber \\ 
&& \times \, P_0(q_1) \ldots P_0(q_r) \ {\rm d}^3\vq_1 \ldots {\rm d}^3\vq_r.
\label{eq:gamexpansion}
\end{eqnarray}
Here $P_0$ denotes the spectrum of perturbations at some initial time and $\Gamma^{(r)}_\delta$ are the multi-point propagators, defined as the (ensemble
averaged) 
variation of late time 
cosmic fields $\Psi_a=(\delta,\theta)$
with respect to the initial conditions $\phi_a \equiv \Psi_a(z_i)$,
\begin{eqnarray}
\frac{1}{r!}\, 
\mg\frac{\delta^r
\Psi_{a}(\vk,z)}{\delta\phi_{b_{1}}(\vk_{1})\dots\delta\phi_{b_{r}}(\vk_{r})}\md
&\equiv&\nonumber\\
&&\hspace{-3cm}\Dirac(\vk-\vk_{1 \ldots r})\ \Gamma^{(r)}_{ab_{1}\dots
b_{r}}\left(\vk_{1},\dots,\vk_{r},z \right),
\label{GammaAllDef}
\end{eqnarray}
where $\vk_{1\ldots r}=\vk_1+\ldots+\vk_r$. In the most general
scenario Eq.~(\ref{eq:gamexpansion}) should allow for arbitrary initial spectra of density and
velocity fields $P^{\rm init}_{ab}$ \citep{2008PhRvD..78j3521B}. But for simplicity 
we have assumed that initial fields are adiabatic and in the growing mode, thus 
$P^{\rm init}_{ab} = u_a u_b P_0(k)$ with $u_a=(1,1)$.
This translates into the more compact expression for the propagators
\beq 
\Gamma^{(r)}_a \equiv \Gamma^{(r)}_{a b_1 \dots b_r} u_{b_1} \dots u_{b_r},
\eeq
used in Eq.~(\ref{eq:gamexpansion}) evaluated for the density fluctuations ($a=1$, or $a=\delta$). 
Equivalent expressions to Eq.~(\ref{eq:gamexpansion}) hold for $P_{\theta \theta}$
(replacing $\Gamma_\delta$ by $\Gamma_\theta$) and 
$P_{\delta \theta}$ (using a cross-term $\Gamma_\delta
\Gamma_\theta$).

Equation~(\ref{eq:gamexpansion}) results from the resummation of a whole set of
(infinite) terms in the standard PT expansion of $P(k)$.
Unlike the standard approach, it is a sum of positive terms each of
which dominates only in a narrow range of scales. Notice that
now each multi-point propagator has contributions to all orders in PT, and
depend also on $P_0$. At low $k$ they can be described perturbatively while
their asymptotic properties at large $k$ can be computed beyond perturbative expansions.
However, to implement in practice Eq.~(\ref{eq:gamexpansion}) one must have a description of multi-point propagators at all scales
(and times), matching the perturbative calculations at low $k$ to the resummed
asymptotic behavior at high $k$. Achieving this in a way that is fast and accurate enough for the needs of  cosmological surveys is the
goal of our work.

The numerical evaluation of increasing orders in Eq.~(\ref{eq:gamexpansion}) becomes very demanding rather quickly. In order to maintain a fast
evaluation time we will concentrate on quasilinear scales and implement Eq.~(\ref{eq:gamexpansion}) up to
$r=3$ for which we need a description of the two, three and four-point propagators.  In what follows we focus in discussing the multi-point propagators in more detail, with emphasis on our particular description of
$\Gamma^{(1)}, \Gamma^{(2)}$ and $\Gamma^{(3)}$. We then discuss in Sec.~\ref{sec:powerspectrum} the computation of the power
spectrum and comparison of our predictions against measurements in simulations.

\section{The multi-point propagators}
\label{sec:MP}

The two-point or nonlinear propagator, first introduced by
\cite{2006PhRvD..73f3519C} 
in the context of Renormalized Perturbation Theory, is the (ensemble
averaged) variation of late time 
cosmic fields $\Psi_a=(\delta,\theta)$
with respect to the initial conditions $\phi_a \equiv \Psi_a(z_i)$,
\begin{equation}
\Dirac(\vk-\vq) \, G_{ab}\left(k,a \right)=\langle \frac{\delta
\Psi_{a}(\vk,a)}{\delta\phi_{b}(\vq)}\rangle,
\label{Gamma2}
\end{equation}
where we adopted the notation $G_{ab} \equiv \Gamma^{(1)}_{ab}$ and
used as time variable the growth factor $a$\footnote{In what follows we will assume the structure of the theory is that of an
Einstein de Sitter universe ($\Omega_m=1$) for which the growth and scale factor coincide.
The validity of the calculations is not significantly affected on
large scales by this assumption if we replace $a$ by the appropriate growth factor $D_+$
for the corresponding $\Lambda$CDM model. We test and
discuss this in detail in Appendix \ref{sec:EdS}. This 
  approach is equivalent to an approximation about the linearly decaying modes, i.e. $D_{-}=D_+^{-3/2}$.}. 
This object emerged from the resummation of an infinite subset of
contributions to the perturbative expansion of the power spectrum. This effectively ``renormalized'' the linear growth factor into
a fully nonlinear and scale-dependent function: the nonlinear propagator
$G_{ab}$. The precise way in which it is a renormalized version of the growth factor can be seen for Gaussian initial conditions, in which case $G_{ab}$ fully describes the cross-correlation between initial and final conditions, i.e. $\langle \Psi_a \phi_b \rangle = G_{ac} \langle \phi_c \phi_b \rangle$. Also, one can easily show from Eq.~(\ref{Gamma2}) that an expansion of cosmic fields in terms
of their initial values (as done in standard PT) leads to
\beq
G_{ab} = g_{ab} +  \mbox{``nonlinear (loop) corrections''} 
\label{eq:loop1}
\eeq
where $g_{ab}$ is the standard linear propagator,
\beq
g_{ab}(a) = \frac{a}{5}
\Bigg[ \begin{array}{rr} 3 & 2 \\ 3 & 2 \end{array} \Bigg] -
\frac{a^{-3/2}}{5}
\Bigg[ \begin{array}{rr} -2 & 2 \\ 3 & -3 \end{array} \Bigg],
\label{eq:linearg}
\eeq
Hence, on large scales were linear PT becomes a good approximation we
recover
\beq
G_{ab}u_b \rightarrow a \ \ {\rm as} \ \ k \rightarrow 0
\eeq
while on small scales nonlinear effects become dominant driving
$G$ to zero, as initial (linear) and final fields become decorrelated. In \cite{2006PhRvD..73f3520C} it was
shown that in this limit it is possible to resum all the dominant perturbative
orders exactly, leading to
\beq
G_{ab}(k,a) \approx g_{ab}(a)\, \exp[-\frac{1}{2} k^2 \sigmav^2] \ \
{\rm as} \ \ k \,\sigmav \gg 1,
\label{eq:highkG}
\eeq 
where the characteristic scale of decay is given by the r.m.s. one-point displacement field that to most growing order
coincides with the amplitude of large-scale velocity flows,
\beq
\sigmav^2 \equiv \frac{(a-1)^2}{3} \int \frac{P_0}{q^2} d^3\vq. 
\label{eq:sigv}
\eeq
\textfb{In \cite{2012PhRvD..85f3509B} it has been explicitly shown that this result can be derived in a very general framework, the eikonal approximation,
irrespectively of the time dependence of the large-scale flows. This partially extends the result of  
\cite{2011JCAP...06..015A} who} were able to resum a sub-leading
set of perturbative contributions, what led to a slight modification of the damping scale $\sigmav$.

The importance of the high-$k$ limit asymptotics in
Eq.~(\ref{eq:highkG}) stands from the fact that in RPT all diagrams
for correlators now have the linear propagator replaced by the
renormalized propagator $G_{ab}$ inside all loops. As a result, after
this resummation, large scale predictions are least sensitive to what is going on at small highly-nonlinear scales, where even the pressure-less perfect fluid approximation (used to derive these results) breaks down. This is in contrast with standard PT where, depending on the number of loops, the propagator has the wrong asymptotics, $G_{ab}(k\to \infty) \to \pm \infty$.  

The two-point propagator (sometimes also called the response function) turned out important not only to RPT but also
in other resummation schemes such as the Path Integral
  approach of \cite{2007JCAP...06..026M}, Closure Theory \citep{2008ApJ...674..617T}, 
large-N expansion \citep{2007A&A...465..725V}, 
 Lagrangian schemes \citep{2008PhRvD..77f3530M} and more recently the
 extended TimeRG of \cite{2012arXiv1205.2235A}. Note, however, that in
 some of these cases the propagator behavior at high-$k$ can be different than in RPT and what we present here (which are in good agreement with simulations). In the Closure and Large-N cases, there are unphysical oscillations superposed with decay, whereas in the Lagrangian approach of \cite{2008PhRvD..77f3530M} part of the propagator remains perturbative, and thus the same issue about violation of high-$k$ asymptotic as in standard PT occurs. 

Another key step forward was achieved in \cite{2008PhRvD..78j3521B}
were it was shown that the concept and results of the two-point propagator could
be extended to an arbitrary number of points,
\begin{eqnarray}
\frac{1}{p!}\, 
\mg\frac{\delta^p
\Psi_{a}(\vk,a)}{\delta\phi_{b_{1}}(\vk_{1})\dots\delta\phi_{b_{p}}(\vk_{p})}\md
&=&\nonumber\\
&&\hspace{-3cm}\Dirac(\vk-\vk_{1 \ldots p})\ \Gamma^{(p)}_{ab_{1}\dots
b_{p}}\left(\vk_{1},\dots,\vk_{p},a \right),
\end{eqnarray}
where $\vk_{1\ldots p}=\vk_1+\ldots+\vk_p$\footnote{Note that $\Gamma^{(p)}$ denotes the {\em $(p+1)$-point
  propagator}, by translation invariance it depends only on $p$
wavenumbers in Fourier space.}. In this case, a relation analogous to Eq.~(\ref{eq:loop1}) is obtained
\beq
\Gamma^{(n)}_a =  {\mathcal F}^{(n)}_a + \mbox{loop corrections} 
\eeq
where on large scales one recovers the well known ${\mathcal F}_a^{(n)} = a^n (F_n,G_n)$ kernels in PT (assuming
growing mode initial conditions and keeping only the fastest growing contribution),
\beq
{\Gamma}_a^{(n)} \sim a^n \left\{F_n(\vk_1 ,\dots,\vk_n),G_n(\vk_1
  ,\dots,\vk_n)\right\}\ \ {\rm as} \ \ k \rightarrow 0	\nonumber
\eeq
for a = 1, 2 (density or velocity divergence fields respectively). Again, on
smaller scales $\Gamma^{(n)}_a$ is expected to be driven to zero, as the $p$-point propagator can be shown to be proportional for Gaussian initial conditions to the cross-correlation $\langle \Psi \phi \ldots \phi \rangle/P_0^p$, a generalization of the two-point result. Remarkably, the multi-point propagators also admit the resummation
of the dominant behavior in the high-$k$ limit, yielding,
\beq
\Gamma^{(n)}_a \rightarrow   {\mathcal F}^{(n)}_a\,  \exp[-\frac{1}{2} k^2 \sigmav^2]
\label{gammanhk}
\eeq
in strict analogy to Eq.~(\ref{eq:highkG}). Because of this, Eq.~(\ref{eq:gamexpansion}) for the power spectrum can be thought of as an expansion in terms of cross-correlations that are i) always positive and each term dominates in a narrow range of scales, ii) are well-behaved even in the nonlinear regime due to Eq.~(\ref{gammanhk}). The convergence of such an expansion in the nonlinear regime has been verified for the case of the Zel'dovich approximation, where the exact result is known and the expansion can be carried out to a large number of loops \citep{Val0712}.

So far we have discussed limiting expressions of the multi-point propagators in the low and high-$k$ regimes. To continue
we need prescriptions valid at all scales that can be integrated over as in Eq.~(\ref{eq:gamexpansion}). This will be the subject of the following sections.

\subsection{Two-point (nonlinear) Propagator}
\label{sec:twopointpropagator}

We now discuss our prescription for
matching the small and large $k$ regimes of $G_{ab}$ and thus reconstruct the full
two-point propagator. The procedure is
similar to that in RPT \citep{2006PhRvD..73f3520C} but is
simplified because we are interested in describing the propagator
at late times (unlike RPT, we will not be doing time integrations over the propagator here). 

The next-to-leading order corrections to the linear propagator are
given by a one-loop computation. If we neglect all sub-leading time
dependencies the result reads (see \cite{2006PhRvD..73f3520C} for the full expression otherwise),
\begin{eqnarray}
\indent \indent\delta G_{11}^{\rm 1loop} &=& \frac{3}{5} a^3 f(k) + {\mathcal O} (a^2)\nonumber \\
\delta G_{12}^{\rm 1loop} &=& \frac{2}{5} a^3 f(k) + {\mathcal O} (a^2)\nonumber \\
\delta G_{21}^{\rm 1loop} &=& \frac{3}{5} a^3 g(k) + {\mathcal O} (a^2)\nonumber \\
\delta G_{22}^{\rm 1loop} &=& \frac{2}{5} a^3 g(k) + {\mathcal O}
(a^2)
\label{eq:Goneloop}
\end{eqnarray}
where
\begin{eqnarray}
f(k)&=&\int\frac{1}{504 k^3 q^5}\left[6k^7q-79k^5q^3+50q^5k^3-21kq^7 \right.
  \nonumber \\
&+&\frac{3}{4}(k^2-q^2)^3(2k^2+7q^2)\ln \frac{|k-q|^2}{|k+q|^2} \, ] P_0(q) \, d^3q, \nonumber \\
g(k)&=&\int\frac{1}{168k^3q^5} \left[6k^7q-41k^5q^3+2k^3q^5-3kq^7 \right.
  \nonumber \\
&+&\frac{3}{4}(k^2-q^2)^3(2k^2+q^2)\ln\frac{|k-q|^2}{|k+q|^2}\,] P_0(q)\, d^3q.    \nonumber 
\label{functionsofk}
\end{eqnarray}
Notice that both $a^2f(k),a^2g(k) \rightarrow -k^2 \sigmav^2/2$ for $k
\,\sigmav\gg 1$. Therefore we can put together Eqs.~(\ref{eq:Goneloop}) and the
most-growing term in Eq.~(\ref{eq:linearg}) and ``exponentiate'' the propagator as,
\begin{eqnarray}
\indent \indent g_{11}+\delta G_{11}^{\rm 1loop} & \rightarrow &G_{11}=\frac{3}{5} a \exp
[a^2 f(k)] \nonumber \\
  g_{12}+\delta G_{12}^{\rm 1loop} &\rightarrow & G_{12}=\frac{2}{5} a \exp [a^2
f(k)] \nonumber\\
 g_{21}+\delta G_{21}^{\rm 1loop} &\rightarrow &G_{21}=\frac{3}{5} a \exp [a^2 g(k)] \nonumber \\
 g_{22}+\delta G_{22}^{\rm 1loop} &\rightarrow &G_{22}=\frac{2}{5} a
 \exp [a^2 g(k)] 
\label{eq:Gfull}
\end{eqnarray}
These expressions recover the correct one-loop result at quasi-linear
scales as well as the dominant large-$k$ asymptotic (at late times).
They do not exactly match the ones given by \cite{2006PhRvD..73f3520C}
for RPT. Those
explicitely preserved sub-leading time dependencies needed to
correctly integrate the evolution from initial conditions.

For initial conditions in the growing mode, $\phi_a \propto u_a=(1,1)$, the relevant 
quantity is the two-point ``density'' propagator $G_\delta = G_{1a}u_a = G_{11}+G_{12}$ that
simply reads,
\beq
G_\delta(k,z) = D_+(z) \exp{\left[f(k) D_+^2(z)\right]}
\label{eq:prop}
\eeq
In Fig.~\ref{fig:prop} we show how this prescription performs against
measurements of the propagator in our
FID ensemble of N-body simulations (top entry in Table \ref{cosmologies})\footnote{The measurements have been performed by cross-correlating the density
field at the desired redshift with itself at the initial conditions
(set in the growing mode), see \cite{2006PhRvD..73f3520C} for a detailed discussion about the estimator.}.
Left, middle and right panels correspond to redshifts
$z=0,0.5$ and $1$ respectively, the model from Eq.~(\ref{eq:prop}) is shown in solid blue while
the high-$k$ limit expression in Eq.~(\ref{eq:highkG}) in dashed
line. Depicted error bars correspond to the variance over the ensemble.
This prescription, although simple, 
describes the measurement at the sub-percent level at all redshifts and
for all scales of interest (bottom panels in Fig.~\ref{fig:prop} show the
corresponding fractional errors). \textfb{In \cite{2012arXiv1208.1191T} a thorough analysis of the performance 
PT predictions at one and two-loop order is performed. It is found that for $z\gapprox 1$ two-loop results can improve
upon one-loop predictions. We stick however to one-loop prediction as given in Eq. (\ref{eq:prop}) 
in this work as it gives predictions of ample precision in all regimes of interest.}
As we will see in
Sec.~\ref{sec:cosmoperformance} this holds
true not only for our fiducial (FID) cosmology but for all the cosmological
models listed in Table~\ref{cosmologies}.  
%
%

\begin{figure*}
\begin{center}
\includegraphics[width=0.245\textwidth]{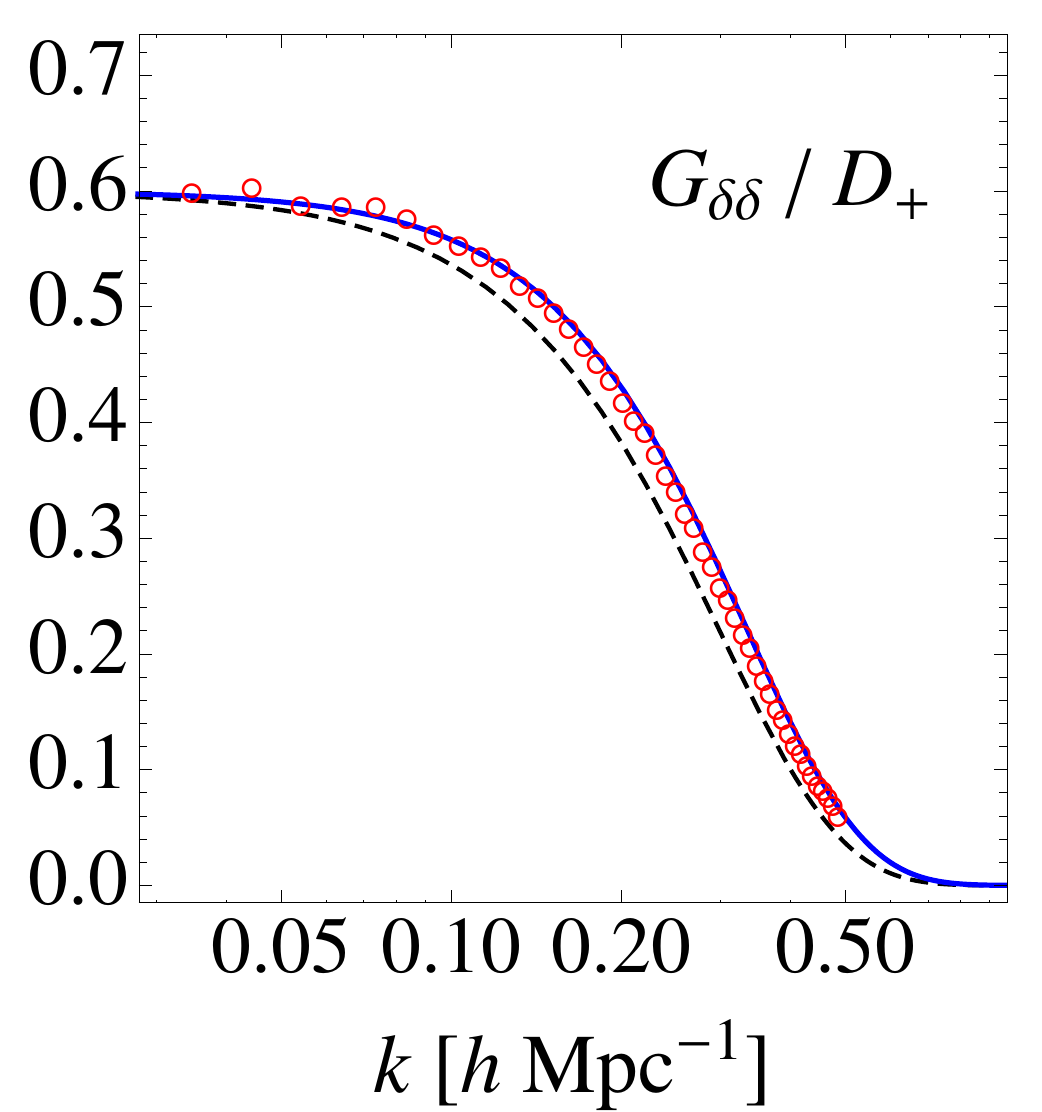}
\includegraphics[width=0.245\textwidth]{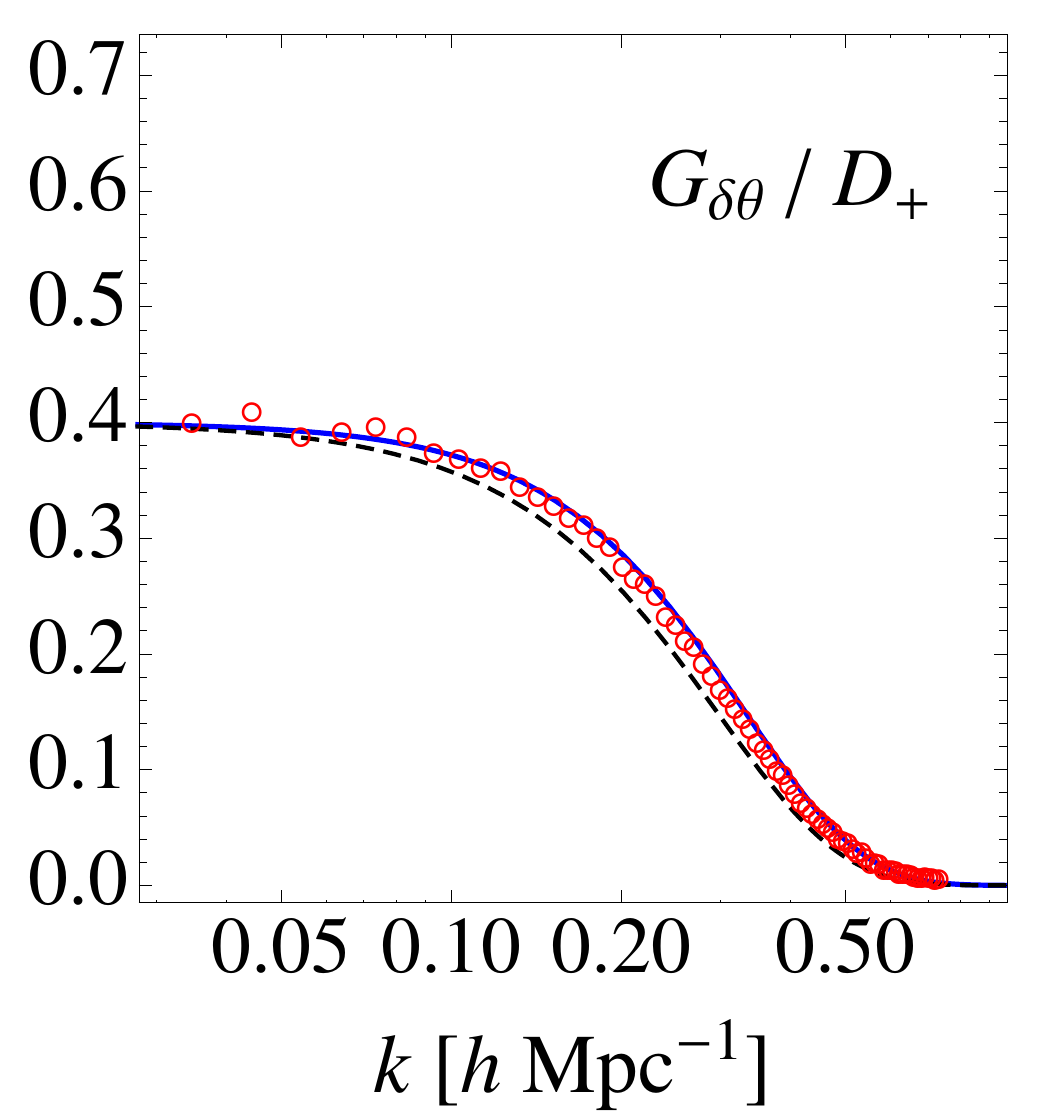} 
\includegraphics[width=0.245\textwidth]{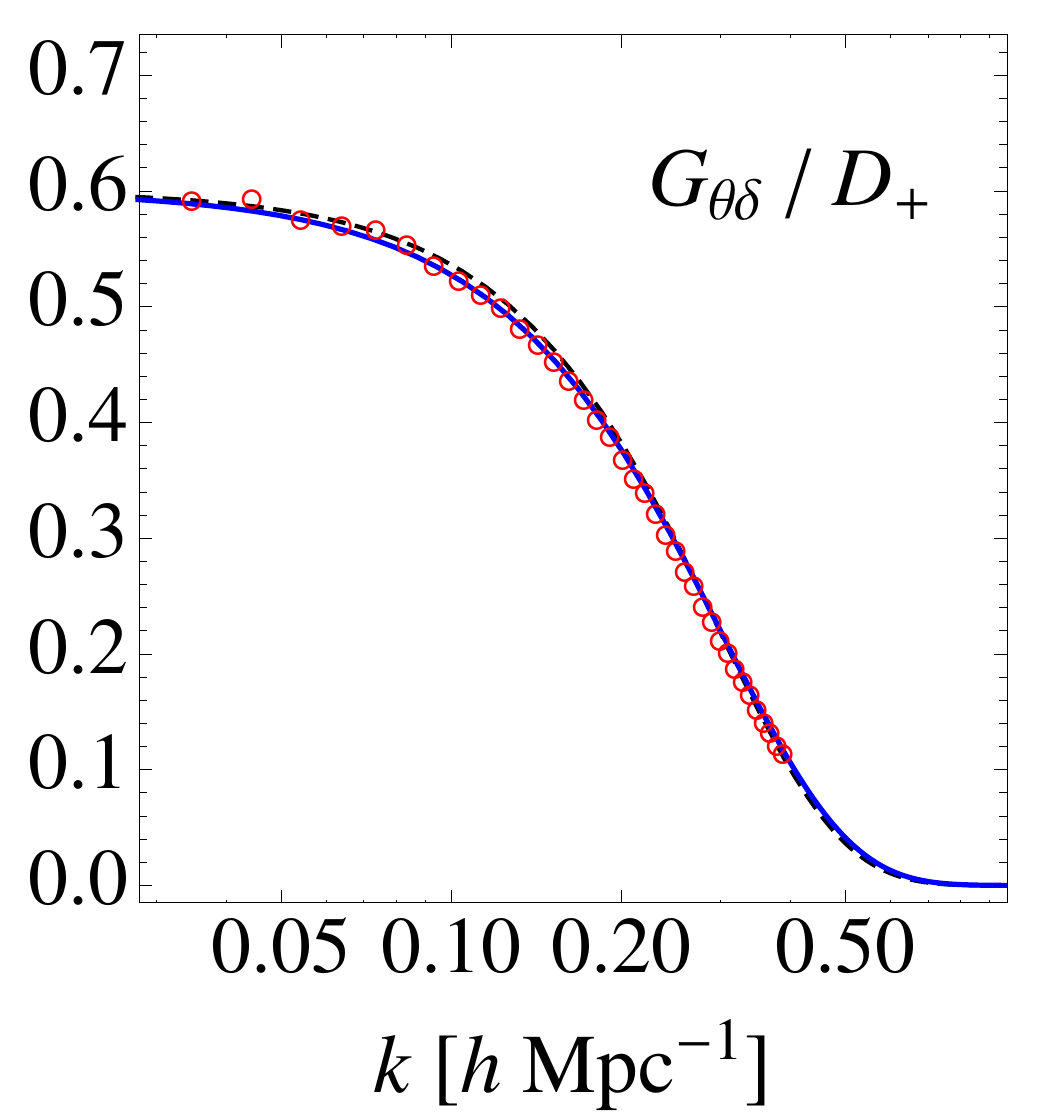} 
\includegraphics[width=0.245\textwidth]{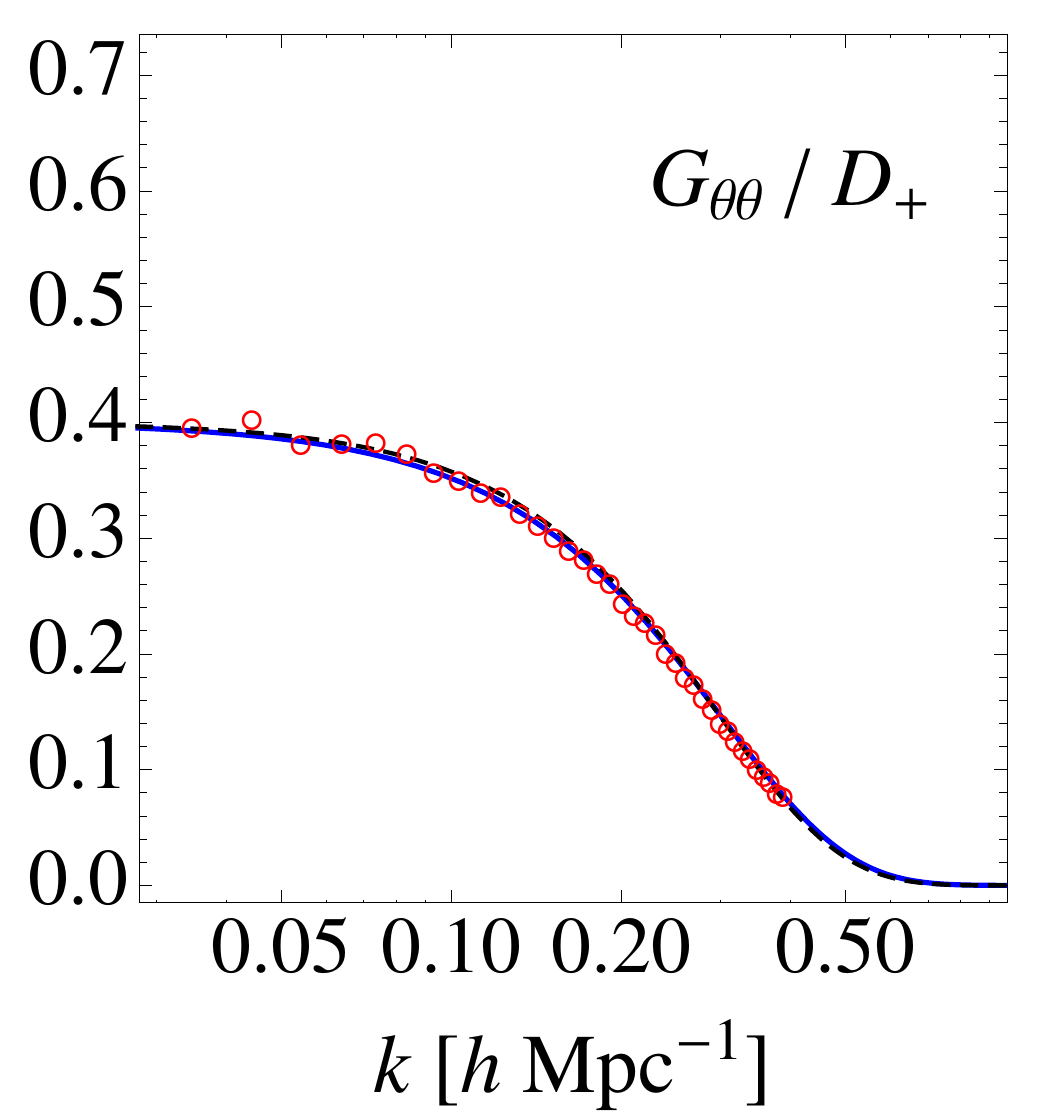} 
\caption{{\it Components of the Nonlinear propagator}. We show for the
  first time the four individual components of the nonlinear propagator
  (normalized to the linear growth factor) measured in dedicated simulations with independent $\delta$ and $\theta$ initial
  conditions. In our model the
  decay of the density propagators $G_{\delta\delta}$ and $G_{\delta\theta}$ is given by
  $\exp[(13/25)f(k) D^2(z)]$ while  the velocity components $G_{\theta\delta}$ and
  $G_{\theta\theta}$ are given by  $\exp[(13/25)g(k) D^2(z)]$, see Eqs.~(\ref{eq:Gfull},\ref{eq:mixmodeIC2}). Dashed lines
  show for 
  reference the decay obtained in the high-$k$ limit $\exp(-k^2 (13/25) \sigmav^2 /2)$ (same for all).}  
\label{fig:propcomponents}
\end{center}
\end{figure*}

\subsection{The Full Matrix Structure of the Propagator}

As discussed in Sec.~\ref{sec:mixmodeIC}, in standard cosmological simulations initialized in the growing mode it is only possible to measure two linear combinations of the two-point propagator. In order to probe the full matrix structure (the four elements of $G_{ab}$) we must initialize simulations with two independent random Gaussian fields for particle positions and velocities, respectively. Let us now discuss how the resummation in the high-$k$ limit is changed when we have independent initial density and velocity fluctuations. 

A good starting point is the analysis of the structure of the one-loop
contribution to $G_{ab}$ calculated from,
\beqa
\delta G^{\rm 1loop}_{ab} &=&\!\!
4 \int_0^\eta ds_1 \int_0^{s_1} ds_2 g_{ac}(\eta-s_1)
\gamma_{cde}(\vk,\vq,\vk-\vq) \nonumber \\   & \times &\!\!
g_{df}(s_1)  g_{eg}(s_1-s_2) 
\times \gamma_{ghi}(\vk+\vq',\vq',\vk) \nonumber \\   &
\times& \!\! g_{hj}(s_2)\lexp \phi_f(\vq) \phi_j(\vq') \rexp  g_{ib}(s_2) \label{gammaintro}
\eeqa 
where $\gamma_{abc}$ is the vertex function in standard PT (see for
instance Eq.~(22) in \cite{2006PhRvD..73f3520C} for a general
derivation of this expression), since this is what then gets exponentiated by
the resummation procedure.

In the standard growing mode case the correlator of initial conditions is,
\beq
\lexp \phi_f(\vq) \phi_j(\vq') \rexp = u_f u_j P_0(\vq) \Dirac(\vq+\vq'),
\eeq
with $u = (1,1)$. For our mixed mode initial conditions we have
instead (see Sec.~\ref{sec:mixmodeIC}),
\beq
\lexp \phi_f(\vq) \phi_j(\vq') \rexp = \delta^{\rm K}_{fj} P_0(\vq) \Dirac(\vq+\vq'),
\eeq
where the Kronecker symbol $\delta^{\rm K}_{fj} = 1$ if $f=j$ and $0$ otherwise.
This means that instead of evaluating 
\beq
g_{df}(s_1)  g_{hj}(s_2) u_f u_j = {\rm e}^{s_1+s_2} u_d u_h
\eeq
in the standard case (with $s=\ln a$), we have to compute
\beq
g_{df}(s_1)  g_{hj}(s_2)  \delta^{\rm K}_{fj}= {13 \over 25} {\rm
  e}^{s_1+s_2} u_d u_h + \mbox{decaying mode},
\eeq
where the decaying mode piece evolves as ${\rm e}^{-3(s_1+s_2)/2}$. Therefore, neglecting the decaying mode contribution we see that the overall effect of using independent random field initial conditions is to renormalize 
\beq
\left\{f(k),g(k)\right\}  \rightarrow {13 \over 25} \left\{ f(k),g(k) \right\} 
\label{eq:mixmodeIC2}
\eeq
in Eqs.~(\ref{eq:Goneloop}). This can be carried out to all orders
leading to the same high-$k$ limit resummation as Eq.~(\ref{eq:prop}) except that
\beq
\sigma^2_d \rightarrow {13 \over 25} \sigma^2_d.
\eeq
Thus our model for the full propagator for independent $\delta$ and $\theta$ initial conditions is simply the
one in Eqs.~(\ref{eq:Gfull}) with the replacement given in Eq.~(\ref{eq:mixmodeIC2}).

In Fig.~\ref{fig:propcomponents} we show the four different components
of the propagator measured in the simulations with independent $\delta$ and $\theta$ initial
conditions against predictions by the above model (in solid blue) and
the high-$k$ asymptotic (in dashed black). 
The propagator was measured following,
\beq
G_{ab}(k) = \frac{\langle \Psi_a(\vk) \phi_b(-\vk) \rangle}{P_0(k)}
\label{eq:propgeneral}
\eeq
where now the initial conditions $\phi_b$ are different for density
($b=1$) or velocities ($b=2$) according to Eq.~(\ref{eq:runsmixmode}).
The velocity divergence fields in Eq.~(\ref{eq:propgeneral}) were estimated following the
  procedure describe in \cite{2004PhRvD..70h3007S}.
Reassuringly all the four components follow the
expected theoretical decay towards small scales. This is an interesting and
important cross-check, in particular for resummation schemes such as
RPT or Closure Theory, that
integrate the individual components separately rather than the  ``density'' and ``velocity''  propagators that until this paper were the only combinations tested
against simulations.

\begin{figure*}
\begin{center}
\includegraphics[width=0.32\textwidth]{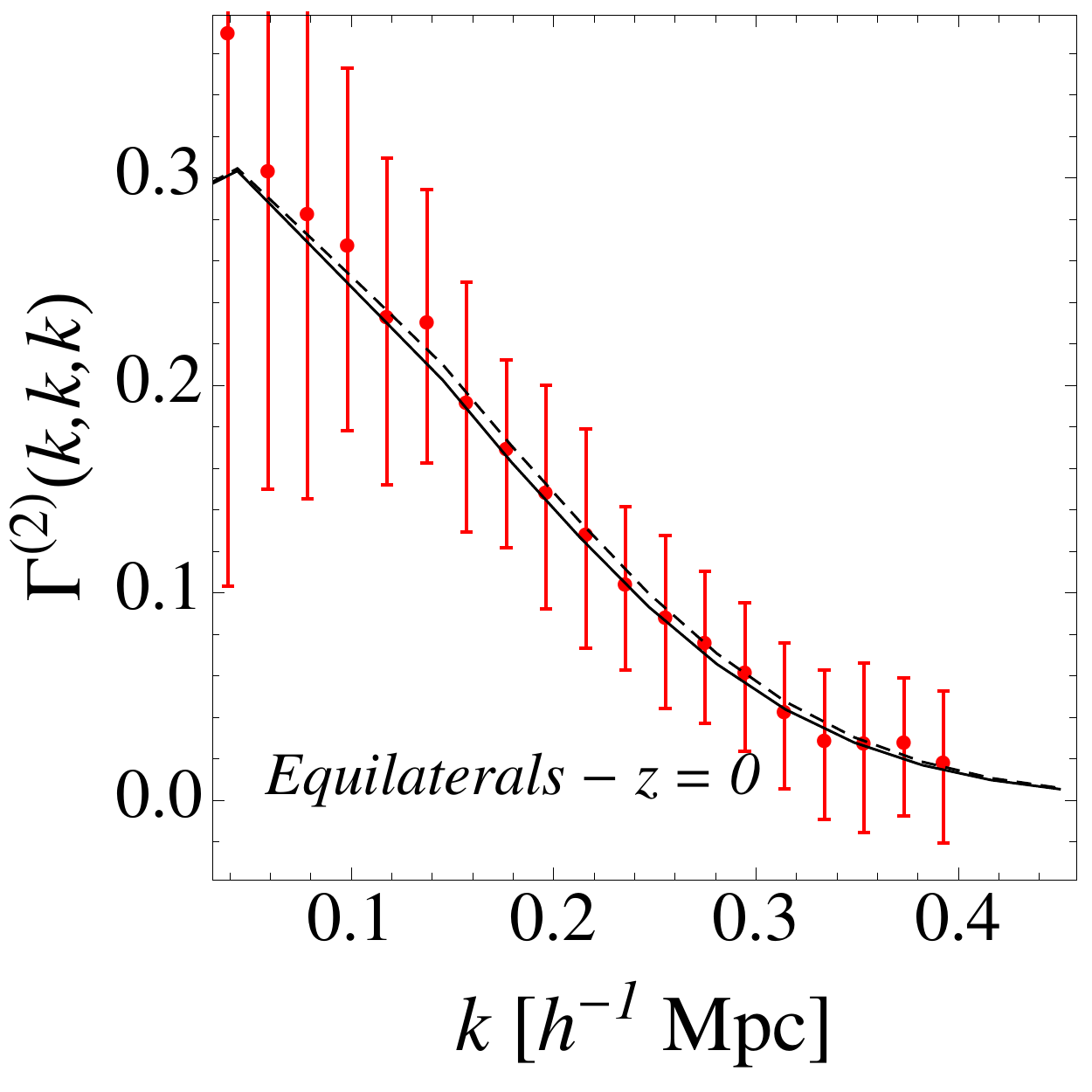}
\includegraphics[width=0.32\textwidth]{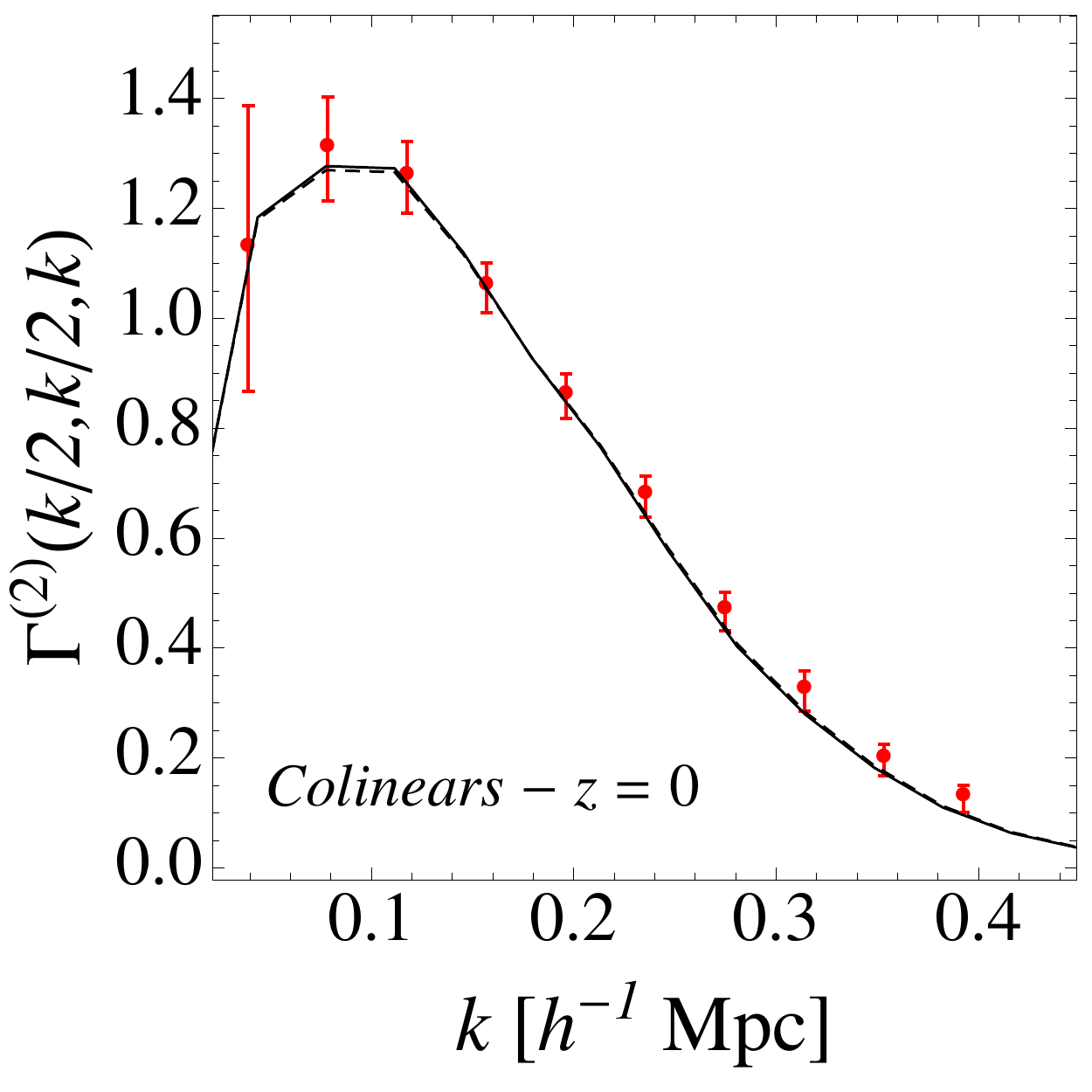} 
\includegraphics[width=0.32\textwidth]{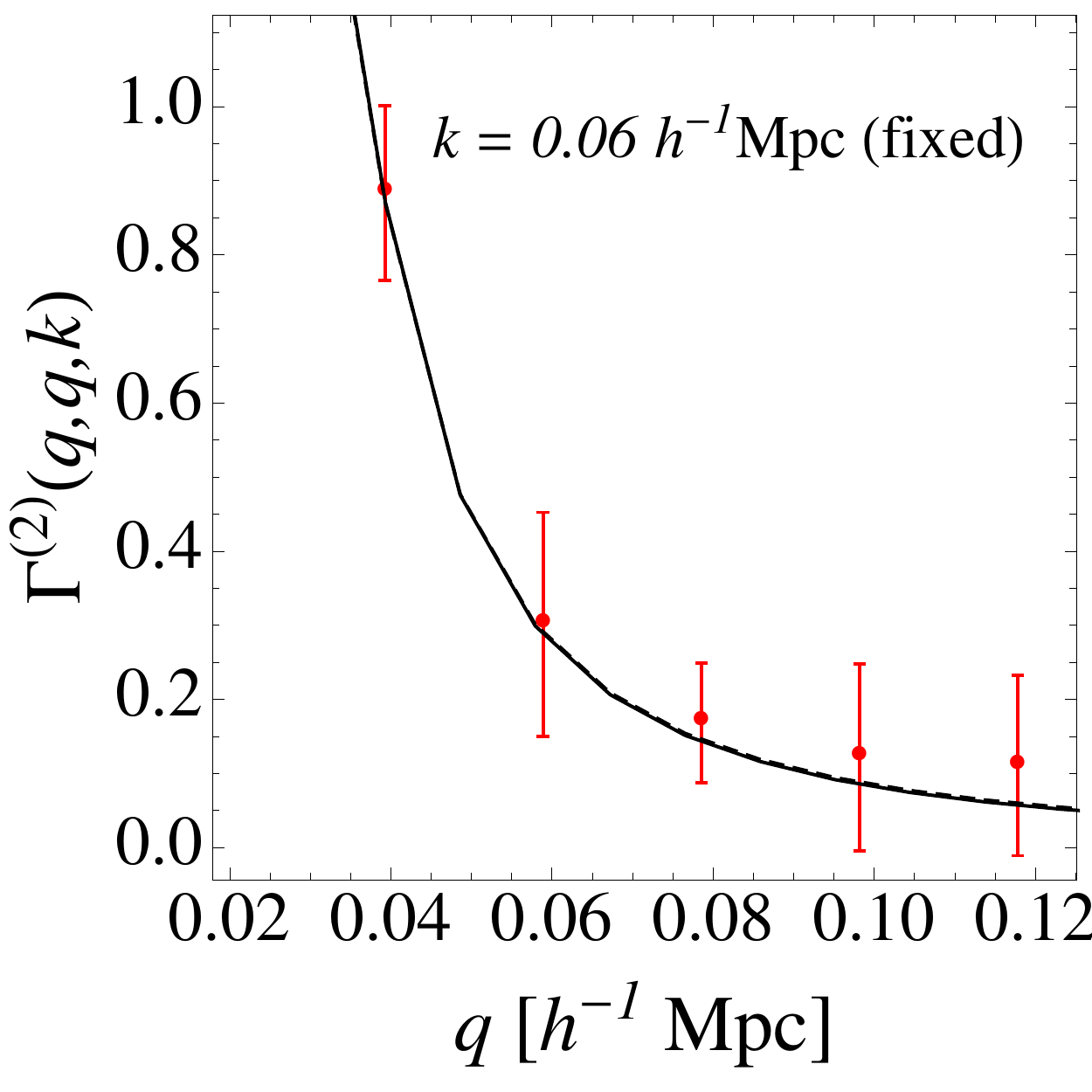} \\
\includegraphics[width=0.32\textwidth]{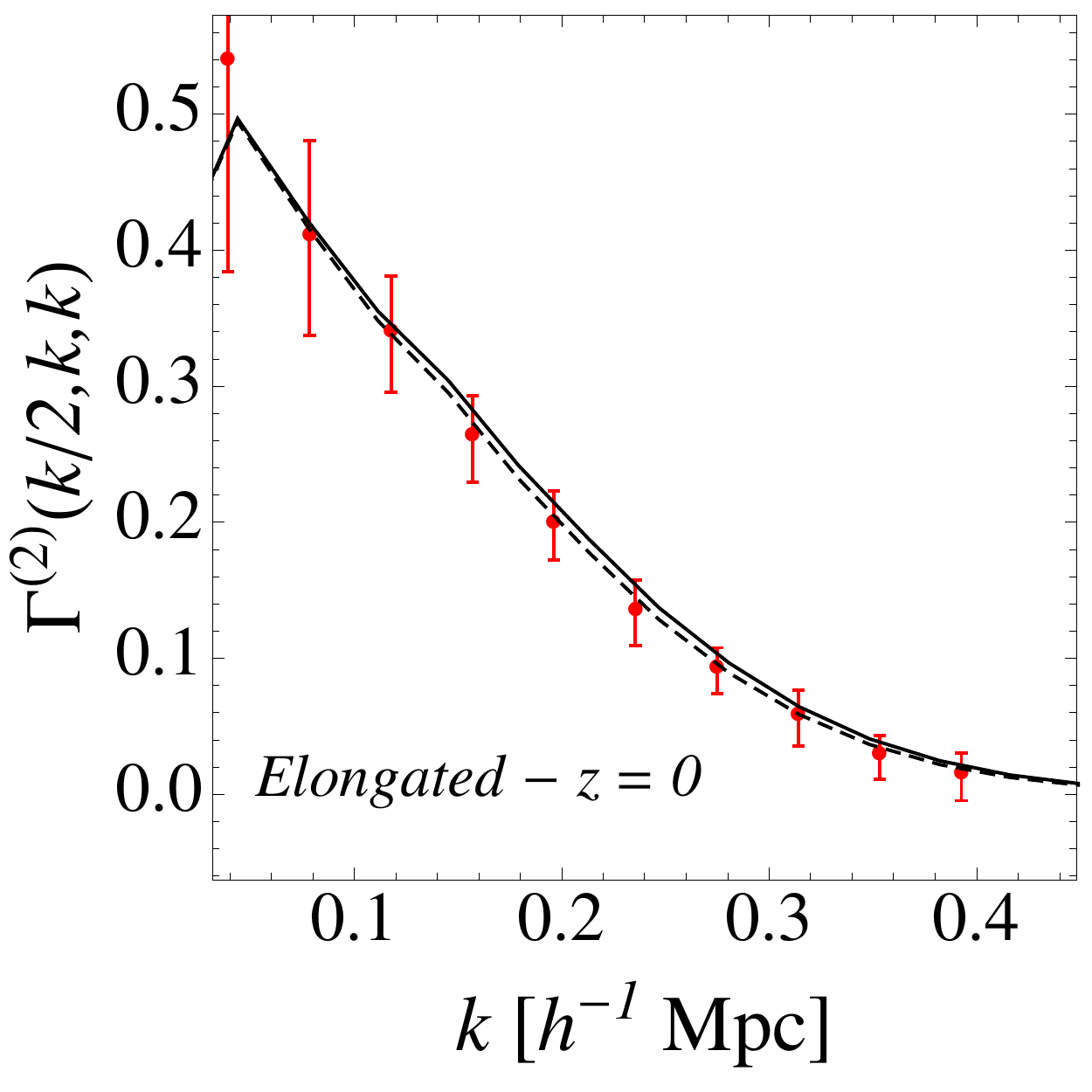} 
\includegraphics[width=0.32\textwidth]{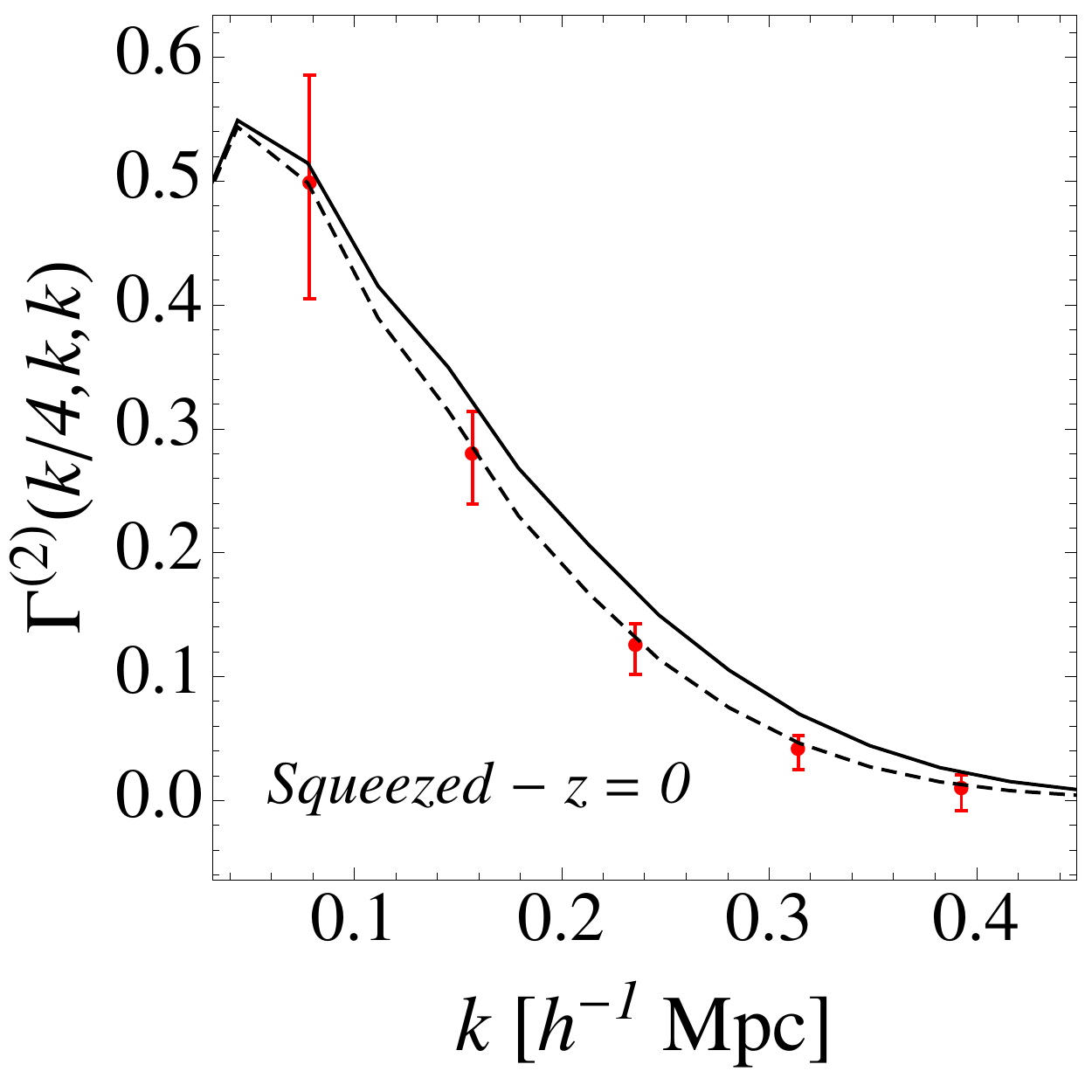} 
\includegraphics[width=0.32\textwidth]{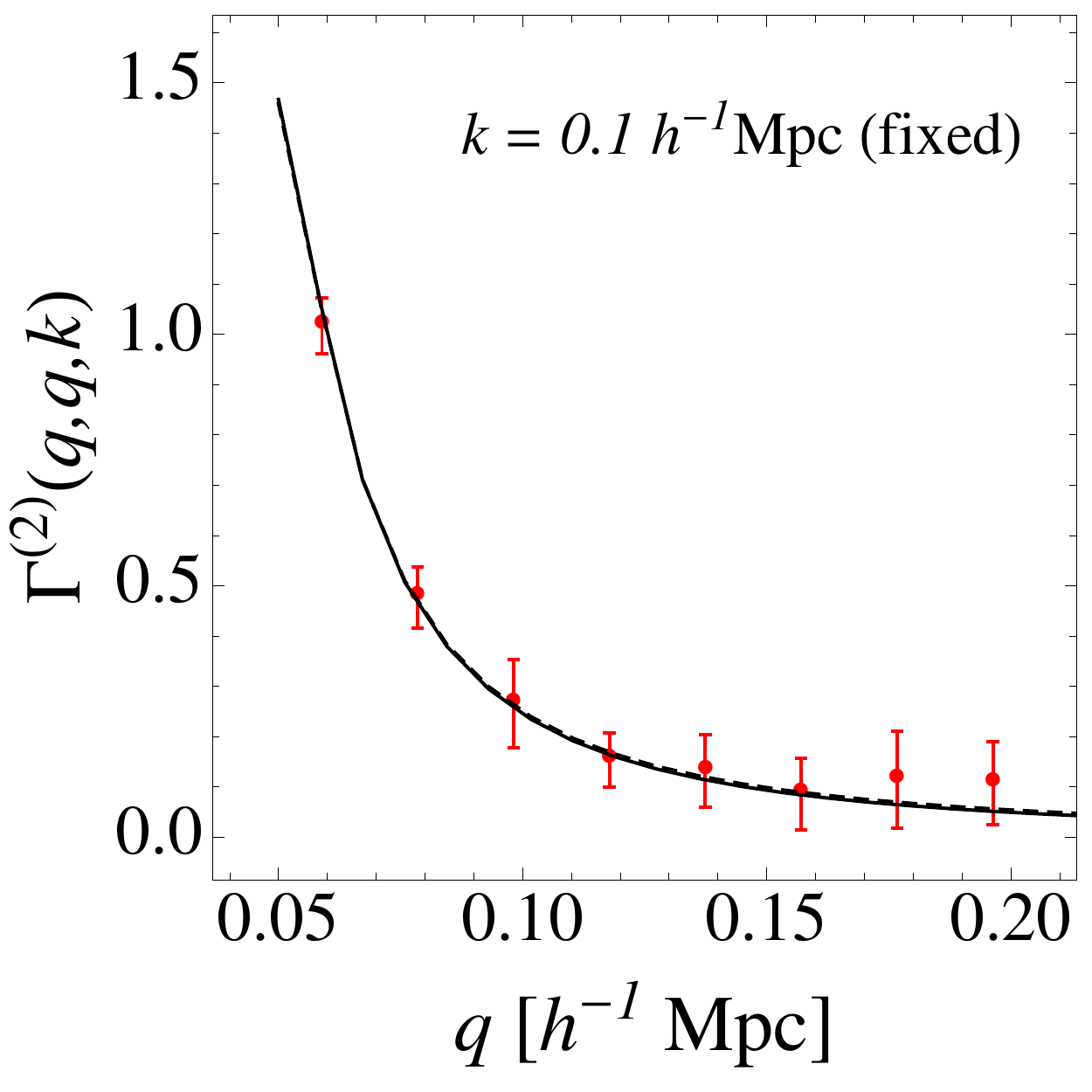} 
\caption{{\it Nonlinear three-point propagator: analytic predictions vs. measurements.}
  Different panels show measurements of $\Gamma^{(2)}(q_1,q_2,q_3)$ in our
 fiducial ensemble of simulations
for different triangular configurations, as indicated in the y-axis label
(see text for details). Solid line corresponds to the interpolation
scheme proposed in this paper, see Eq.~(\ref{eq:g2}). Dashed line to the one
introduced in Bernardeau et al. (2011) ({\RegPT}), see Eq.~(\ref{eq:g2regPT}). They are mostly indistinguishable for
almost all configurations and agree with the measurements remarkably
well. In particular for right panels where we show
the configuration that contribute the most to the one-loop power spectrum
at (fixed) wavenumber $k = 0.06\Mpc$ ($0.1\Mpc$). Error bars
correspond to the variance over the ensemble and results (in left and
middle panels) are plotted against $k_3=k$. } 
\label{fig:gamma2}
\end{center}
\end{figure*}

\subsection{Three and Four-Point Propagators}

The three-point propagator was introduced by \cite{2008PhRvD..78j3521B}
and recently studied in detail in \cite{2012PhRvD..85l3519B}. In the
later work a general scheme (called {\RegPT}) to interpolate between small and large
scale information for any propagator was proposed.  In particular, the {\RegPT} prescription for the three-point propagator is
given by,
\begin{eqnarray}
\Gamma^{(2) \rm RegPT}_a(\vk_1,\vk_2) \!\!\!\!\!
&=& \!\!\!\!\!\left[\Gamma^{(2) \rm Tree}(\vk_1,\vk_2) +
\delta \Gamma^{(2) \rm one-loop}(\vk_1,\vk_2) \right. \nonumber \\ 
&+& \left. \! \! \frac{1}{2} k^2 \sigmav^2\,
\Gamma^{(2) \rm Tree}(\vk_1,\vk_2) \right]\exp(-k^2 \sigmav^2/2)
\nonumber \\
&& \label{eq:g2regPT}
\end{eqnarray}
where $\vk=\vk_1+\vk_2$. Here $\Gamma^{(2) \rm Tree}$, $\delta
\Gamma^{(2)\rm one-loop}$ and $\sigmav$ depend on time. The one-loop term in this expression 
is described in detail in
\cite{2012PhRvD..85l3519B} and involves one integral over $P_0(q)$ for
each triangle configuration $(\vk_1,\vk_2)$. Although Eq.~(\ref{eq:g2regPT}) gives a
very good agreement with measurements in simulations we have found
that its usefulness to compute the one-loop $P(k)$ is limited because
it takes a long time to evaluate. We hence seek an alternative prescription.

We have found that, in analogy to Eq.~(\ref{eq:prop}), the following
expression ($\vk=\vk_1+\vk_2$)
\beq
\Gamma^{(2)}_\delta(\vk_1,\vk_2;z) 
 = D_+^2(z) \, F_2(\vk_1,\vk_2) \, \exp [f(k) D_+^2(z)]
\label{eq:g2}
\eeq
yields very similar results to that in Eq.~(\ref{eq:g2regPT}) and virtually the same
one-loop power spectrum after the corresponding momentum integration. 
Figure~\ref{fig:gamma2} shows measurements of $\Gamma^{(2)}$ for different
triangle configurations together with the prediction by
the model in Eq.(\ref{eq:g2}) in solid black (used throughout this paper) and
{\RegPT} from Eq.~(\ref{eq:g2regPT}) in dashed black. Left and Middle panels correspond
to {\it equilateral} with $k_1=k_2=k_3=k$ (top left); {\it colinear}
with $k_1=k_2=k/2$ and $k_3=k$ (top center); {\it elongated} (bottom left) with $k_1=k/2$ and $k_2=k_3=k$; and {\it squezeed} (bottom center) $k_1=k/4$ $k_2=k_3=k$ configurations\footnote{We assume the final (nonlinear)
  density field has wave-vector $k_3$ (last argument). Hence the
  three-point propagator is only symmetric with respect to the 1st and
2nd indices that corresponds to the initial (linear) fields.}.

Notice that the theory is binned in the same way as the data, this is essential to
recover the correct asymptotic behavior at low-$k$ (see \cite{2012PhRvD..85l3519B} for
details on the $\Gamma^{(2)}$ estimator and the binning correction).

From these panels it is clear that both models perform very well for
all these configurations with a slight over-prediction by Eq.~(\ref{eq:g2}) for squeezed configurations.

In addition the right panels of Fig.~(\ref{fig:gamma2}) show the same
comparison for the configuration that
would yield the dominant contribution to the one-loop computation of $P(k)$. From Eq.~(\ref{eq:gamexpansion}) we see that the one-loop power spectrum is of the form
\begin{eqnarray}
\indent P^{\rm 1loop}(k) &\sim& \frac{4\pi}{k} \int P_0(q_1) q_1 dq_1
\int P_0(q_2) q_2 dq_2 \nonumber \\ 
 &\times& \left[ \Gamma^{(2)}_\delta(q_1,q_2,k)\right]^2
\end{eqnarray}
Hence by symmetry reasons the most relevant configuration for a given
$k$ is roughly $\Gamma^{(2)}(q,q,k)$. Figure~(\ref{fig:gamma2}) shows this
configuration for $k=0.06\kvecMpc$ (top right panel) and
$k=0.1\kvecMpc$ (bottom right panel). Here again the model in Eq.(\ref{eq:g2})
describes the N-body results remarkably well yielding the same answer as
{\RegPT} (notice the dashed and solid line are on top of each other).

The four-point propagator is basically a measure of the trispectrum
between final and initial density fields. Thus it is  difficult to
measure from the N-body simulations. Nonetheless from
theoretical grounds we do know the behavior at low and high-$k$, and
have no reason to expect a different behavior at intermediate scales
from the one already probed for the two and three-point
propagators. Hence we will adopt the following prescription  ($\vk=\vk_{123}$),
\begin{eqnarray}
\Gamma^{(3)}_\delta(\vk_1,\vk_2,\vk_3;z) \!\!
 & = & \!\! D_+^3(z) F_3(\vk_1,\vk_2,\vk_3) \, \exp [f(k) D_+^2(z)], \nonumber \\
&&\label{eq:g3}
\end{eqnarray}
for the four-point propagator, in full analogy to Eqs.~(\ref{eq:prop}) and (\ref{eq:g2}).
This prescription will then satisfy the low-$k$ and high-$k$ asymptotics.

Provided with prescriptions for the propagators up to four points, we are now ready to discuss the multi-point expansion for the power spectrum.

\begin{figure}
\begin{center}
\includegraphics[trim = 0cm 1.2cm 0cm 0cm, width=0.35\textwidth]{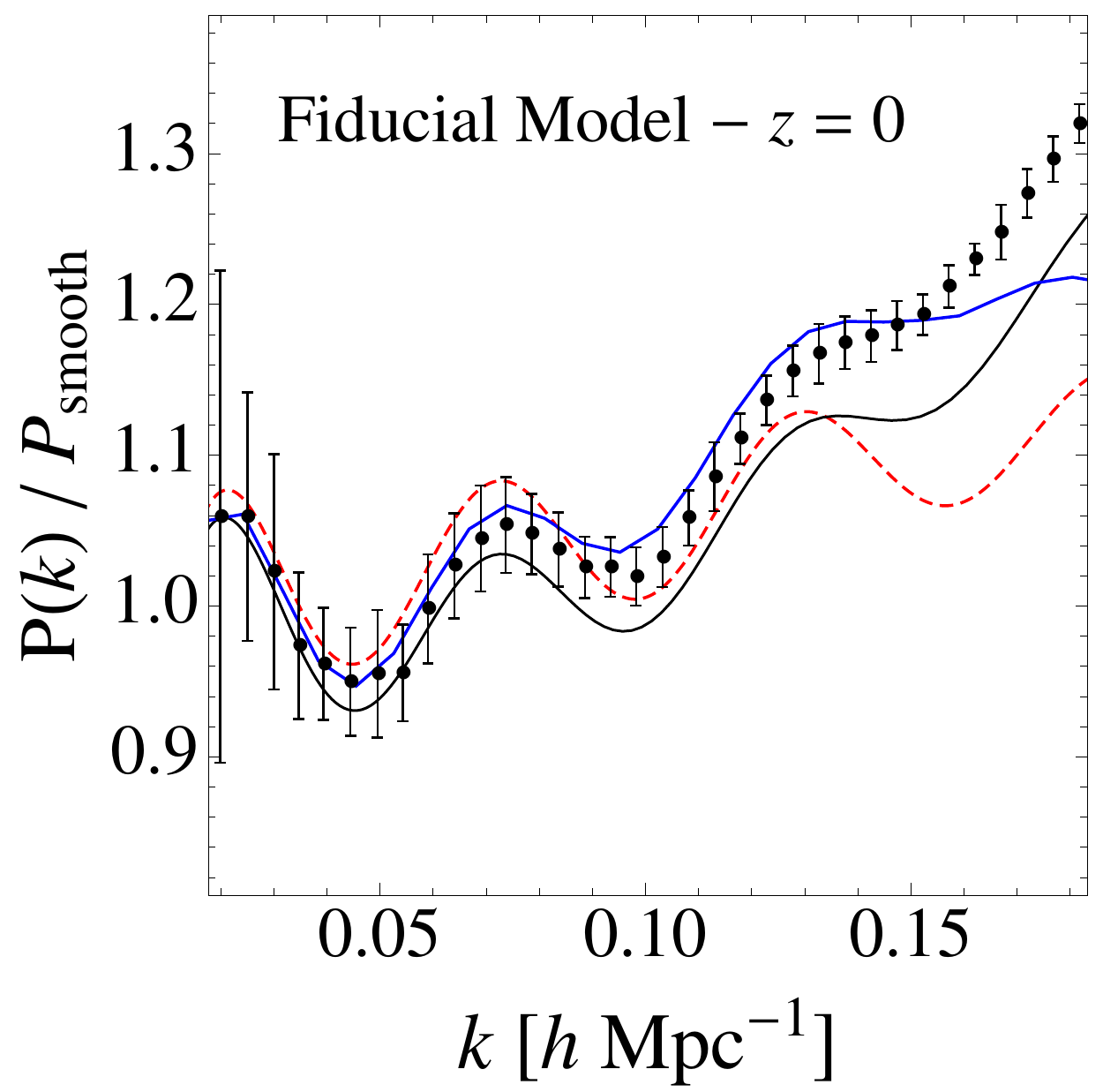}
\includegraphics[trim = 0cm 1.2cm 0cm 0cm, width=0.35\textwidth]{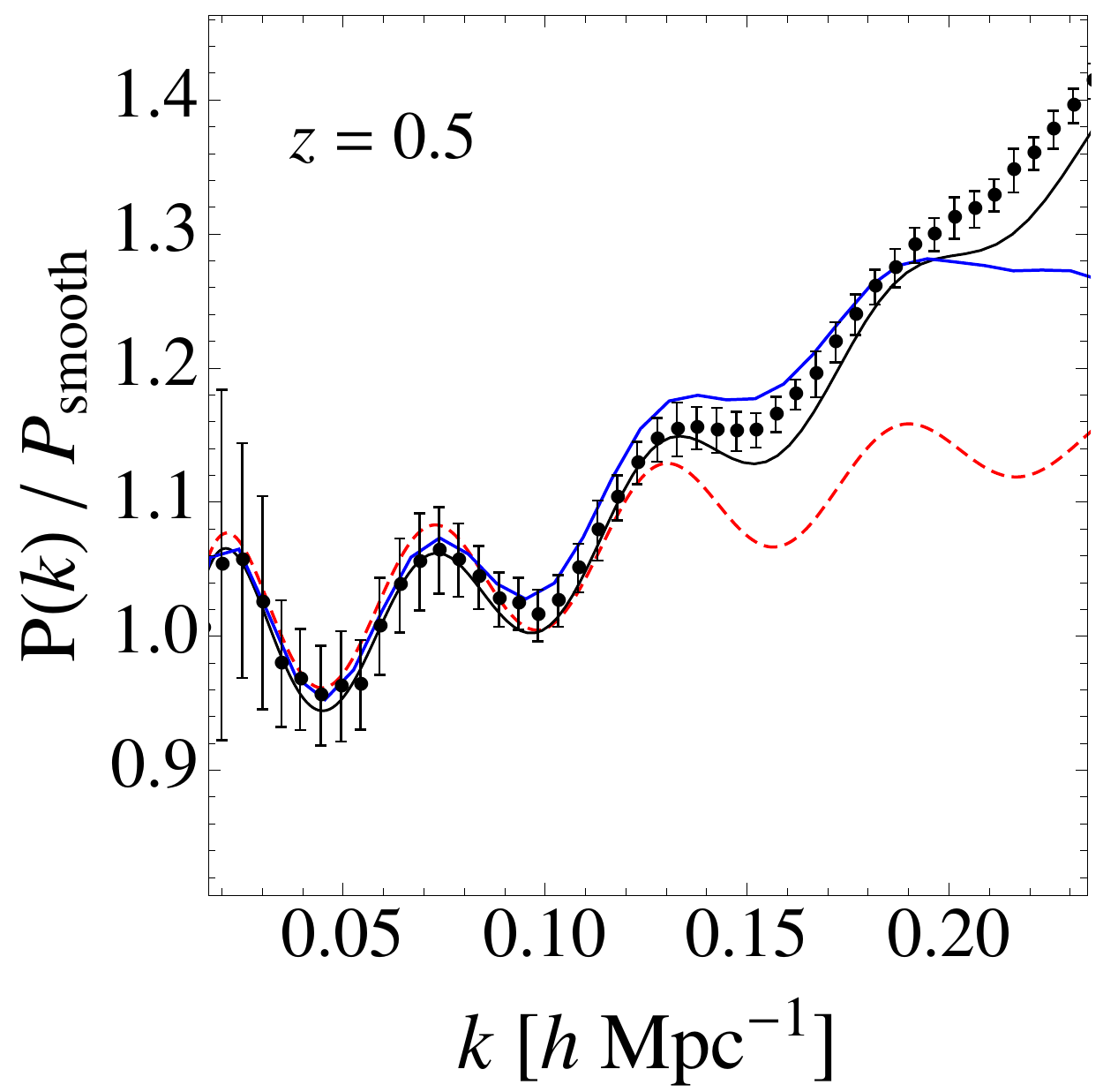}
\includegraphics[trim = 0cm 0cm 0cm 0cm, width=0.35\textwidth]{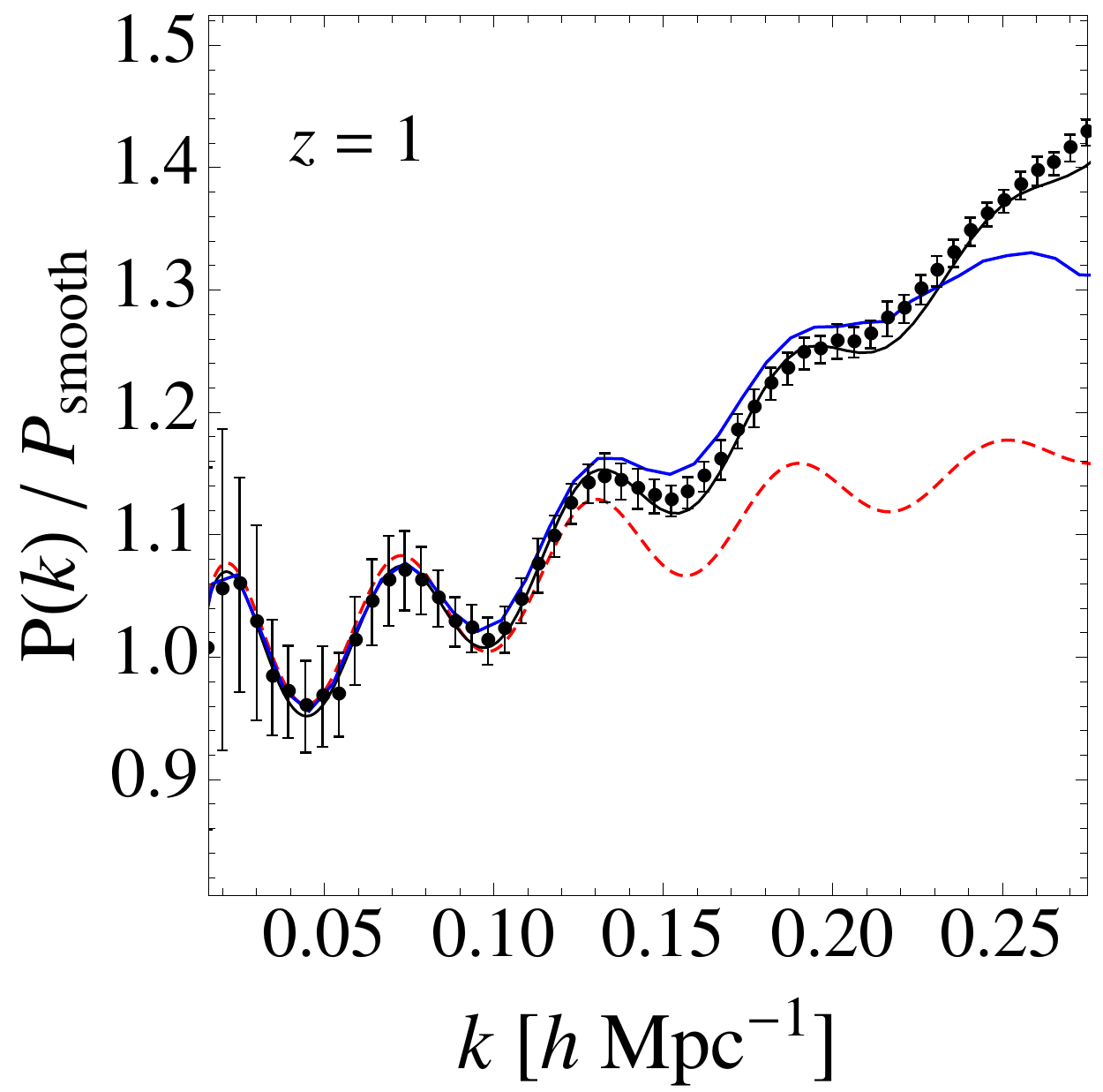}
\caption{The multi-point propagator expansion presented
  in this paper (solid blue line) against measurements of $P(k)$
    in our FID ensemble of N-body simulations 
    (top entry in Table~\ref{cosmologies}) at $z=0$, $0.5$ and $1$. The dotted red line is linear theory and
    solid black is {\tt halofit}. The evaluation time of the multi-point
    expansion shown in each panel is at most five seconds.} 
\label{fig:power}
\end{center}
\end{figure}

\section{Power spectrum }
\label{sec:powerspectrum}

\textfb{In this section we present the prescriptions we adopt to do the actual computations of the power spectra. 
We call {\MPTbreeze} this implementation and comment at the end of the
section on possible alternative approaches.} 

To describe the power spectrum at mildly nonlinear scales we implemented
the first three terms in the expansion in Eq.~(\ref{eq:gamexpansion}).
In diagrammatic
language they correspond to renormalized versions of tree level, one and two loops respectively 
(by renormalized we mean that the renormalized propagators include themselves loops to all orders). The tree-level term is simply given by 
\beq
P_{\rm tree} = [\Gamma^{(1)} (k,z)]^2 P_0
\label{eq:tree}
\eeq
and coincides with the propagator renormalization term of RPT \citep{2006PhRvD..73f3519C}. Here $\Gamma_\delta^{(1)}$
is given by Eq.~(\ref{eq:prop}) and $P_0$ is the linear, post-recombination, spectrum of fluctuations. 
This term is the one that contain most information on narrow band features of
the primordial perturbations such as the Baryon Acoustic
Oscillations. The ``mode-coupling'' contributions start with the next 
term in Eq.~(\ref{eq:gamexpansion}) that reads,
\beq
P_{\rm 1-loop}(k,z) = 2\!\! \int d^3\vq [\Gamma^{(2)}_\delta(\vk-\vq,\vq;z)]^2
P_0(|\vk-\vq|) P_0(q)
\label{eq:oneloop}
\eeq
Assuming $\vk$ along the $z$-axis it can be easily turn into 
\beq
P_{\rm 1-loop}(k,z) = 4\pi\!\!\int_{-1}^{1} dx \int
\!\! dq [\Gamma^{(2)}_\delta(p,q,y;z)]^2
P_0(p) P_0(q)
\nonumber
\eeq
where $x = \vk \cdot \vq / (k\,q)$ and we have introduced $\vp = \vk
-\vq$ and $y = \vp \cdot \vq / (p\,q)$ for clearness. Here $\Gamma^{(2)}$ is given
by Eq.~(\ref{eq:g2}) with $F_2 (p,q,y) = \frac{5}{7}+\frac{y}{2}(\frac{q}{p}+\frac{p}{q})+\frac{2y^2}{7}$ the standard 2nd-order PT kernel
(e.g. \cite{2002PhR...367....1B}). 
We are then left with an integration in 2-dimensions which can be
rapidly evaluated with a standard Gaussian quadratures routine.

The third term in Eq.~(\ref{eq:gamexpansion}) is slightly more involved but can be treated
similarly, we now have
\begin{eqnarray}
P_{\rm 2-loop}(k,z) \! & \! = \! & \! 6\!\! \int d^3\vq_1 \int d^3\vq_2 
\, [\Gamma^{(3)}_\delta(\vk-\vq_{12},\vq_1,\vq_2;z)]^2  \nonumber \\
&& P_0(|\vk-\vq_{12}|) P_0(q_1) P_0(q_2). \label{eq:twoloop}
\end{eqnarray}
Here we can take $k$ along the $z$-axis and $\vq_1$ in the
$x-z$,
\begin{eqnarray}
\indent\indent \vk \,\,  &=& k \,\, (0,0,1) \nonumber \\
\vq_1 &=& q_1 (\sin \theta, 0, \cos \theta) \nonumber \\
\vq_2 &=& q_2 (\sin \phi \sin \alpha, \cos \phi \sin \alpha, \cos
\alpha) , \nonumber 
\end{eqnarray}
so we are then left with the following 5-dimensional integral
\begin{eqnarray}
P_{\rm 2-loop}(k,z) \! & \! = \! & \! 12 \pi \!\! \int dq_1 \int dq_2 \int
dx \int dy \int d\phi 
\,  \nonumber \\ && \!\!\!\!\!\!\! [\Gamma^{(3)}_\delta(\vq_3,\vq_1,\vq_2;z)]^2 
P_0(q_3) P_0(q_1) P_0(q_2) \label{eq:2loops}
\end{eqnarray}
were we introduced $\vq_3 = \vk - \vq_{12}$; $x = \cos \theta = \vk \cdot \vq_1 / (k\,q_1)$ and $y = \cos \alpha = \vk \cdot \vq_2 / (k\,q_2)$
are integrated in $[-1,1]$ and $\phi$, the azimuthal orientation of
$\vq_2$, in $[0,2\pi]$. The propagator $\Gamma^{(3)}_\delta$ is given by Eq.~(\ref{eq:g3}) in
terms of $f(k)$, Eq.~(\ref{functionsofk}), and $F_3$, which in turn is solved iteratively in
terms of $(F_2,G_2)$ (see e.g. \cite{2002PhR...367....1B}). 

\begin{figure*}
\begin{center}
\includegraphics[trim = 0cm 1cm 0cm 0cm, width=0.325\textwidth]{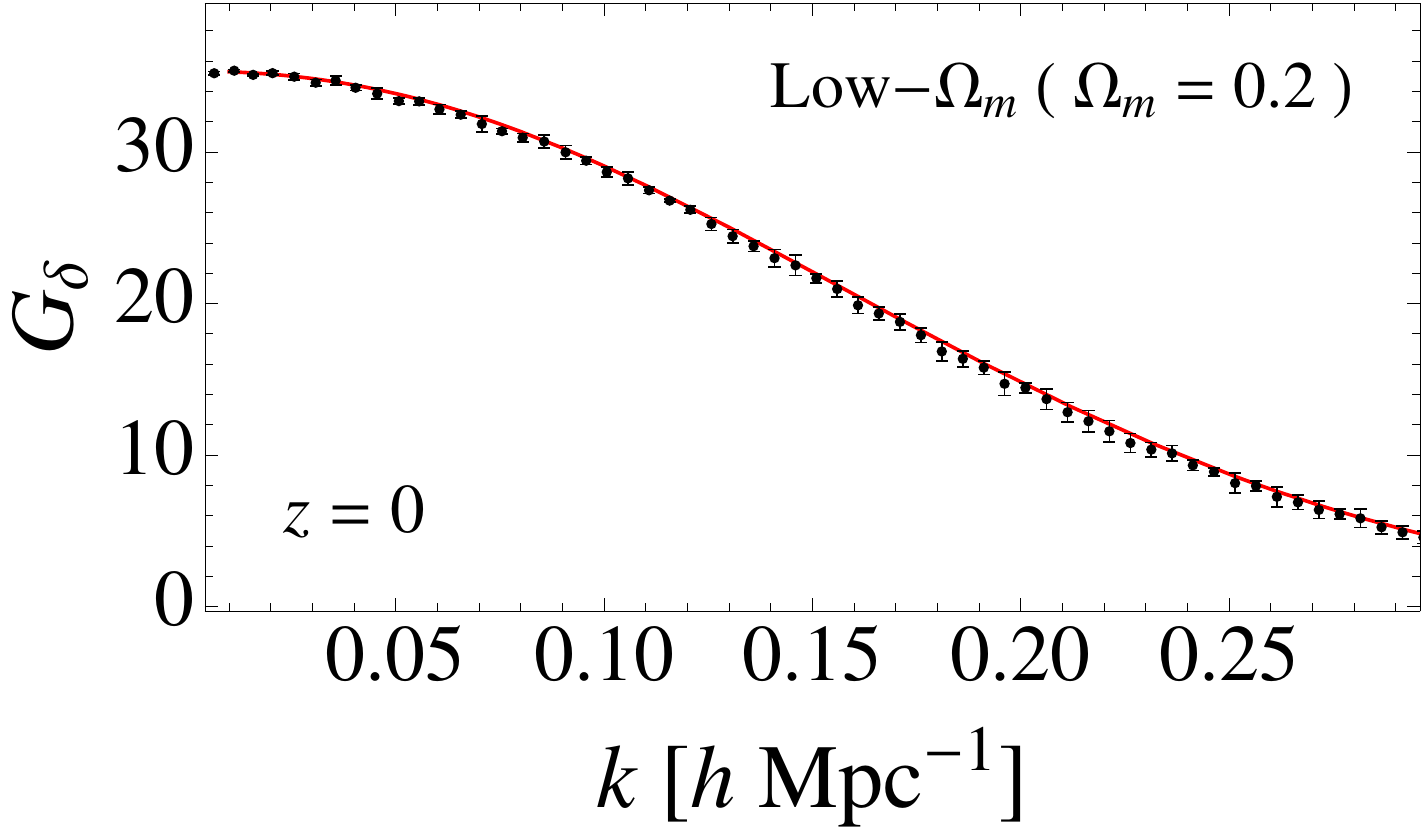} 
\includegraphics[trim = 0cm 1cm 0cm 0cm, width=0.325\textwidth]{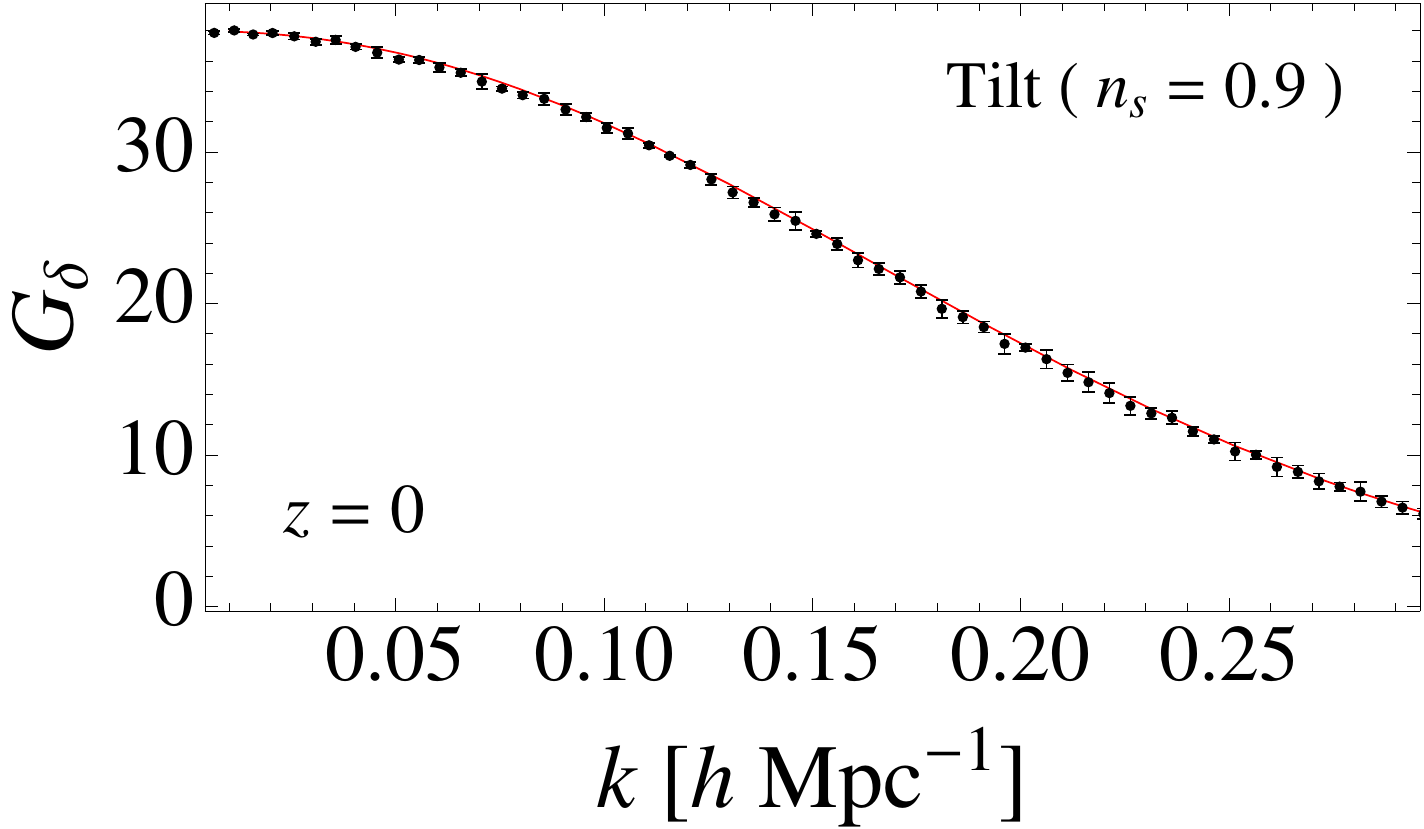} 
\includegraphics[trim = 0cm 1cm 0cm 0cm, width=0.325\textwidth]{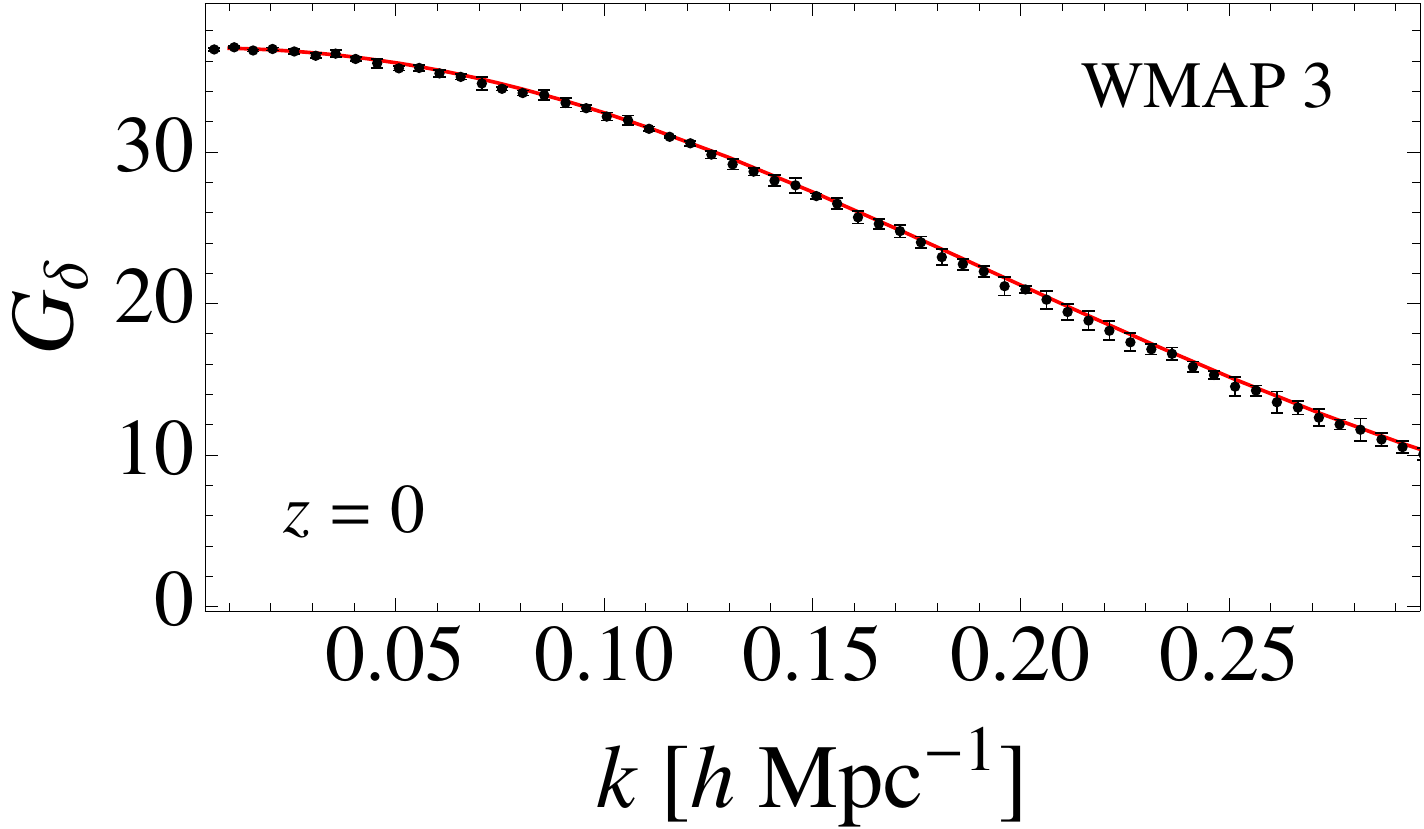} \\
\includegraphics[trim = 0cm 0cm 0cm 0cm, width=0.325\textwidth]{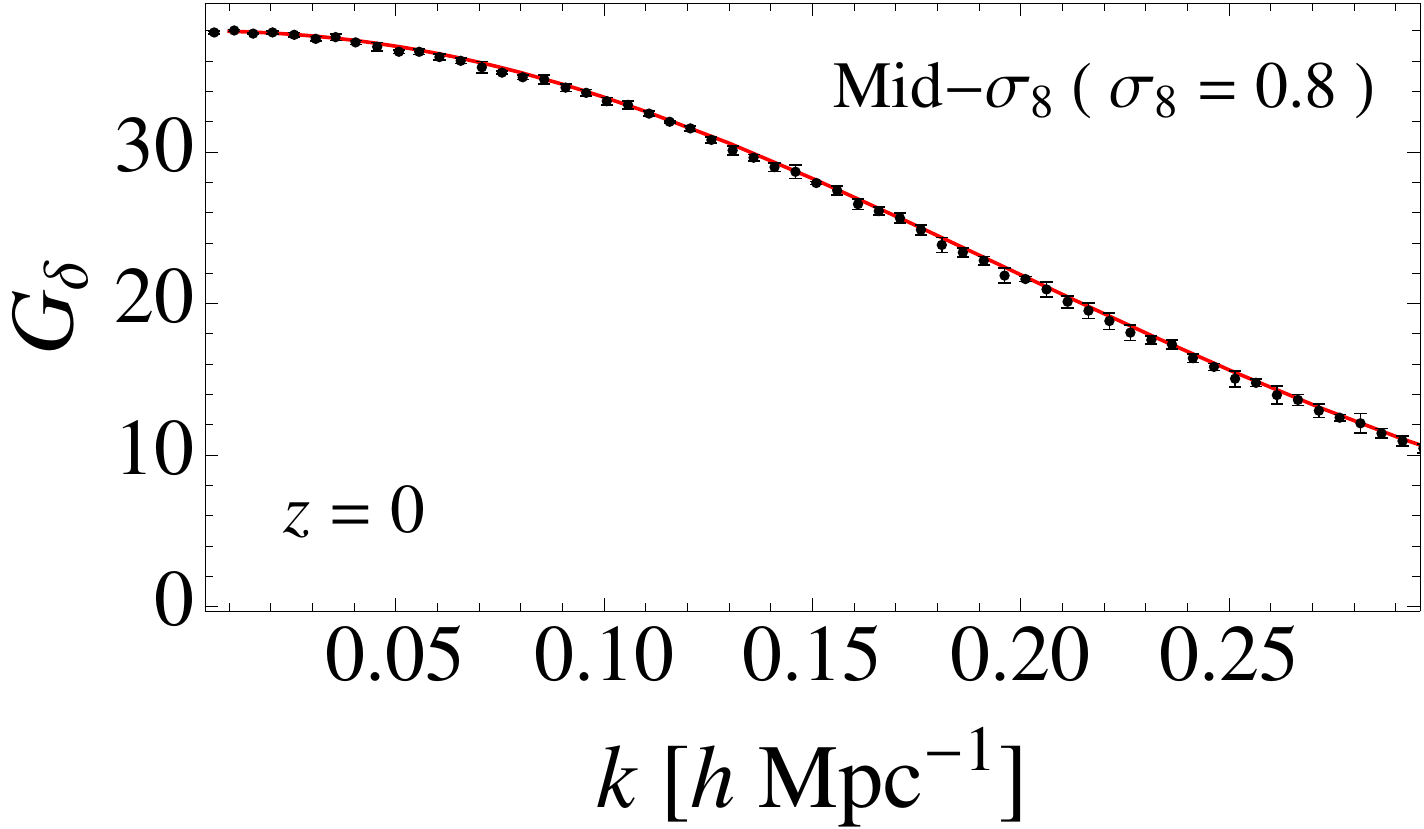} 
\includegraphics[trim = 0cm 0cm 0cm 0cm, width=0.325\textwidth]{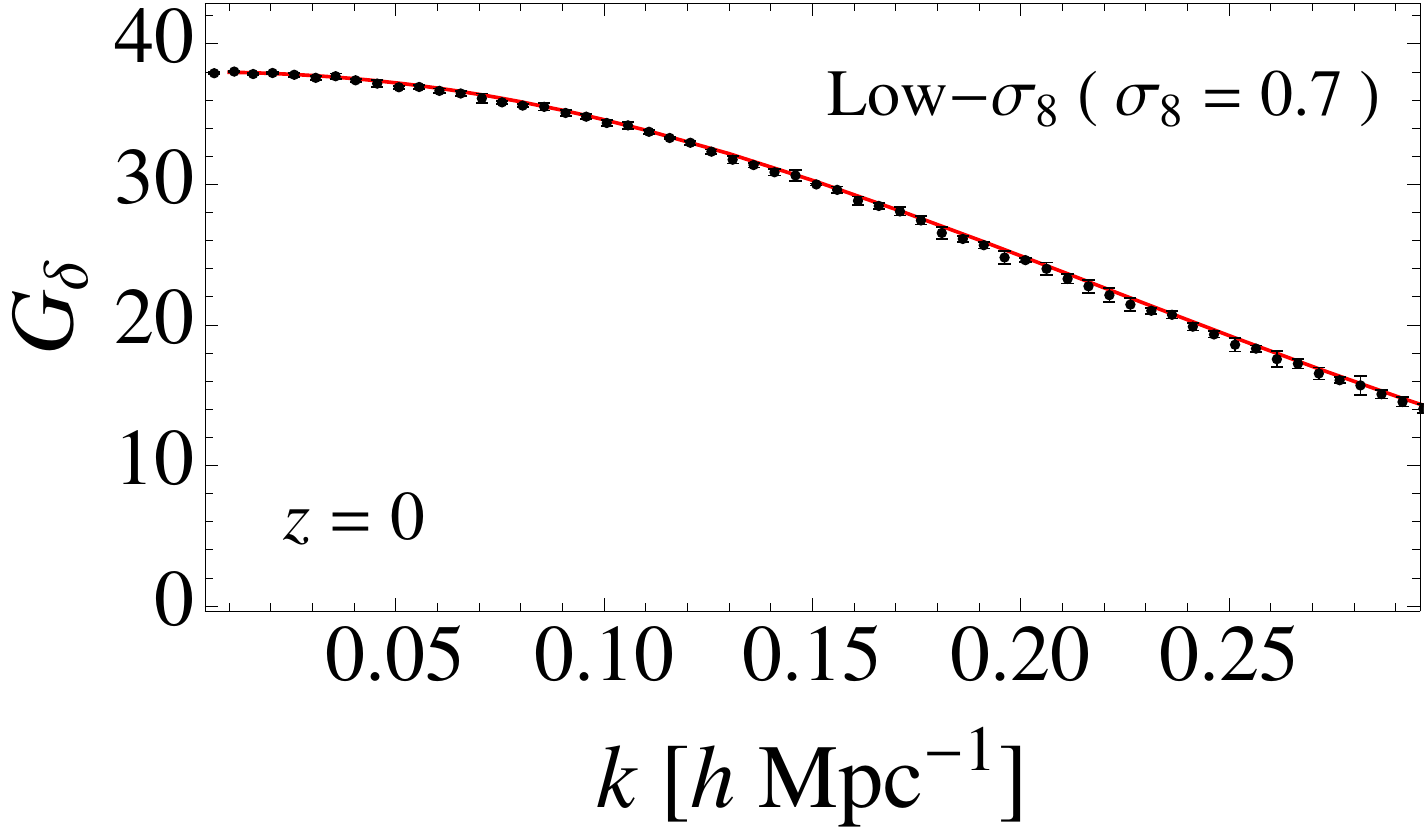} 
\includegraphics[trim = 0cm 0cm 0cm 0cm, width=0.325\textwidth]{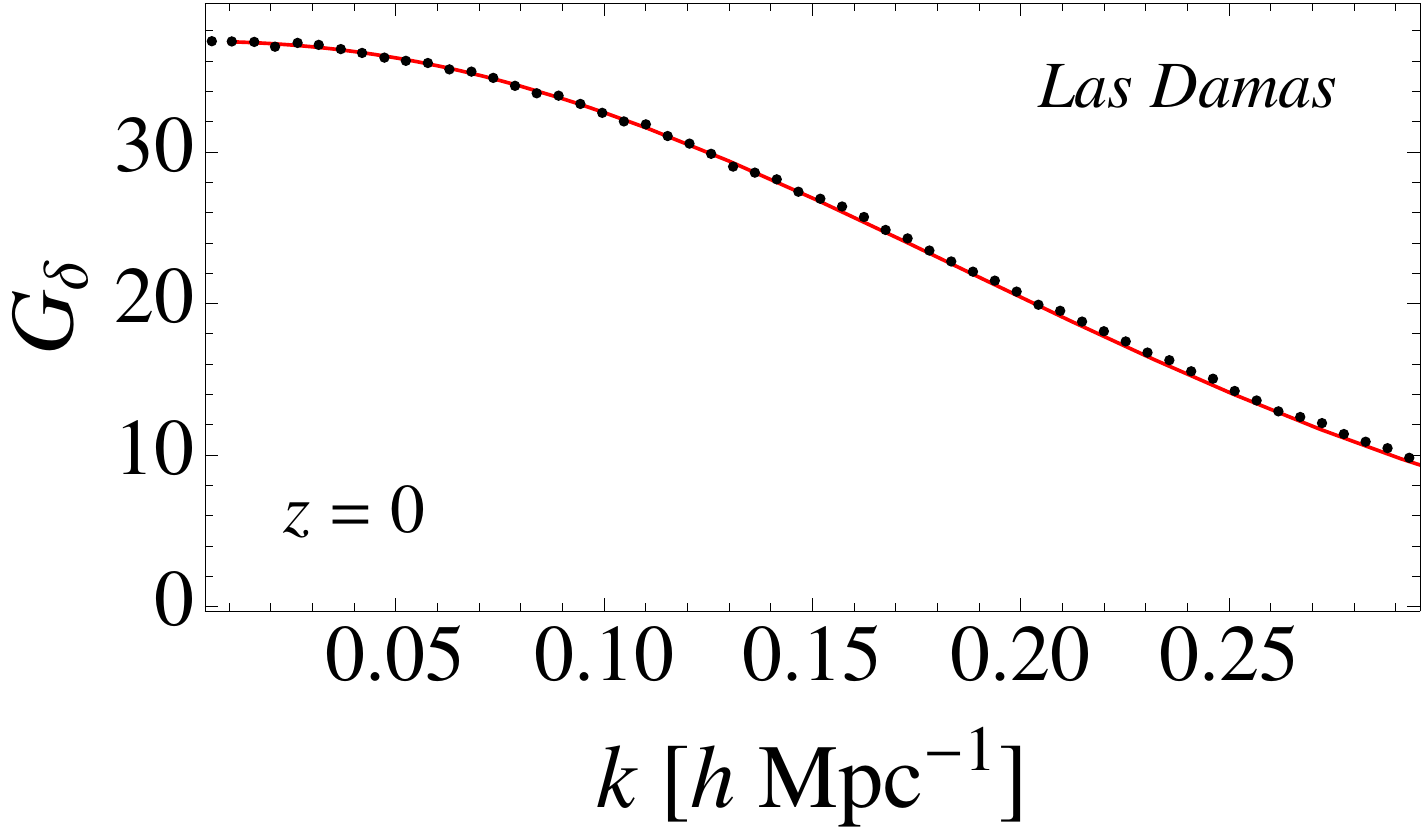} 
\caption{{\it Nonlinear two-point propagator at
    z = 0 for different cosmological models.} In each panel symbols with error
  bars correspond to the measurements of the propagator over four 
  simulations of the given cosmology. In solid red lines  we show the prescription 
  used throughout this paper corresponding to the exponentiation of
  the most-growing one-loop contribution:
  $D_+ \exp{\left[D_+(z)^2f(k)\right]}$. It agrees with the
  measurements at the sub-percent level for all the scales of interest
  in all cases studied.} 
\label{fig:cosmologies-prop-z0}
\end{center}
\end{figure*}

It should be noted that all the required integrals safely converge for any realistic shape of the matter power spectrum. The regularization scale
$\sigmav$ is finite as soon as the power spectrum is steeper than
$k^{-1}$ in the low wave-mode limit ($n > -1$). All terms involved in the computation 
of the power spectrum actually require the same conditions in the IR domain. The convergence 
in the UV domain is more diagram dependent. The condition for the existence of $\sigmav$ demands that $n<-1$ if $n$
is the power spectrum index at large wave-modes. The convergence properties of $P_{\rm 1-loop}(k,z)$ and $P_{\rm 2-loop}(k,z)$
are all determined by the behavior of the $\gamma_{abc}(\vk,\vq,\vk-\vq)$ (symmetrized) vertex  functions, introduced in Eq. (\ref{gammaintro}).  
They obey the scaling relation,
\begin{equation}
\gamma_{abc}(\vk,\vq,\vk-\vq)\sim \frac{k^{2}}{q^{2}},
\end{equation}
when $q\gg k$. As in the expressions of $P_{\rm 1-loop}(k,z)$ and $P_{\rm 2-loop}(k,z)$, two such factors are introduced 
in the high $q$ limit, the convergence of the former is secured as soon as $n<1/2$ and that of  the latter
when $n<-2/3$. It therefore does not lead to new constraints besides the existence of $\sigmav$. This is at variance with 
the convergence properties of the multi-loop corrections to the two-point propagators. They indeed lead
to much stringent constraints in the UV domain as discussed in \cite{2012arXiv1208.1191T}. We will comment further on the consequences 
of those properties in Sec.~\ref{sec:regPT}.

To perform the integration of Eq.~(\ref{eq:2loops}) we employ a 
MonteCarlo algorithm routine called {\tt Vegas} 
\citep{1978Lepage,1980Lepage}, included within the
{\tt v1.4 CUBA} library of multidimensional numerical integration routines
described in  \cite{2005CoPhC.168...78H}\footnote{Publicly available
  at http://www.feynarts.de/cuba/}. This is discussed in more detail
in Sec.~\ref{sec:performance}.

We are now ready to compute the model prediction and compare it
with measurements in N-body simulations.
Figure~\ref{fig:power} shows the two-loops multipoint expansion 
in solid blue line, linear theory in dotted red and {\tt halofit} \citep{2003MNRAS.341.1311S} in solid
black, at $z=0,0.5$ and $1$ (top to bottom panels).
Symbols with error bars are the corresponding measurements of
$P(k)$ in our fiducial cosmological model (error-bars correspond to the
variance over the ensemble of $50$ simulations, see Table~\ref{cosmologies}). All lines are divided by a smooth
broad-band linear spectrum to reduce the $y$-axis dynamic
range. Overall, the multi-point expansion match the measurements at
the $2\%-3\%$ level up to scales $k \sim 0.16 \kvecMpc, 0.19 \kvecMpc, 0.23 \kvecMpc$
at $z=0,0.5$ and $1$ respectively. These values coincide roughly with
$\sigmav^{-1}$ at the given redshift, with $\sigmav$ given by
Eq.~(\ref{eq:sigv}). Notice also that these scales are slightly beyond
the Baryon Acoustic Oscillations region. Improving upon this $\sim 2\%$ accuracy
requires a more precise ansatz for the multi-point
propagators, in particular for $\Gamma^{(3)}$ because it dominates
within this region ($k \gtrsim 0.1 \kvecMpc$).
 
We have also investigated whether the addition of the next term in the
expansion of Eq.~(\ref{eq:gamexpansion}), that corresponds to a 3-loop computation,
extends the agreement to higher wave-modes. The result is that it
does but only slightly because of the already strong exponential
suppression of the corresponding five-point propagator. In turn, the
numerical evaluation of the 3-loop integral becomes very
lengthy. An alternative path is to combine our PT approach
with halo model prescriptions, in the spirit of \cite{2011A&A...527A..87V}. This will be
the subject of future work.

For our FID cosmology we have checked that the recent ansatz by
\cite{2012JCAP...04..013T} 
under-estimates the N-body measurements by $4\%$ at $z=0$, 
gives a $\sim 1\%$ match at $z=0.5$ and over-estimates $P(k)$
by $\sim 2\%$ at $z=1$ (at BAO scales). This ansatz is 
based upon a split between long and short wave-modes, which was somewhat
arbitrarily choosen to be a sharp $k$-cutoff at $\Lambda=k/2$.
The strength of nonlinear
corrections in the model, hence also the above residuals, depends
(systematically with redshift) on $\Lambda$.

\subsection{\MPTbreeze : Code performance and convergence}
\label{sec:performance}

The major advantage of the method presented in the previous section is
that the evaluation time is of the order of a few seconds, comparable to that of linear Boltzman
codes used to compute the transfer function of different species. This
is opposed to resummation techniques such as RPT and Closure Theory that take significantly longer times to compute. Hence our approach it is very well suited for sampling the large-scale structure likelihood of present and future datasets at BAO scales.

In addition note that due to the structure of the expansion the same
evaluation done for a given redshift can be properly re-scaled to
another one at almost no extra cost, just recomputing $G$ and the
corresponding growth factor.

It seems then appropriate to discuss the performance of the code in more
detail and the numerical convergence. Clearly, the component that is numerically
intensive to evaluate is the 5 dimensional ``2-loop'' integration
given in Eq.~(\ref{eq:2loops}). As mentioned before we have perform it
with the Monte Carlo
algorithm {\tt Vegas} that uses importance sampling as a variance-reduction
technique. 

The accuracy of the integration, and hence the evaluation time, is controlled by
setting the required absolute $\varepsilon_{\rm abs}$ and relative
$\varepsilon_{\rm rel}$ errors\footnote{Convergence is
  achieved once the estimation $\hat{I}$ of the integral $I$ satisfies
$|\hat{I}-I|\le {\rm max}(\varepsilon_{\rm abs},\varepsilon_{\rm rel}
I)$} \citep{2005CoPhC.168...78H}.

For $\varepsilon_{rel}=1\%$ (with $\varepsilon_{\rm abs}=0$) the
full computation of the $z=1$ power
spectrum including tree, one and two loops in steps of $\delta
k = 0.005\kvecMpc$ (roughly the fundamental mode in our simulations) up to $k_{max}=\sigmav^{-1}=0.244 \kvecMpc$ 
($49$ bins) {\it evaluates in only $5$ secs}. The same integration with $\varepsilon_{rel}=0.5\%$
requires $10$ secs and roughly $3$ minutes for
$\varepsilon_{rel}=0.1\%$.
Moreover, we find that generally the three accuracies yield the same
power spectrum to a sub-percent level thus $\varepsilon_{\rm rel}=1\%$ (the
fastest) seems
a reasonable choice.

A similar test at $z=0.5$ is even quicker because the validity of the
expansion is more limited. Now $k_{max}=\sigmav^{-1}=0.195 \kvecMpc$
($39$ bins) {\it evaluates in just $\sim 3$ secs.} setting
$\varepsilon_{\rm rel}=1\%$ ($7$ secs. with $\epsilon_{rel}=0.5\%$).
Notice that all the timings reported are for single CPU (and the code
compiled with the Intel compiler {\it ifort}).

In all cases the numerical integration in momentum space is done from
a fixed $q_{min}\sim 10^{-4}$ up to some cut-off scale $q_c$. A priori
the box-sizes of our simulations are large enough that wave-modes
longer than $L_{box}$ have negligible amplitude to alter the measured power
spectrum in the simulations. Hence we assume no finite box-size effect
and take $q_{min}$ arbitrarily small. On the other hand we find that
we need $q_c \sim 1 \kvecMpc$ in order for the integrals to
converge within $0.5\%$. This mild sensitivity to the ultraviolet (UV) cut-off is due to
the particular prescription for MP adopted here, with the decay standing
as an overall multiplicative function 
($\Gamma^{(n)}=G(k) \times F_n$) that factors out of loop integrals
such as Eqs.~(\ref{eq:oneloop},\ref{eq:twoloop}). This is unlike
approaches such as RPT where the full
nonlinear propagator is integrated inside all loops, what
screens much strongly the UV regime where e.g. shell-crossing enters
(and convergence is achieved already for $q_c \sim 0.5 \kvecMpc$).

Lastly we have also performed our integrals with other Monte Carlo algorithms implemented
within the {\tt CUBA} library, reaching always the same answer 
obtained with {\it Vegas} but employing more time to converge. For
example {\it Suave} (importance sampling with globally adaptive
subdivisions) employs 1 minute 25 seconds to evaluate the $z=1$
spectrum for $\varepsilon_{\rm rel} = 0.005$. {\it Divonne} \citep{1981aFriedman,1981bFriedman},
which uses stratified
sampling, requires 61 seconds for
the same task. As described above, {\it Vegas} is substantially
faster requiring only 10 seconds (or less for higher $\varepsilon_{\rm rel}$).

\begin{figure}
\begin{center}
\includegraphics[trim = 0cm 1cm 0cm 0cm, width=0.36\textwidth]{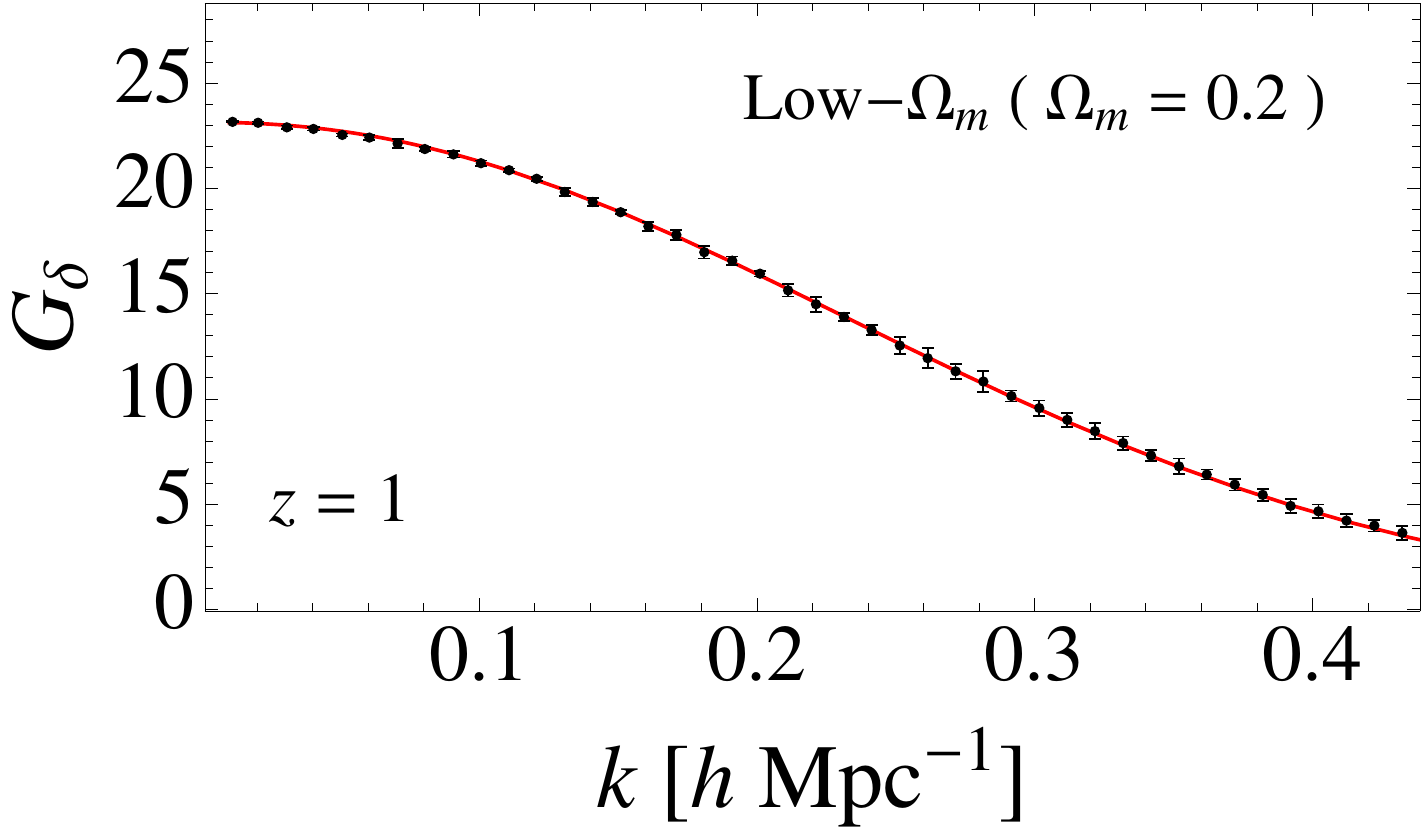} \\
\includegraphics[trim = 0cm 1cm 0cm 0cm, width=0.36\textwidth]{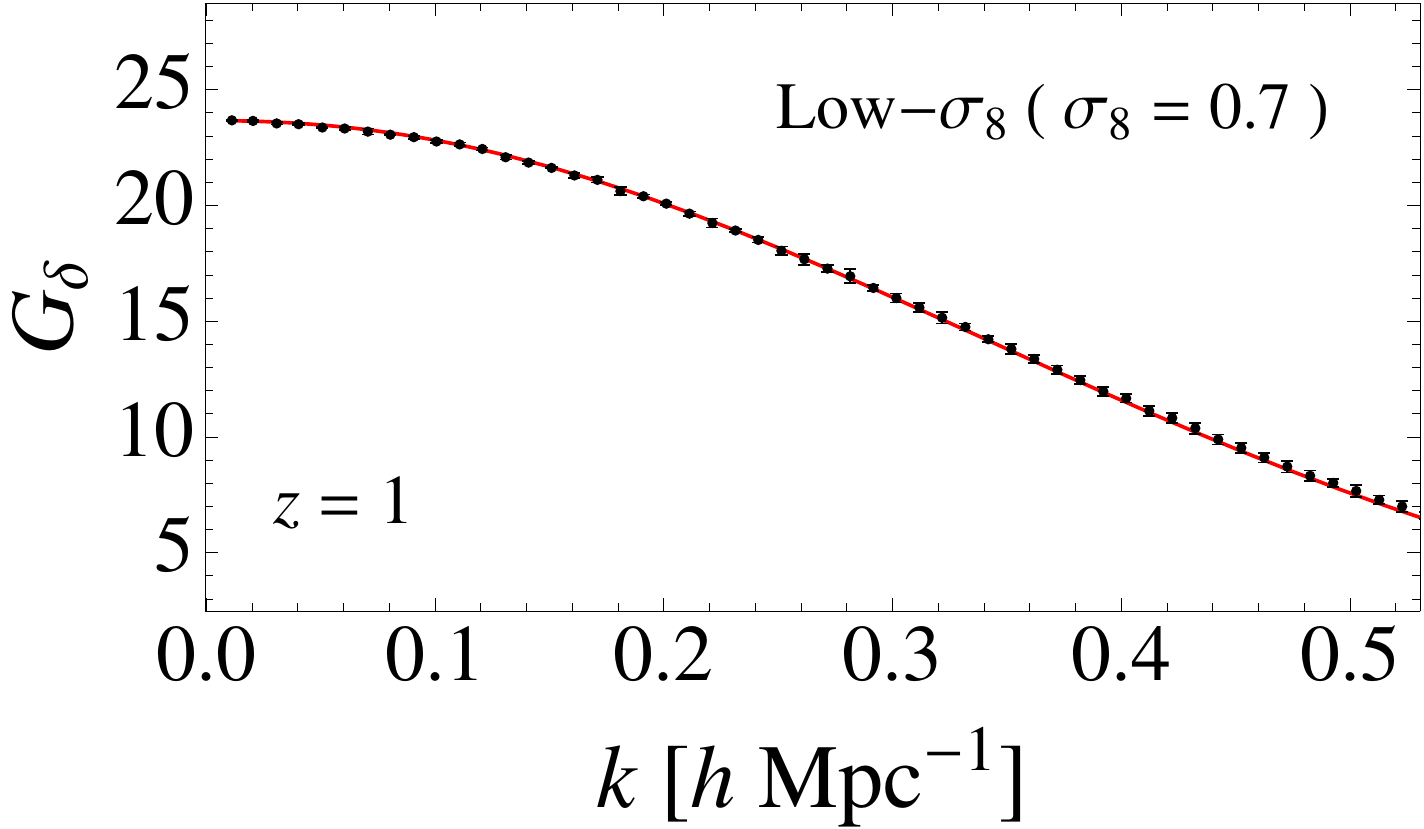} \\
\includegraphics[trim = 0cm 0cm 0cm 0cm, width=0.36\textwidth]{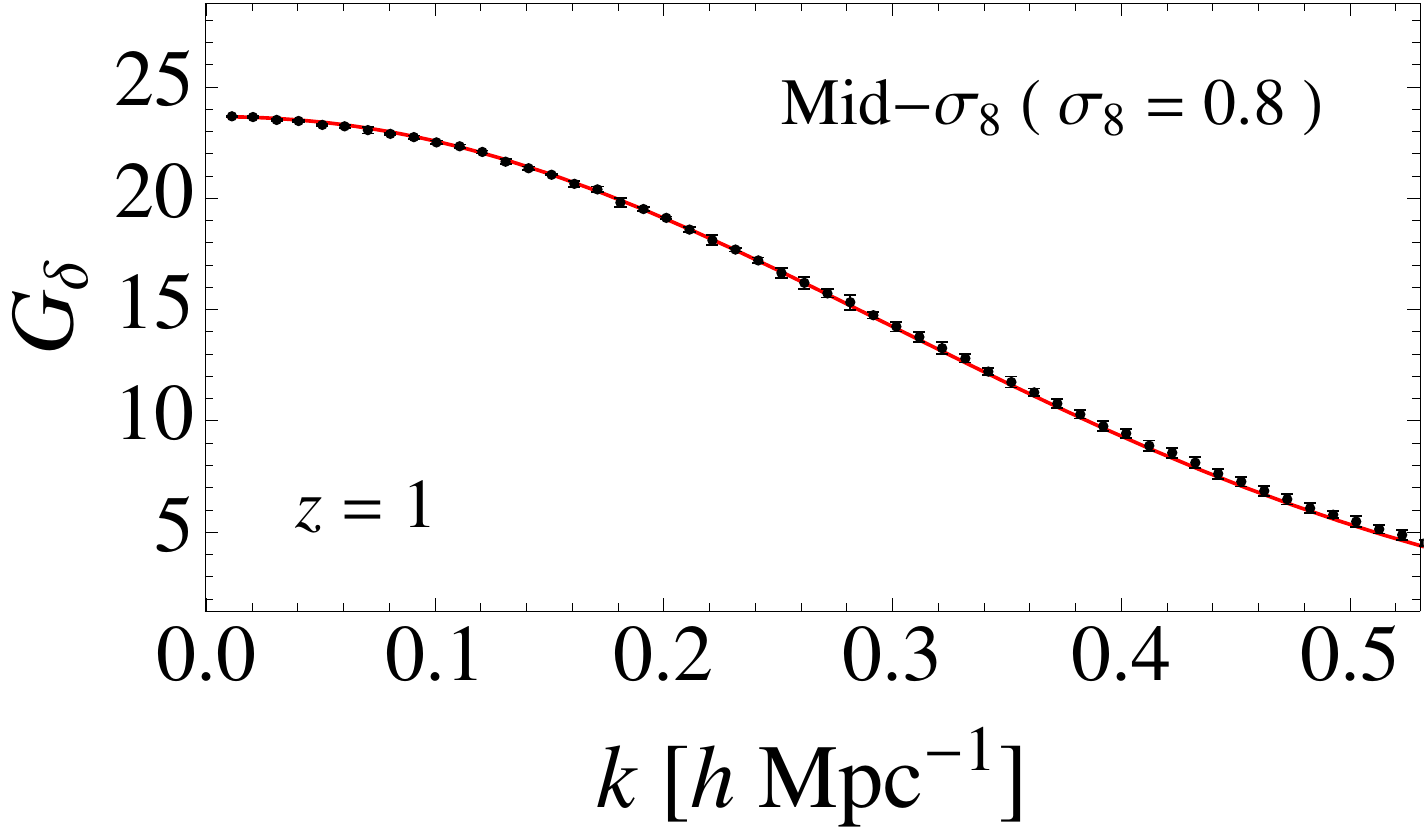} 
\caption{{\it Same as Fig.~\ref{fig:cosmologies-prop-z0} but at z = 1}
  (and for three cosmological models). The simple model for $G_\delta$
  (solid red line) 
  still performs at the sub-percent level in all scales and models.} 
\label{fig:cosmologies-prop-z1}
\end{center}
\end{figure}

\subsection{{\MPTbreeze} compared to {\RegPT}}
\label{sec:regPT}

\begin{figure*}
\begin{center}
\includegraphics[trim = 0cm 1.3cm 0cm 0cm, width=0.32\textwidth]{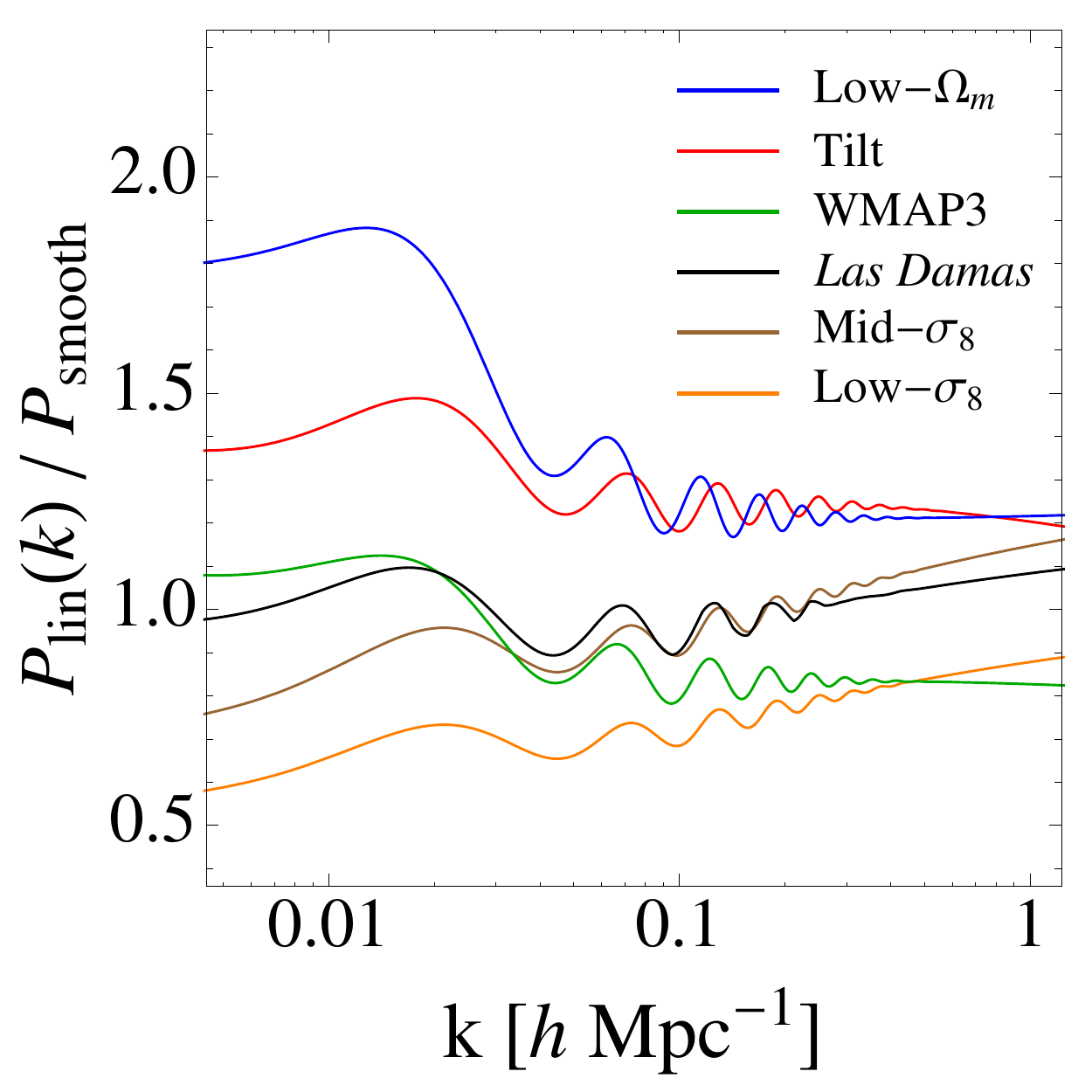}
\includegraphics[trim = 0cm 1.3cm 0cm 0cm, width=0.32\textwidth]{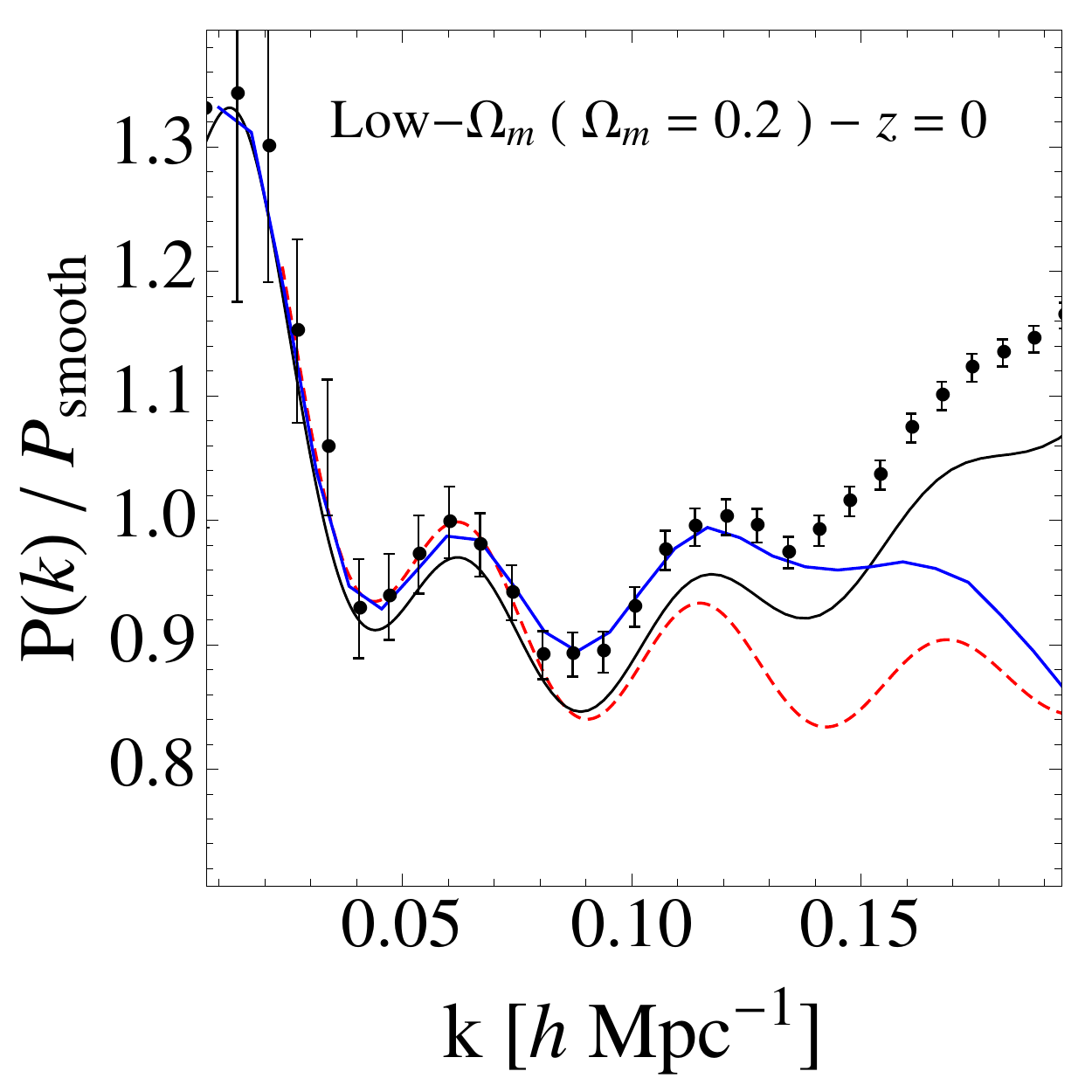}
\includegraphics[trim = 0cm 1.3cm 0cm 0cm, width=0.32\textwidth]{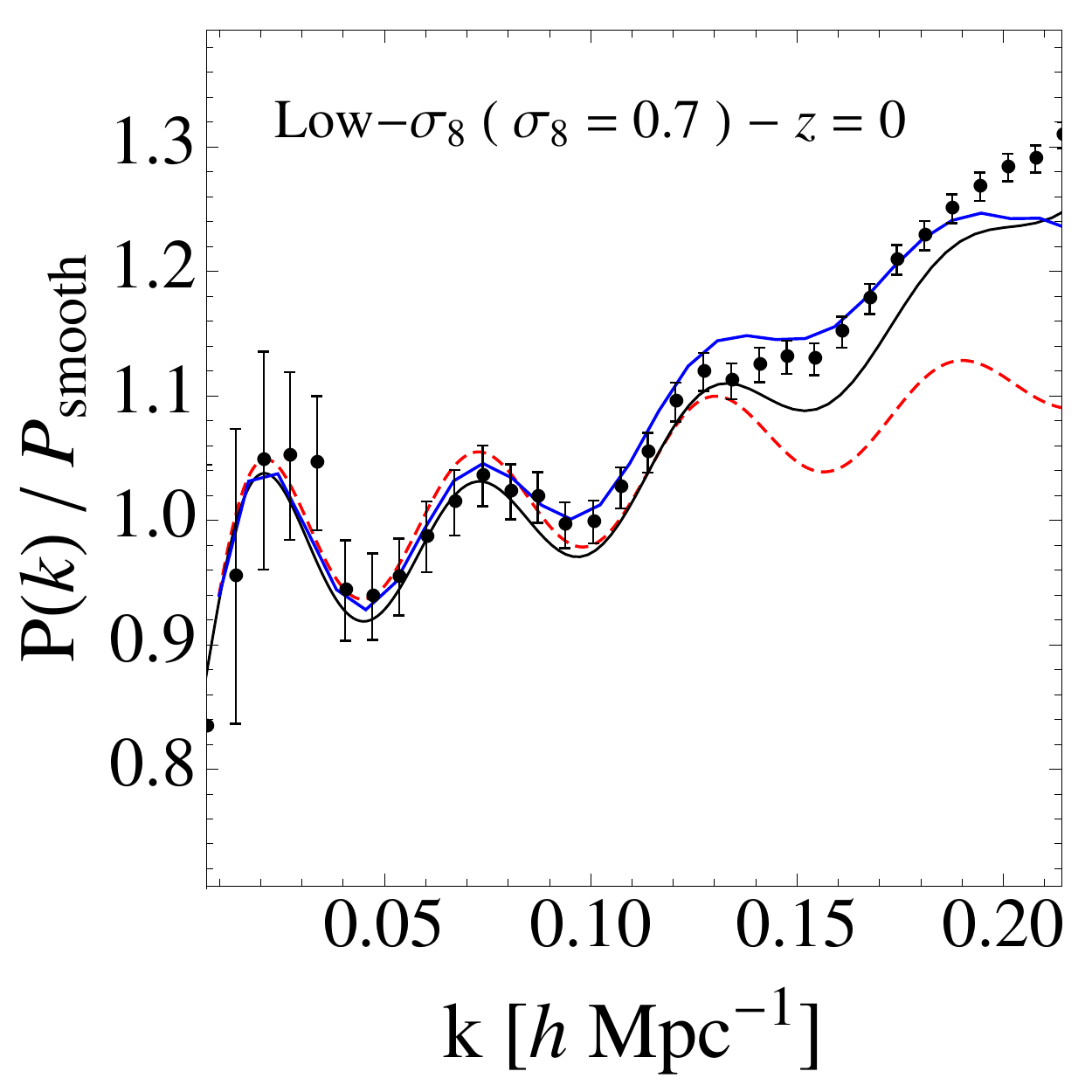} \\
\includegraphics[trim = 0cm 0cm 0cm 0cm, width=0.32\textwidth]{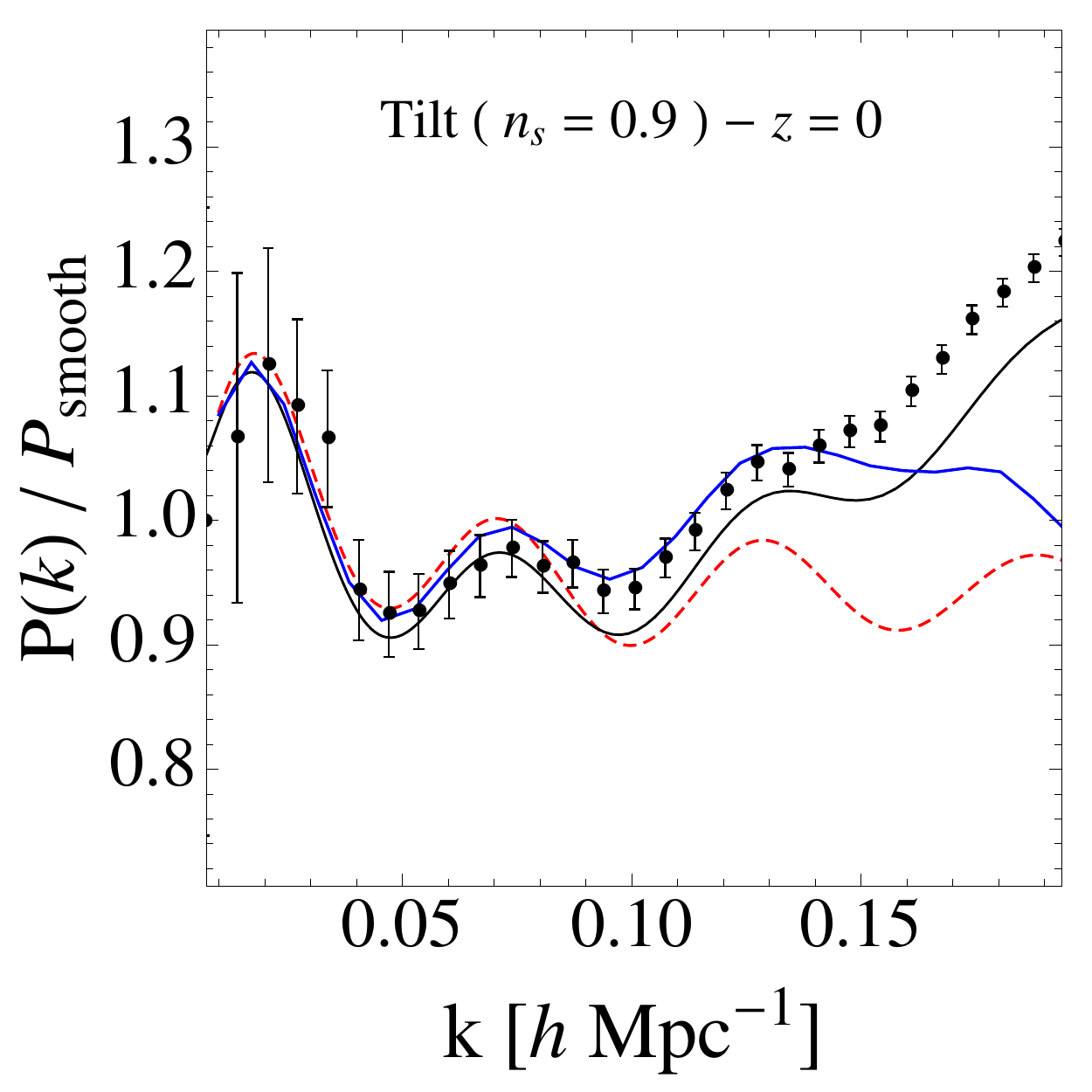}
\includegraphics[trim = 0cm 0cm 0cm 0cm, width=0.32\textwidth]{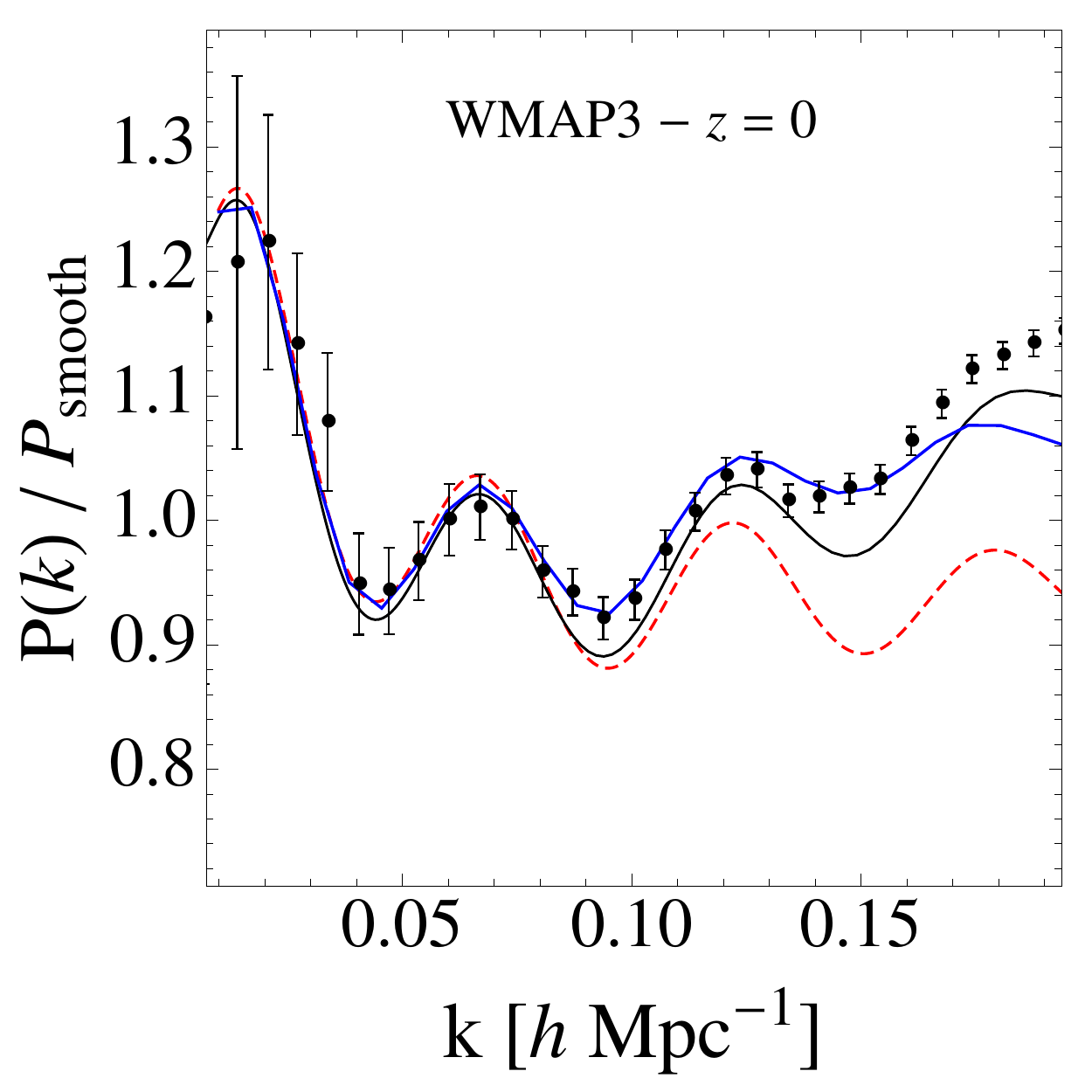}
\includegraphics[trim = 0cm 0cm 0cm 0cm, width=0.32\textwidth]{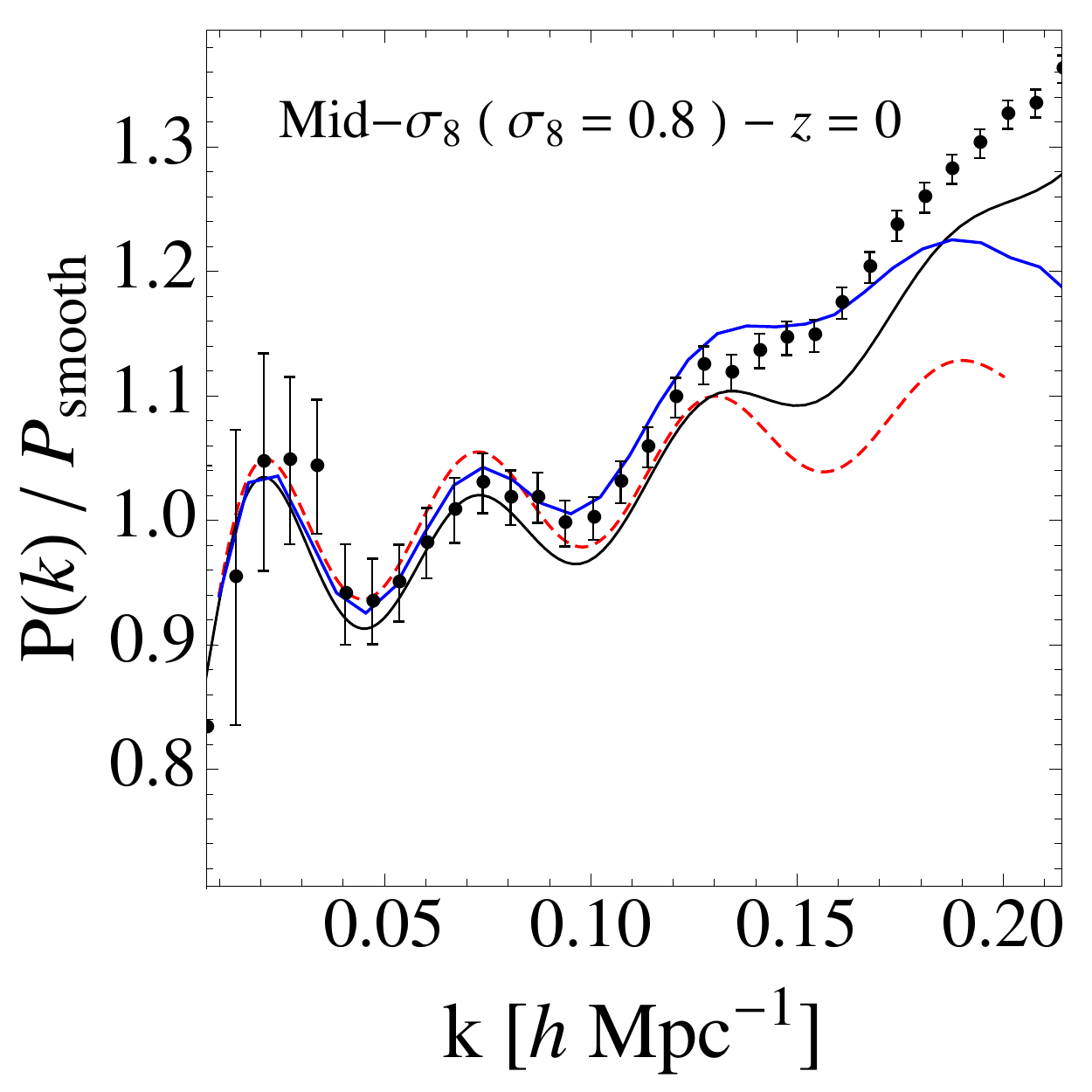}
\caption{{\it Power spectrum performance for different cosmological
    models at z=0
  .} Top left panel shows the linear power spectra in each
  of the simulated cosmologies divided by a smooth
  broad-band power. 
  Remaining panels show measurement of $P(k)$ over four independent simulations for each model.
  We used the same $P_{\rm smooth}$ (except for its normalization) for all panels. 
  In these panels dashed red lines correspond to linear theory, black
  solid lines to {\tt halofit} and solid blue lines to the the multi-point expansion
  presented in this work (see text for more details). The later is
  found accurate at the $\sim 2\%$ level on BAO scales throughouth all
cosmologies investigated.} 
\label{fig:cosmologies-power-z0}
\end{center}
\end{figure*}

It is possible to construct alternative implementations of MP resummations. {\RegPT} is an alternative proposition that relies on
slightly different choices for the computation of the multi-point functions
and on numerical implementations \citep{2012arXiv1208.1191T}. 
Let us first stress that to a large extent both approaches share the same advantages and disadvantages: the power spectrum is constructed out of a sum of positive terms and each of these terms could be computed on its own; the resulting power spectrum exhibits a large-$k$ cutoff which signals the limit of the validity range of the computation.
The main difference in the implementations comes from the fact that, in the {\RegPT} case, all two loop order terms are taken into account following the prescription proposed in \cite{2012PhRvD..85l3519B} in the computation of the propagators. This is not the case in this work where the loop corrections to the propagators are computed with a more phenomenological approach. 

Because of these differences, the computational difficulties of the two approaches vary. The computation of the expression (\ref{eq:oneloop})
when the one-loop correction to $\Gamma^{(2)}$ is taken into account
reveals quite costly for a direct computation. The {\MPTbreeze}
implementation avoids this difficulty. Even with direct Monte-Carlo
integrations it is then possible to obtain the results within a very
short time\footnote{In \cite{2012arXiv1208.1191T}, it will be shown that it is 
possible to considerably shorten the CPU time required to compute the
diagram involved in the {\RegPT} prescription with the use of a ``fast'' algorithm. The latter is based 
on the use of precomputed kernel functions making possible the
computation of diagrams in $\sim$ 0.01 secs. per mode. 
We note that such a procedure can also be applied in the context of {\MPTbreeze}.}.

The other main difference concerns the converging properties of the
involved diagrams. As shown in Sect. \ref{sec:powerspectrum}, diagrams
involved in the {\MPTbreeze} computations have good converging properties. In practice, for wave modes of the order of $0.1 - 0.5\ h/$Mpc
the results depend on wave modes that are of comparable scales.
This is not necessarily the case for the {\RegPT} implementation. As it
will be stressed in a forthcoming paper
, the
two-loop corrections to the two-point propagator are sensitive in
particular to wave-modes well above a few $\kvecMpc$. The use of these
two-loop corrections improves upon the two-point propagator predicted
value at least for $z\gapprox 1$ so that {\RegPT} predictions can
be potentially more precise. But it also makes the predictions less
robust with regards to the overall spectrum as it makes the
results more sensitive to nonlinear scales where baryon physics and shell crossings are likely to significantly affect the growth of structure.

\begin{figure*}
\begin{center}
\includegraphics[width=0.3\textwidth]{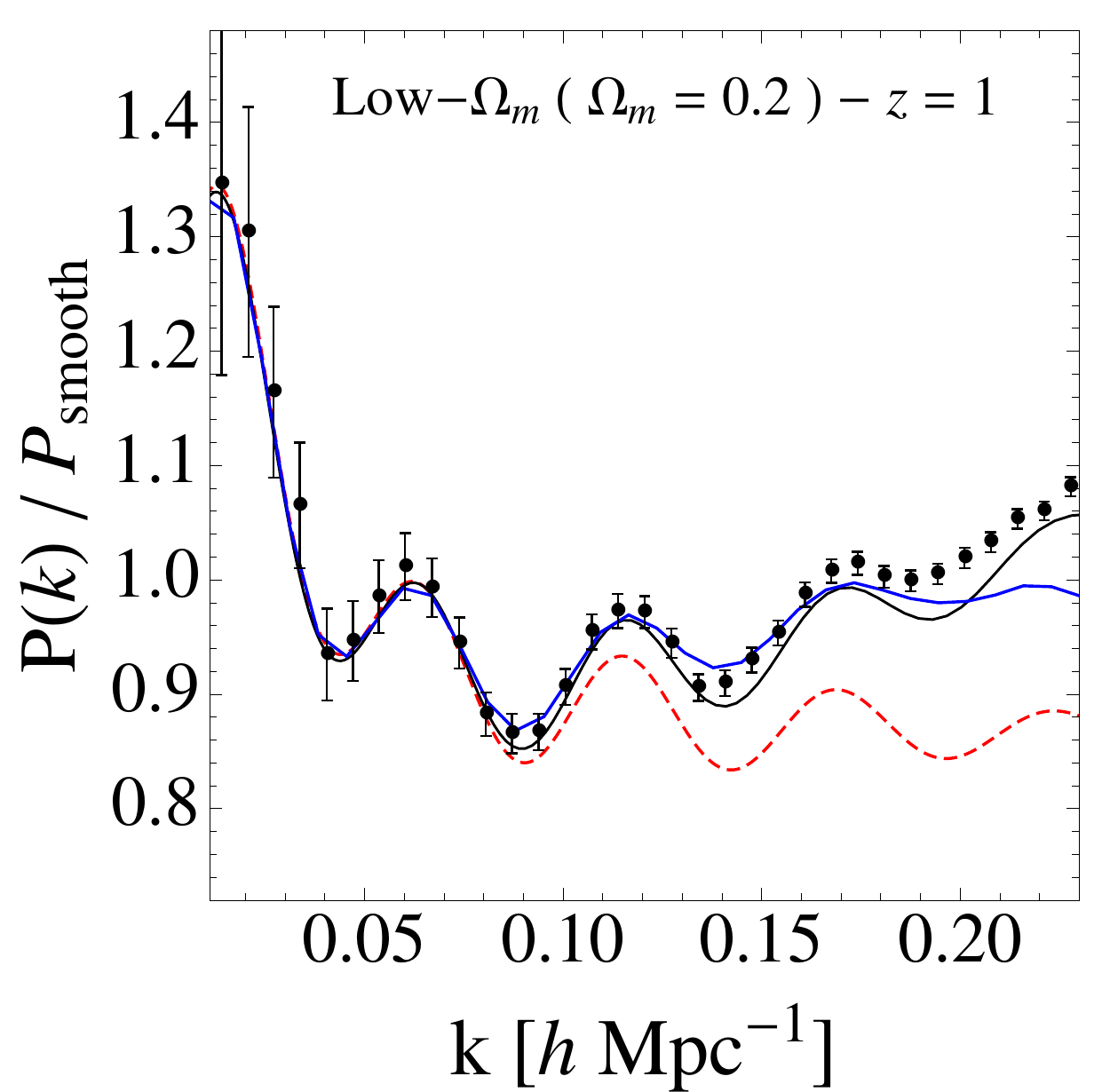}
\includegraphics[width=0.3\textwidth]{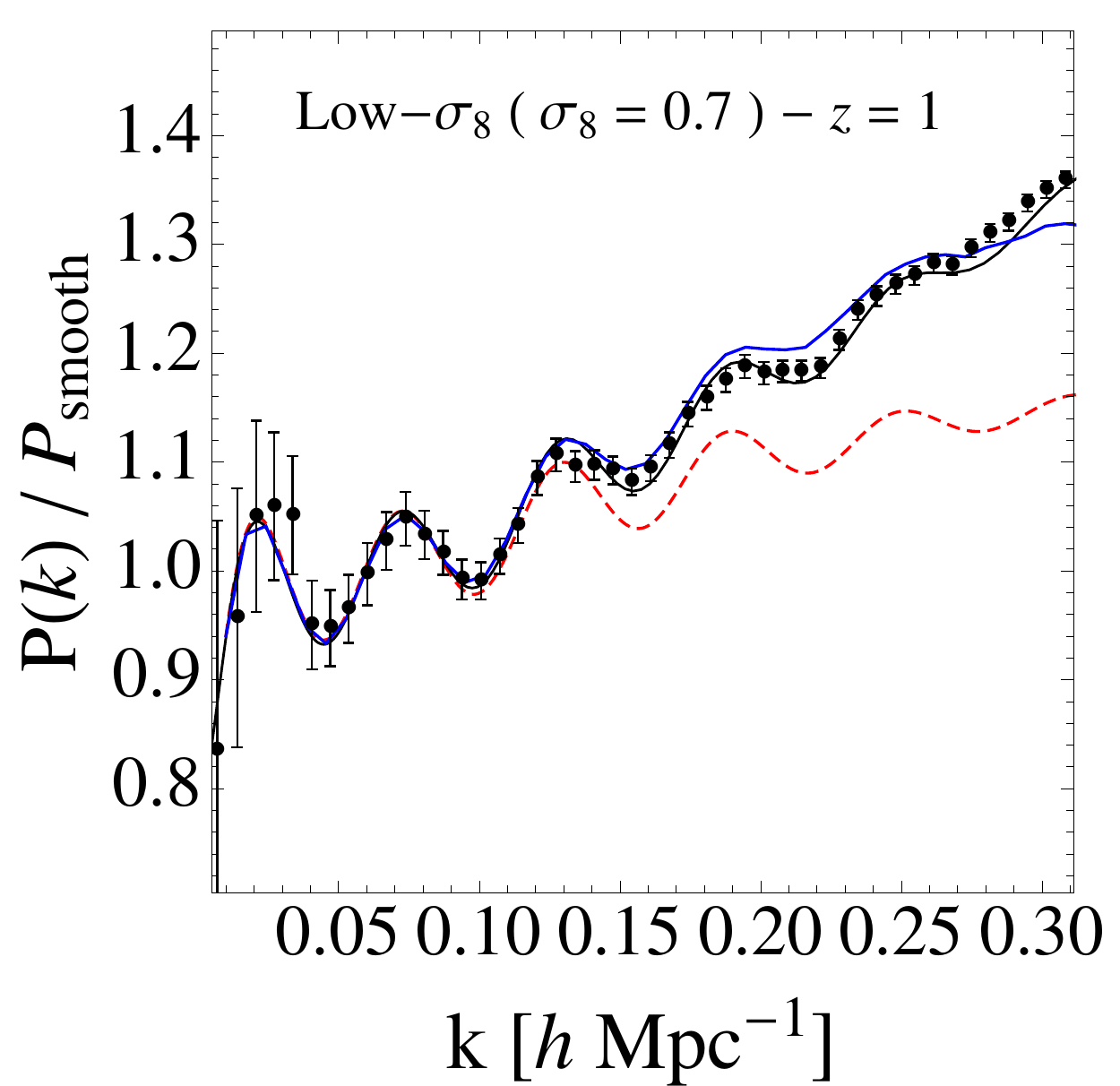} 
\includegraphics[width=0.3\textwidth]{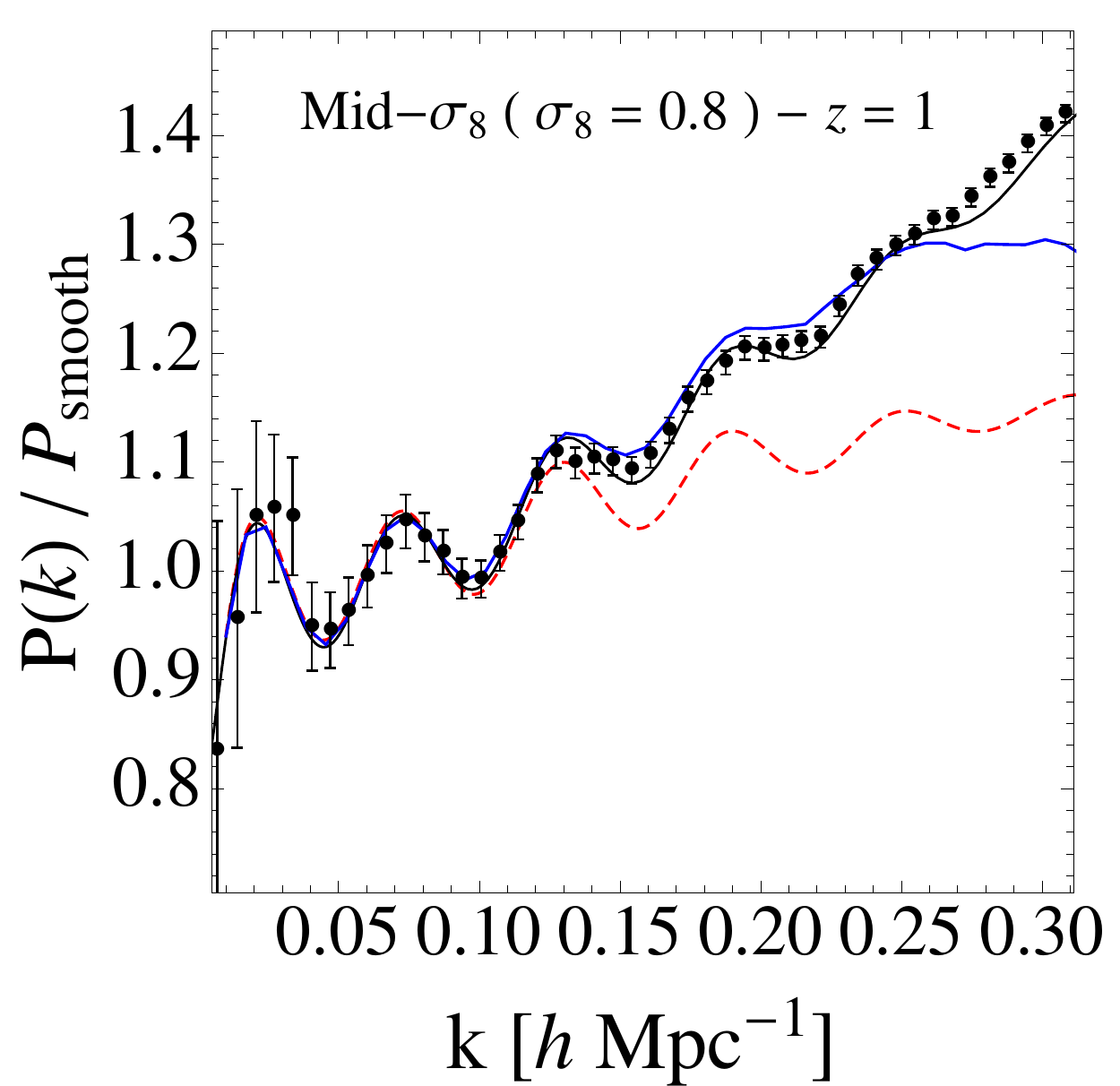}
\caption{{\it Power spectrum performance for different cosmologies at
    $z = 1$.} Symbols correspond to simulation measurements,
  dashed red lines to linear theory, black
  solid to {\tt halofit} and blue solid to the multi-point
  expansion, which remains accurate at $\lesssim 2\%$ on at least all scales showing
  Baryon Acoustic Oscillations.} 
\label{fig:cosmologies-power-z1}
\end{center}
\end{figure*}

\begin{figure*}
\begin{center}
\includegraphics[trim = 0cm 1.3cm 0cm 0cm, width=0.32\textwidth]{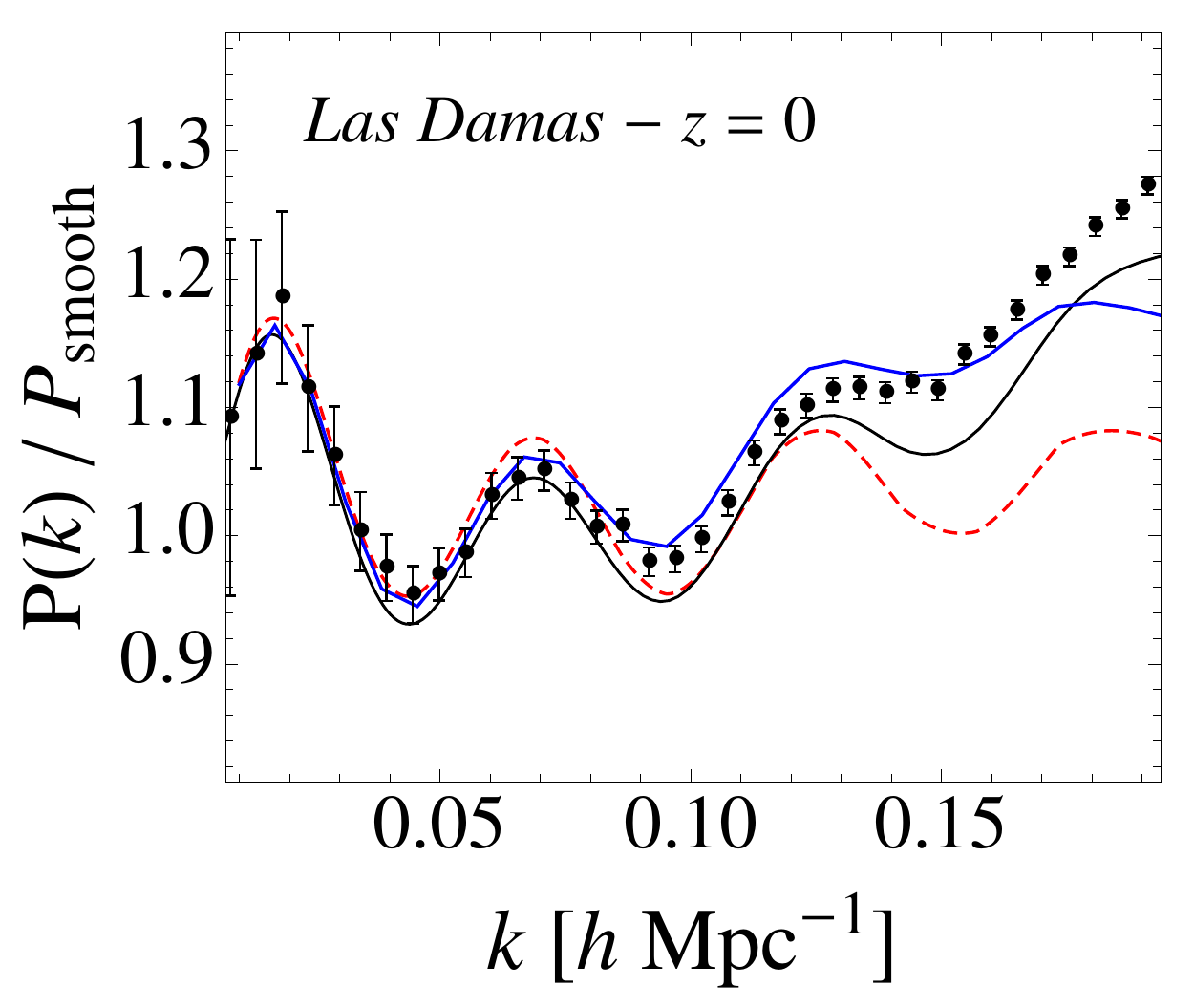}
\includegraphics[trim = 0cm 1.3cm 0cm 0cm, width=0.32\textwidth]{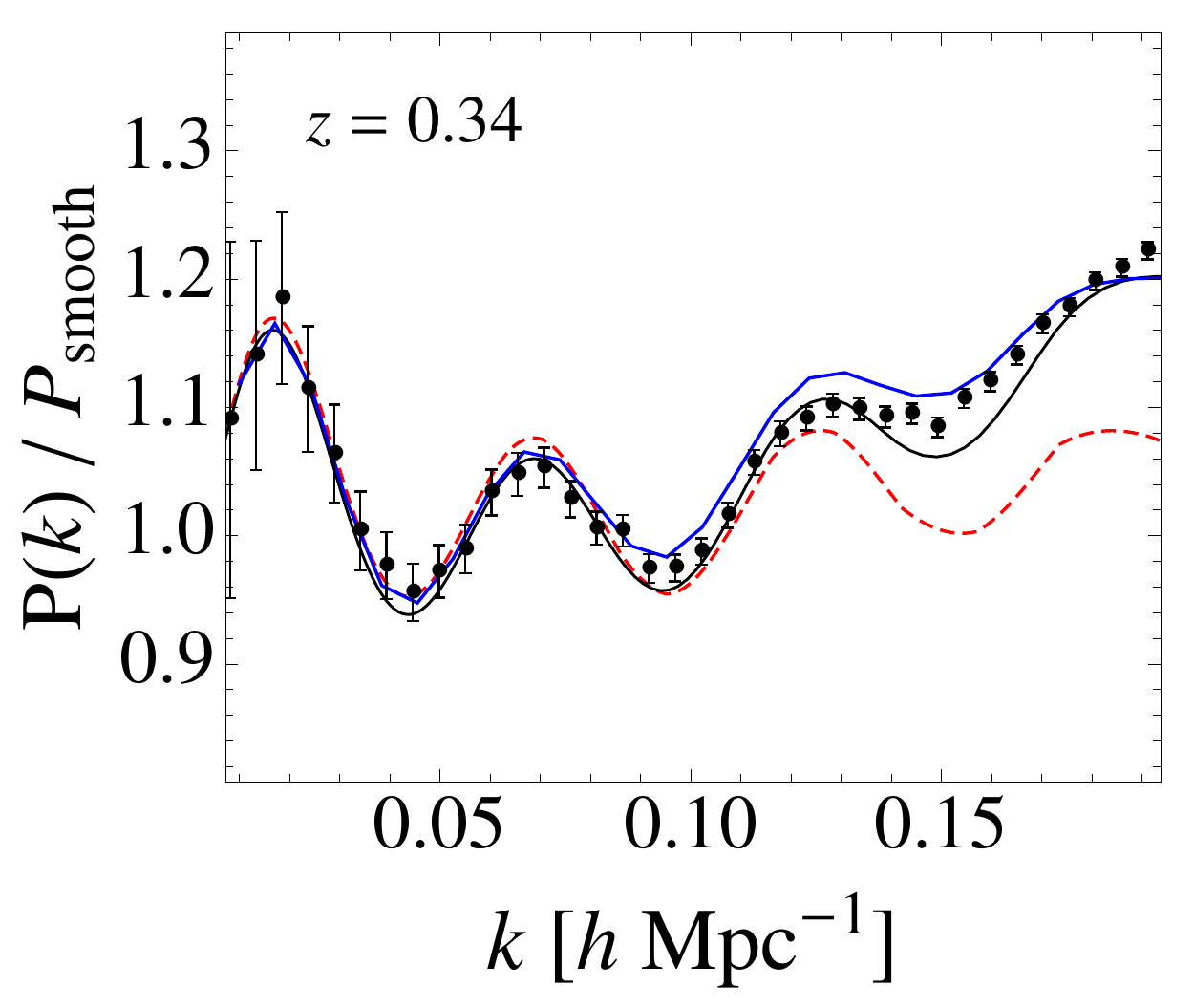}
\includegraphics[trim = 0cm 1.3cm 0cm 0cm, width=0.32\textwidth]{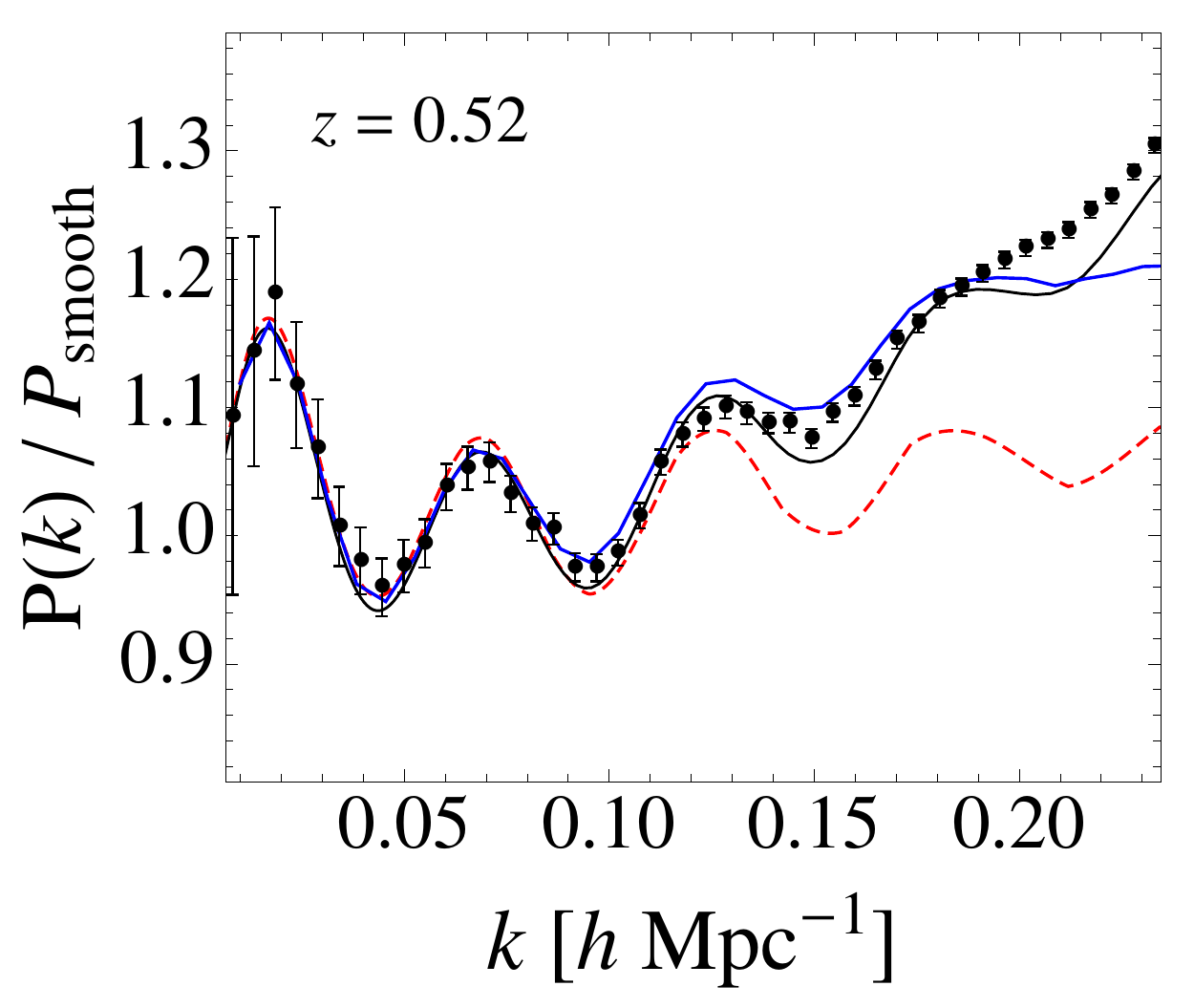} \\
\includegraphics[trim = 0cm 0cm 0cm 0cm, width=0.32\textwidth]{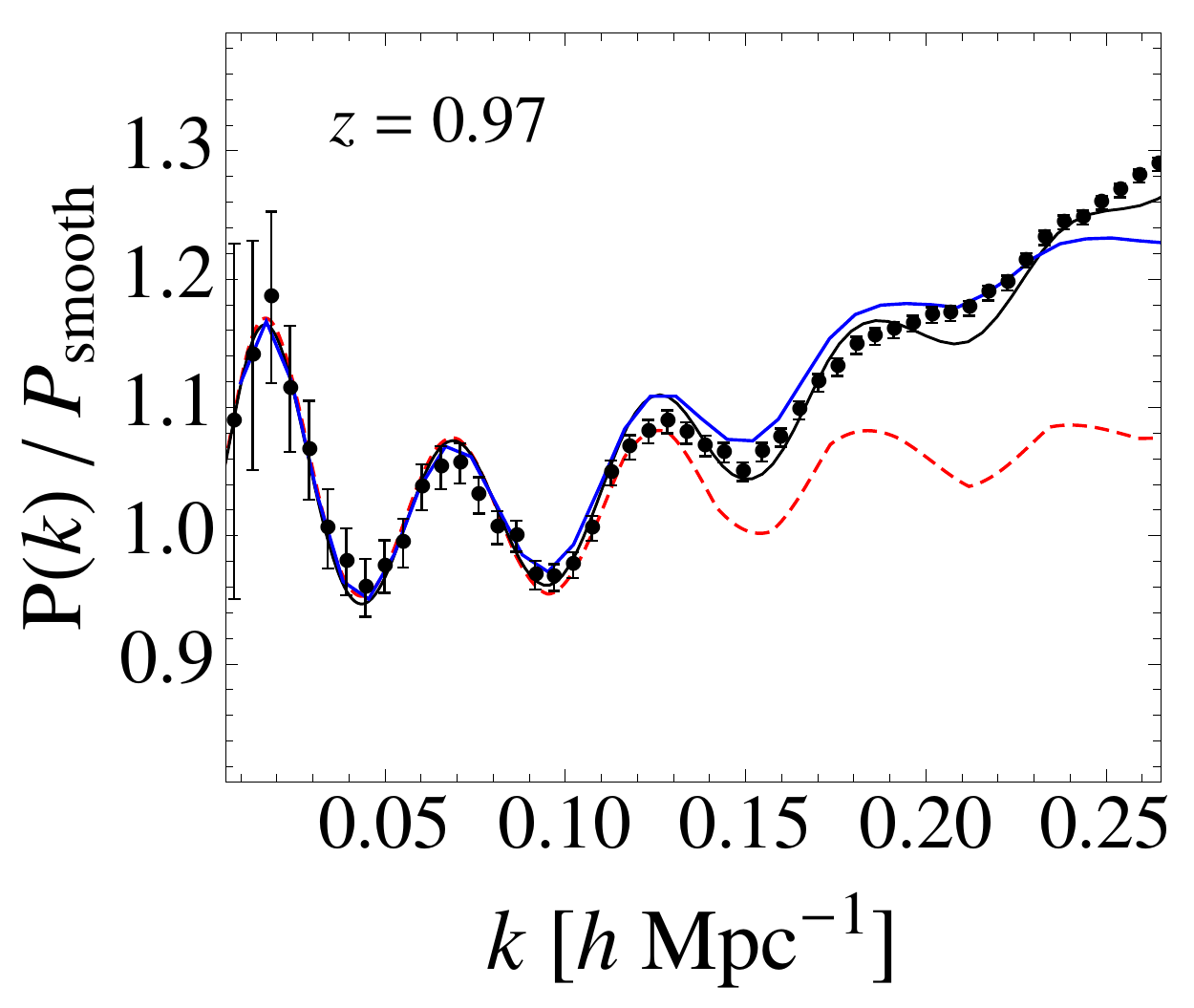}
\includegraphics[trim = 0cm 0cm 0cm 0cm, width=0.32\textwidth]{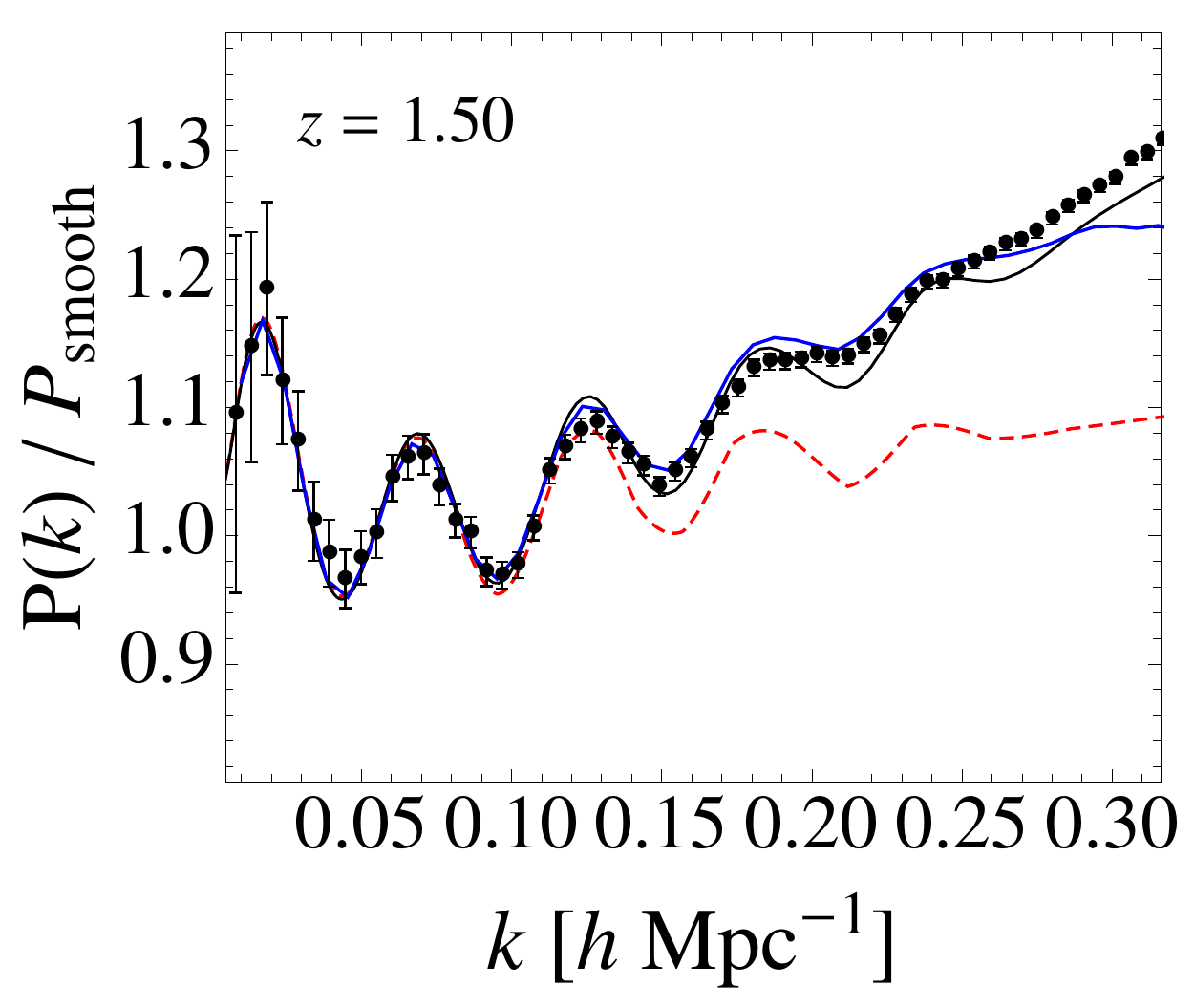}
\includegraphics[trim = 0cm 0cm 0cm 0cm, width=0.32\textwidth]{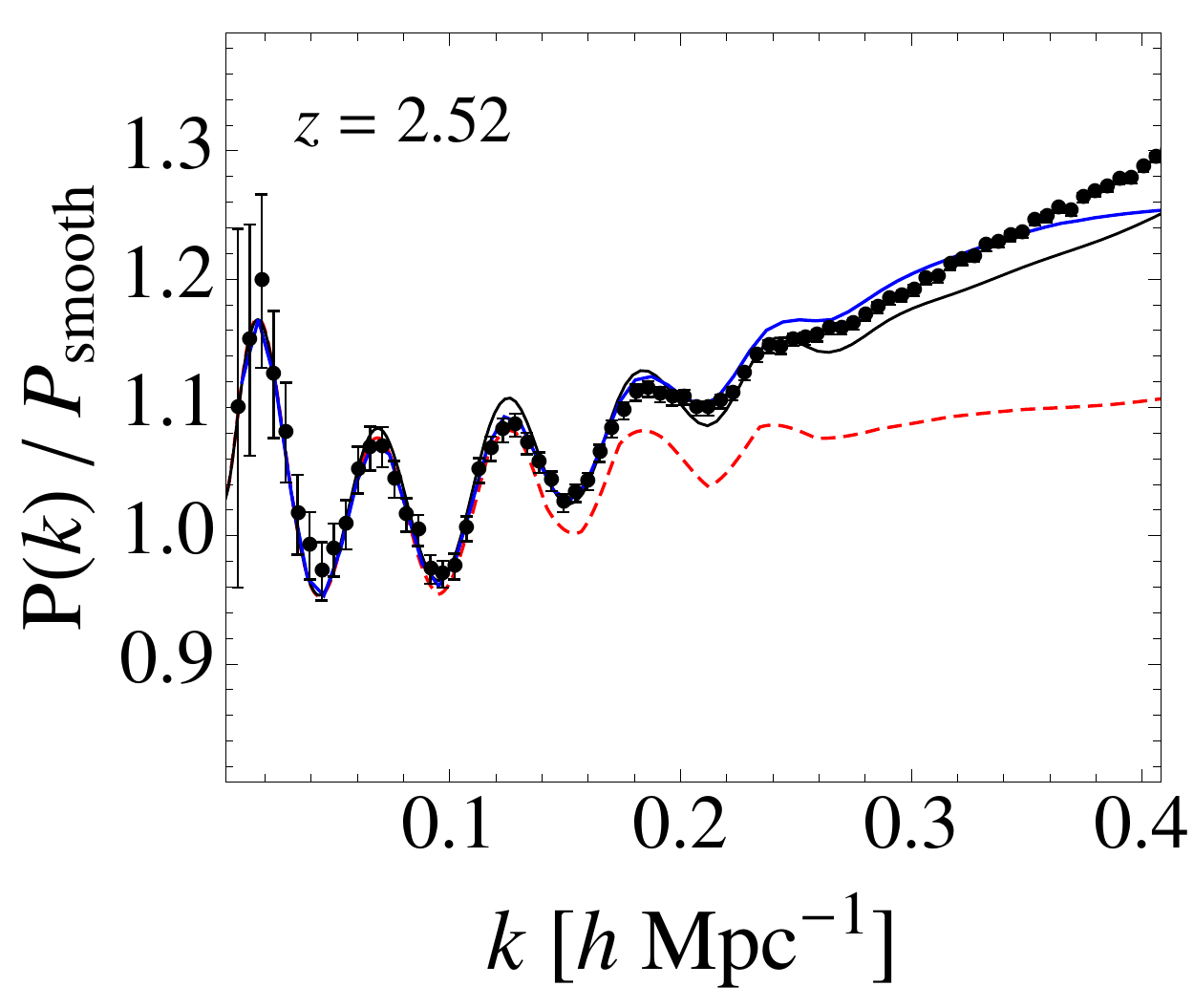}
\caption{Performance of the model presented in this paper for {\tt Las
    Damas} cosmology (detailed in Table~\ref{cosmologies}) as a function of redshift.
    Data points with error bars are the measurements in three runs of
    Oriana. Solid blue are the predictions
    by the multi-point expansion (this work), while red
    dashed and solid black corresponds to linear theory and {\tt
      halofit} respectively.} 
\label{fig:cosmologies-power-las-damas}
\end{center}
\end{figure*}

\section{Multi-point expansion vs. N-body for varying
  cosmological models}
\label{sec:cosmoperformance}

In this section we extend the testing of the multi-point expansion
beyond the fiducial cosmological model used so far to measurements in our
dedicated set of N-body simulations of various cosmologies described in Table \ref{cosmologies}.
This kind of study is important given the number of assumptions
leading to, say,
Eqs.~(\ref{eq:tree},\ref{eq:oneloop},\ref{eq:twoloop}). 
  It also helps setting in robust terms the validity and
usefulness of our approach. Our results are summarized in Figs.~\ref{fig:cosmologies-prop-z0}, \ref{fig:cosmologies-prop-z1}, \ref{fig:cosmologies-power-z0} and~\ref{fig:cosmologies-power-z1}. 

A clear picture of the different scenarios investigated is given in the
top-left panel of Fig.~\ref{fig:cosmologies-power-z0} where we show the ratio of linear spectra of
the simulated cosmologies to a reference smooth power\footnote{The
  smooth baseline used  throughout the paper is a 
  BBKS transfer function \citep{1986ApJ...304...15B} of shape $\Gamma
  = 0.135$ with tilt $n_s=0.99$ and
arbitrary normalization.}. 
The amplitude of the longest wavelength modes (i.e. $P^{1/2}$) varies by up to
$70\%$, and a similar spread is found for the effective tilt at $k
\sim 0.01\kvecMpc$.

Let us first concentrate in the prediction for the two-point propagator
given in Eq.~(\ref{eq:prop}) for the different cosmological
models and its comparison to simulation measurements. This is depicted in Fig.~\ref{fig:cosmologies-prop-z0} for
propagators measured at $z=0$ and Fig.~\ref{fig:cosmologies-prop-z1}
for those at $z=1$.
Remarkably in all cases the simple model given in Eq.~(\ref{eq:prop})
performs at the sub-percent level for all the scales of interest, at
least up to $z=1$.

We now turn to the analysis of power spectrum prediction using
multi-point expansion for all the cosmological models detailed in
Table~\ref{cosmologies}. Figures \ref{fig:cosmologies-power-z0} and
\ref{fig:cosmologies-power-z1} show the measured power spectrum at
$z=0$ and $z=1$ respectively.
Solid blue line is the two-loop model from Eqs. (\ref{eq:tree}-\ref{eq:twoloop}),
solid black corresponds to {\tt halofit} and dashed red to
linear theory. 

At both $z=0$ and $z=1$ the multi-point expansion perform as for the
FID case, that is, it matches the measurements up to $k \sim
\sigmav^{-1}(z)$ (for the given cosmology) at the $\lesssim 2\%$ level. This means, broadly speaking, up to $k =
0.15\kvecMpc$ at $z=0$ and $k = 0.25\kvecMpc$ at $z=1$.
In turn {\tt halofit} works at the $\lesssim 6\%$ level. In some cases,  
however, this departure appears at rather low $k$ (e.g. for low-$\Omega_m$).

In Fig.~\ref{fig:cosmologies-power-las-damas} we concentrate in the performance of our
implementation of the multi-point expansion as a function of redshift,
from low $z$ to high $z$ using {\tt LasDamas} measurements as a benchmark.
{\tt Halofit} seems to perform best around $z\sim 1$ but for smaller
and higher redshifts departs more substantially from the
measurements. For example, at $z=0$ {\tt Halofit} is suppressed
compared to simulations by $5\%$ at $k=0.15\kvecMpc$ and $8\%$ at $
k=0.3\kvecMpc$. Using {\tt Halofit}  as a benchmark for PT
calculations is therefore not accurate enough for present-day
work~\citep{2008PhRvD..77b3533C}. It may be for this reason that
\cite{CarHerSen1206} find a strong effect due to shell crossing at
weakly nonlinear scales, after checking their effective stress tensor
reproduces {\tt Halofit}  for the {\tt LasDamas} cosmology. In fact,
in earlier work \cite{PueSco0908}  characterized the impact of a
non-zero stress tensor and concluded that (extrapolating to the {\tt
  LasDamas} cosmology at $z=0$), shell crossing effects are about
$1\%$ ($2\%$) at $k=0.2~(0.3) \kvecMpc$ (see also \cite{2012JCAP...01..019P}). This justifies ignoring these effects in the calculations we present here at the scales of validity of $\MPTbreeze$ (note that as shown in \cite{PueSco0908} the effects of shell-crossing are strongly redshift-dependent, thus at $z>0$ they are less of a concern).

It is worth stressing here that in all cases the performance of our code is maintained at a few seconds, as in the FID
case described in Sec.~\ref{sec:performance}.

Error bars shown in all power spectrum figures correspond to the expected statistical error
(variance of the band power spectra) for the
given simulation box-size. We have found that using only four realizations to
estimate this error can be very unreliable. One needs at least
ten or more runs to estimate this robustly (we have perform this
testing using our 50 FID runs). Instead we found that the FKP expression
\citep{1994ApJ...426...23F} below matches the resulting ensemble error in our large FID ensemble very well
on the scales we are interested ($k \lesssim 0.4 \kvecMpc$)
\beq
\frac{\sigma_P}{P} = \sqrt{\frac{2}{(4 \pi k^2 \delta k)/k_f^{3}}},
\label{eq:FKP}
\eeq
where $\delta k$ is the particular binning used in the $P$ estimation
and $k_f = 2\pi / L_{\rm box}$.
Hence we use Eq.~(\ref{eq:FKP})
to depict error estimates in
Figs.~\ref{fig:cosmologies-power-z0}, \ref{fig:cosmologies-power-z1}
and \ref{fig:cosmologies-power-las-damas}.

In Figs.~(\ref{fig:power},\ref{fig:cosmologies-power-z0},\ref{fig:cosmologies-power-z1}
\ref{fig:cosmologies-power-las-damas}) we have decided to show, besides {\MPTbreeze} results,
those from tools with comparable functionality such as {\tt
  halofit}. We have nonetheless tested that
RPT, as detailed in \cite{2008PhRvD..77b3533C}, is also accurate at
the percent level but in an broader range
of scales. However this is at the expense of longer evaluation times
(hours as opposed to seconds).

In addition 
we note that 
a fitting formulae for the nonlinear power spectrum that updates  {\tt
  halofit} is available, the Coyote
Universe emulator \citep{2010ApJ...713.1322L}. However, this emulator can only be
used within a specific region of parameter space
which unfortunately exclude all the cosmologies investigated in our paper, even
the one of {\tt LasDamas}. 
This is mostly because the Coyote emulator does not treat $h$ as an
independent variable, but rather as one
fixed by the distance to the large scattering surface as measured by CMB. 
Hence  in Figs.~\ref{fig:cosmologies-power-z0},
\ref{fig:cosmologies-power-z1} and
\ref{fig:cosmologies-power-las-damas} we can only show the estimates
from {\tt halofit}.

\section{Conclusions}
\label{sec:conclusions}

With {\MPTbreeze} we have implemented a renormalized perturbative
approach for the nonlinear power spectrum, the so called multi-point propagators expansion.
We put particular emphasis on the description of the BAO range of scales
from low to high redshift. The main advantage over other techniques
already present in the literature is that the evaluation time is of the
order of $5$-$10$ seconds (in a single CPU), as discussed in Sec.~\ref{sec:performance}. 

Our implementation is based on a phenomenological description of the multi-point
propagators themselves. In particular how their scale-dependence interpolates
between their perturbation theory forms at low-$k$ and their large-$k$
behavior obtained from non-perturbative re-summations. For the
two-point (nonlinear) propagator at late times there is an unambiguous
way to do this interpolation through well-known one-loop results,
this is discussed in Sec.~\ref{sec:twopointpropagator}.

We compared the adopted prescription with propagator measurements in N-body simulations
for seven different cosmological models (listed in Table
\ref{cosmologies} and discussed in more detail below). 
Remarkably we find it always gives a sub-percent agreement for all
scales of interest, as shown in 
Figs.~\ref{fig:prop}, \ref{fig:cosmologies-prop-z0} and \ref{fig:cosmologies-prop-z1}.
In addition, we developed simulations with
independent initial positions and velocities. This allowed us to test for the first time
the full matrix structure of the two-point propagator. We found that
all individual components show an exponential suppression towards small
scales, in turn very well reproduced by our simple prescription (see Fig.~\ref{fig:propcomponents}).

We then moved to the three-point propagator $\Gamma^{(2)}$ that is a function of
triangle configurations. Hence its description is a priori much more
complex. Recently \cite{2012PhRvD..85l3519B} put forward an interpolation scheme that
respects the low-$k$ at arbitrary loop order and the high-$k$ limit. 
Evaluating this would require one-loop calculations of $\Gamma^{(3)}$,
which we find slow to integrate numerically.
Instead, we found that the decay of the three-point propagator when
compared to measurements in N-body simulations is to
a large extent the same as for the two-point case described above
(in turn much faster to evaluate). In particular for configurations 
relevant for $P(k)$ calculations, see Fig.~\ref{fig:gamma2}.

Provided with these results for the multi-point propagators we implemented 
the MP expansion in Eq.~(\ref{eq:gamexpansion}) for the nonlinear power spectrum. In
parallel we developed a set of large N-body simulations of different
cosmological models. Given the number of approximations in 
implementing a practical approach, testing against different cosmological
models seems the only route to establish robust conclusions. The
top-left panel of Fig.~\ref{fig:cosmologies-power-z0} gives an idea of the different slopes and
amplitudes of the linear spectrum simulated in this paper. 

Comparison of $P(k)$ measurements at various redshifts and
cosmological models and our theory modeling is presented in
Figs.~\ref{fig:cosmologies-power-z0}, \ref{fig:cosmologies-power-z1}, \ref{fig:cosmologies-power-las-damas}. Overall we can establish the accuracy of our approach at the
$2\%$ level on weakly nonlinear scales (roughly up to
$\sigmav^{-1}$, given in Eq.~\ref{eq:sigv}). It always improves over standard PT and {\tt
  halofit}. As mentioned
above the evaluation time scale remains around 5 to 10 secs. depending
on redshift, integration accuracy, etc. 

Probably the main disadvantage of our implementation is that the
range of validity does not extend much beyond BAO scales, particularly
a low-$z$. This can be overcome by interpolating weakly nonlinear
scales, as described by our approach, with high-$k$ asymptotic given
by halo models \citep{2011A&A...527A..87V} or $P(k)$ resummation techniques
\citep{2012arXiv1205.2235A}. We leave this extra construction for
future work.

The code used to compute the multipoint expansion presented in this work is
publicly available at {\tt http://maia.ice.cat/crocce/mptbreeze/}.

\section{Acknowledgements}

Some of the simulations presented here are part of the {\tt LasDamas}
collaboration suite  and were run thanks to a Teragrid allocation and
the use of RPI and NYU computing resources.  
Funding for this project was partially provided by the Spanish
Ministerio de Ciencia e Innovacion (MICINN), project AYA2009- 13936,
Consolider-Ingenio CSD2007- 00060, European Com- missions Marie Curie
Initial Training Network CosmoComp (PITN-GA-2009-238356), research
project 2009-SGR-1398 from Generalitat de Catalunya and the Juan de la
Cierva MEC program. RS acknowledges support by grants NSF AST-1109432 and NASA NNA10A171G
and
FB acknowledges support by the French \textsl{Programme National de Cosmologie et Galaxies}.
We thank Pablo Fosalba for comments on the draft.

\bibliography{gexpdensity3}

\appendix

\section{The EdS approximation}
\label{sec:EdS}

The derivation of the multi-point expansion as presented in \cite{2008PhRvD..78j3521B}
assumes that the cosmological model is such that $\Omega_m(\tau)/ f_+(\tau)^2 = 1$, where $f_+(\tau) = d \ln D_+(\tau) /
d \ln a$ and $D_+(\tau)$ is the growing mode solution for the density
contrast of the {\it linearized} equations of motion: 
$\delta(\vk,\tau) = D_+(\tau) \delta_0(\vk)$.

In this approximation the equations of motion simplify
considerably reducing basically to the ones in an Einstein de Sitter background
($\Omega_m=1$) with a factorized linear growth
factor, i.e. replacing the one in EdS, $a(\tau)$, for the
corresponding $D_+(\tau)$  (see \cite{1998ApJ...496..586S}).
Then  at each perturbative order the solution becomes
separable in $\tau$ and $\vk$ with the corresponding most growing term
satisfying $D_n = (D_+)^n$ and the PT kernels reducing to the ones in 
an EdS Universe.

This approximation is known to be very accurate because for
most of the cosmic evolution $\Omega_m\sim 1$. It is usually followed in standard perturbation
theory as well in resummed approaches such as those in \cite{2006PhRvD..73f3519C}, \cite{2007A&A...465..725V}, 
\cite{2007JCAP...06..026M}, \cite{2008ApJ...674..617T} and \cite{2008PhRvD..77f3530M}.

\cite{2008JCAP...10..036P} investigated its limitations by
numerically integrating a system of coupled differential equations 
involving the power spectrum and bispectrum (see also \cite{2009PhRvD..79j3526H}). Here we follow an
alternative approach implementing numerical simulations to directly
address the validity of this approximation, in particular at BAO scales
where we expect our analytical model to yield percent-level predictions.  A different testing using numerical simulations is presented in~\cite{McDTraCon0602}. 

From the theoretical point of view the approach described above amounts to say
that all the (nonlinear) cosmological evolution is set by the linear
growth factor $D_+$. Hence two cosmological models with the same linear
spectrum (normalized at say $z=0$) will show the same nonlinear $P(k)$ at times $\tau_1$ and
$\tau_2$ such that $D_{+,\rm model\,1}(\tau_1)=D_{+,\rm model\,2}(\tau_2)$.

Hence we did the following numerical experiment: provided with our
fiducial LCDM run for $\Omega_m=0.27$ started at $z^{\rm LCDM}_i=49$. We implemented a complementary
CDM simulation ($\Omega_m=1$) started at a time that matched the
growth from the initial conditions in the LCDM case. This yielded $z^{\rm CDM}_i=37$. We chose outputs
for the LCDM at $z=0, 0.3$ and $1$ with corresponding growth from the
initial redshift $D_+=38, 32.78$ and $23.66$. Thus we output the $\Omega_m=1$ CDM run at times
that matched these growth factors, $z=0, 0.159$ and $0.605$.
We then compared the measured power spectrum and two-point propagator in the LCDM run with the
corresponding ``matched growth'' one in the CDM case.
If the approximation were perfect these ratios would be $1$.

\begin{figure}
\begin{center}
\includegraphics[width=0.4\textwidth]{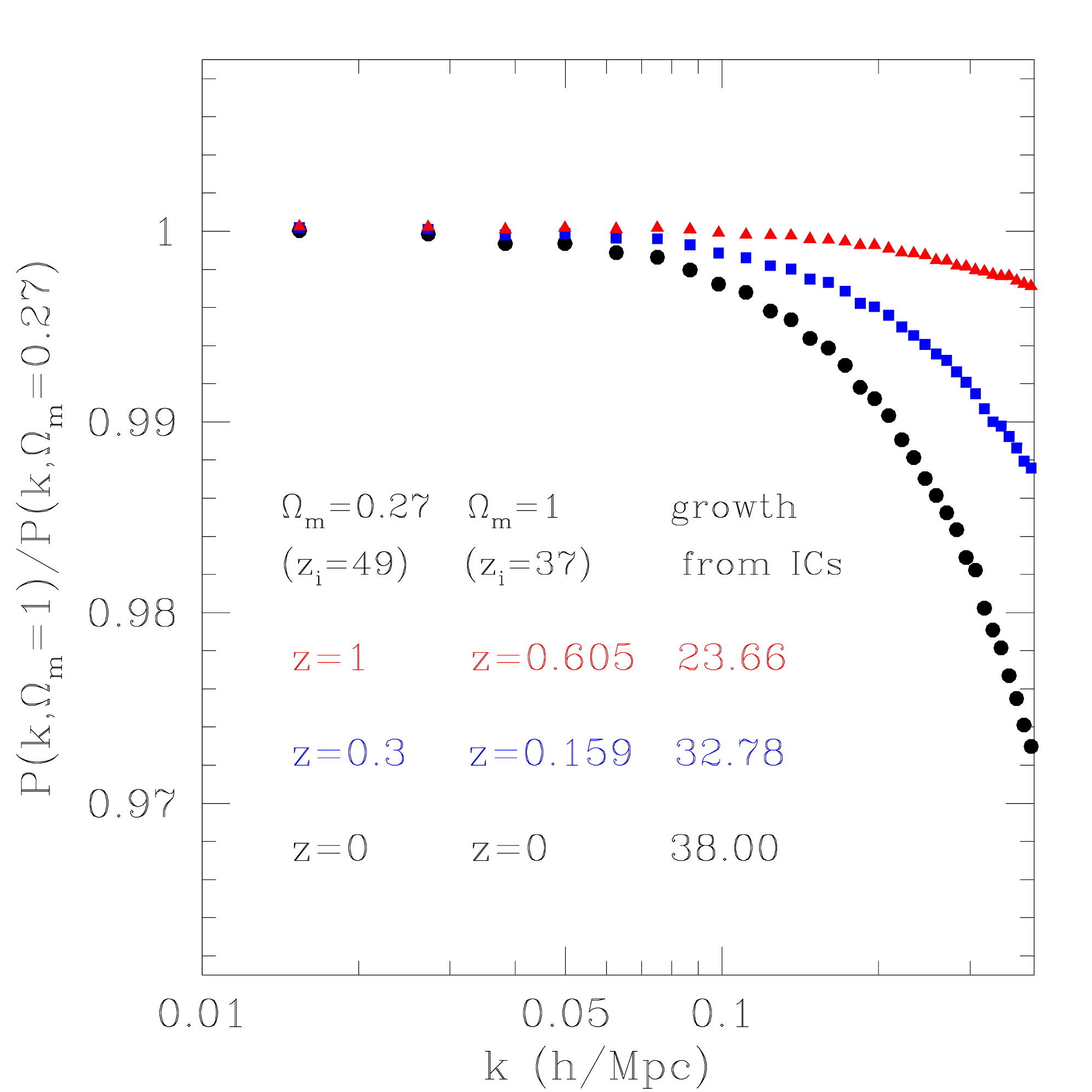}  
\includegraphics[width=0.4\textwidth]{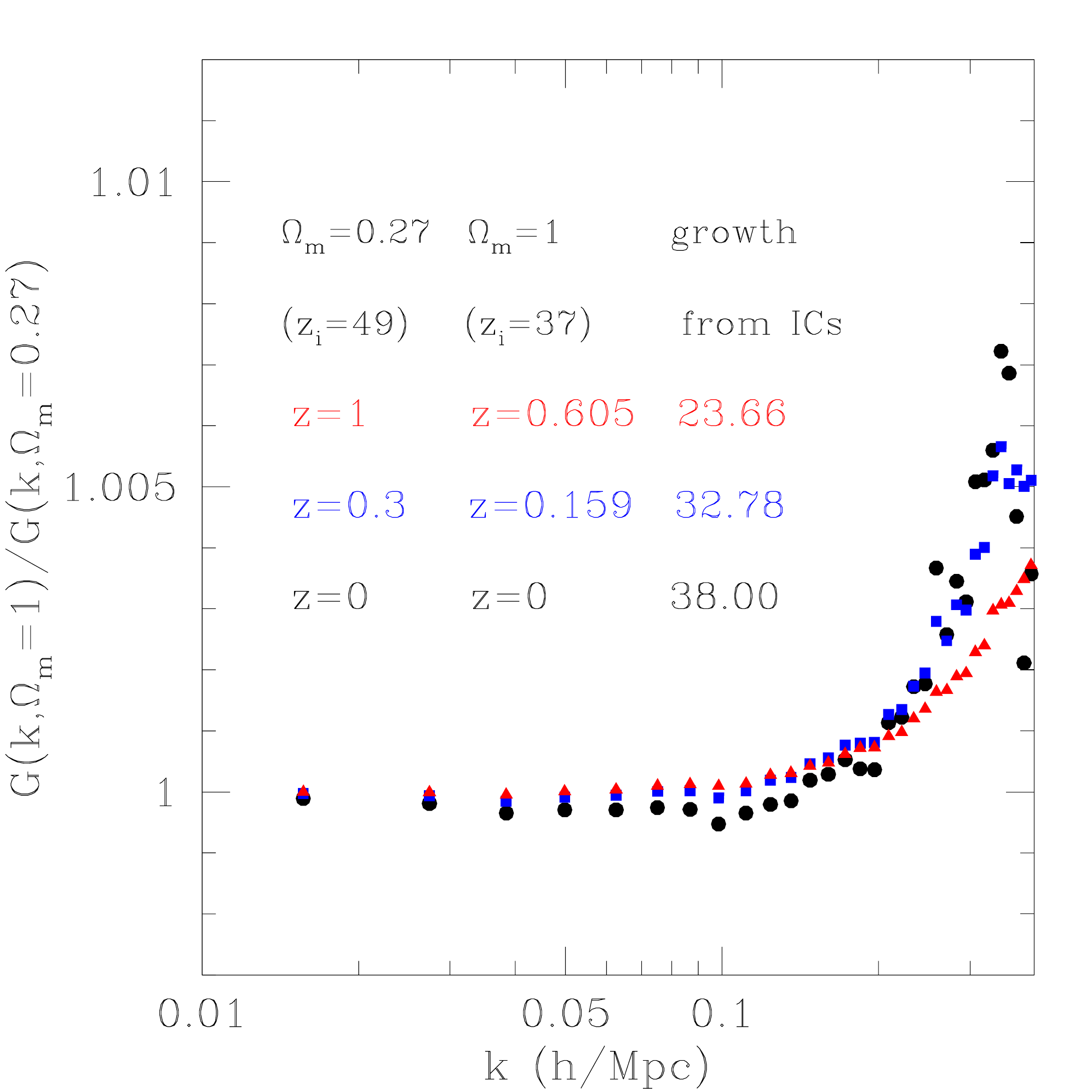}
\caption{An approximate solution of the PT equations of motion (at
  each order) for an arbitrary $\Lambda$CDM background is obtained by solving
  the CDM case with a time dependence given by the linear growth of
  the $\Lambda$CDM model, i.e. changing $a(\tau) \rightarrow D_+(\tau)$. We test this
  approximation here by running ``growth matched'' simulations of similar
  $\Lambda$CDM and $\Omega_m=1$ CDM models (except by their value of $\Omega_m$). The top panel
shows the relative difference in the measured power spectra and the bottom panel 
the ratio of nonlinear propagators. The approximation is weakest at low-$z$ but
it is always never worse than $1\%$ at BAO scales.} 
\label{fig:eds-aprox}
\end{center}
\end{figure}

Results are shown in Fig.~\ref{fig:eds-aprox} for the redshifts
mentioned above. Top panel corresponds to the power spectrum
comparisons and bottom to the nonlinear two-point propagator.

As expected on the largest scales (where linear theory applies) the
ratio is indeed unity. At high $z$ (where $\Omega_m$ is closer to $1$)
the ratio is still unity within $0.2\%$ for $k \lesssim 0.4 \kvecMpc$
In turn, at low $z$ the approximation does break down towards small
scales. Nonetheless BAO scales are mostly unaffected.
For instance, at $k \lesssim 0.2\Mpc$ and 
$z=0.3$ ($z=0$) the CDM power is $\lesssim 0.5\%$ ($\lesssim 1\%$) smaller than the
LCDM one. \cite{2008JCAP...10..036P} finds similar deviations but a
factor of two smaller ($0.5\%$ error at $z=0$ and $k \sim 0.2\kvecMpc$).

In any case the approximation is found accurate at the sub-percent
level on large BAO scales at low-$z$ with extended validity at higher
$z$ and smaller scales.

\end{document}